\documentclass[%
 aip,
 amsmath,amssymb,
 reprint,
]{revtex4-1}

\usepackage{graphicx}
\usepackage{dcolumn}
\usepackage{bm}

\usepackage{hyperref}
\hypersetup{hidelinks}
\usepackage[utf8]{inputenc}
\usepackage[T1]{fontenc}
\usepackage{mathptmx}
\usepackage{etoolbox}
\usepackage[textsize=tiny]{todonotes}
\usepackage{outlines}
\usepackage{bbm}
\usepackage{booktabs}
\usepackage{mathtools}
\usepackage{siunitx}

\DeclareMathAlphabet{\mathcal}{OMS}{cmsy}{m}{n}

\newcommand{\rev}[1]{#1}

\makeatletter
\def\@email#1#2{%
 \endgroup
 \patchcmd{\titleblock@produce}
  {\frontmatter@RRAPformat}
  {\frontmatter@RRAPformat{\produce@RRAP{*#1\href{mailto:#2}{#2}}}\frontmatter@RRAPformat}
  {}{}
}%
\makeatother

\definecolor{darkred}{rgb}{.7,0,0}

\definecolor{darkgreen}{rgb}{.15,.55,0}

\definecolor{darkblue}{rgb}{0,0,0.7}

\begin{document}

\preprint{AIP/123-QED}

\title{Multitask methods for predicting molecular properties from heterogeneous data}

\author{K.~E.~Fisher}
\email{kefisher@mit.edu}
\affiliation{Department of Aeronautics and Astronautics, Massachusetts Institute of Technology}

\author{M.~F.~Herbst}
\affiliation{Mathematics for Materials Modelling, Institute of Mathematics \& Institute of Materials,  École Polytechnique Fédérale de Lausanne, 1015 Lausanne, Switzerland}
\affiliation{National Centre for Computational Design and Discovery of Novel Materials (MARVEL), École Polytechnique Fédérale de Lausanne, 1015 Lausanne, Switzerland}

\author{Y.~M.~Marzouk}
\affiliation{Department of Aeronautics and Astronautics, Massachusetts Institute of Technology}

\date{\today}

\begin{abstract}
Data generation remains a bottleneck in training surrogate models to predict
molecular properties. We demonstrate that multitask Gaussian process regression
overcomes this limitation by leveraging both expensive and cheap data sources.
In particular, we consider training sets constructed from coupled-cluster (CC)
and density functional theory (DFT) data. We report that multitask surrogates
can predict at CC-level accuracy with a reduction to data generation cost by
over an order of magnitude. Of note, our approach allows the training set to
include DFT data generated by a heterogeneous mix of exchange-correlation
functionals without imposing any artificial hierarchy on functional accuracy.
More generally, the multitask framework can accommodate a wider range of
training set structures---including full disparity between the different levels
of fidelity---than  existing kernel approaches based on $\Delta$-learning,
though we show that the accuracy of the two approaches can be similar.
Consequently, multitask regression can be a tool for reducing data generation
costs even further by opportunistically exploiting existing data sources.
\end{abstract}

\maketitle

\section{\label{sec:introduction}Introduction}

    \begin{table*}
    \begin{center}
        \rev{
        \begin{tabular}{ lc|cl }
        \toprule
        \emph{model}    &&& a data-based surrogate for quantum chemical prediction \\[2pt]
        \emph{training} data &&&  possibly heterogeneous data used to construct a model  \\[2pt]
        \emph{testing}  data &&&  data used to evaluate prediction error of a model; disjoint from \emph{training} data  \\[2pt]
        \emph{level of theory} &&& the first principles method used to generate quantum chemical data (e.g., CCSD(T), DFT with PBE, \ldots) \\[2pt]
        \emph{high fidelity} &&& an accurate, expensive method such as~CCSD(T) \\[2pt]
        \emph{low fidelity} &&& a method, such as DFT,
        which is cheaper but less accurate than the \emph{high fidelity} method \\[2pt]
        \emph{task}      &&&  a \emph{model} component trained with homogeneous data
        \\[2pt]
        \emph{primary}  &&& the \emph{task} associated with the main prediction objective of the \emph{model} \\[2pt]
        \emph{secondary} &&& supplemental \emph{tasks} which are included in a \emph{model} to support prediction of the \emph{primary task}  \\[2pt]
        \emph{core (C)} &&& molecular configurations for which \emph{training} data is available for the \emph{primary task} \\[2pt]
        \emph{additional (A)} &&&  molecular configurations for which \emph{training} data is only available for \emph{secondary tasks}\\[2pt]
        \emph{target (T)} &&& molecular configurations for which the \emph{model} will make predictions (i.e., the \emph{testing} set) \\[2pt]
        \bottomrule
        \end{tabular}
        \caption{ Definitions of key terms. }}
     \label{tab:key_terms}
    \end{center}
    \end{table*}

Predicting molecular properties based on first-principle approaches drives
innovation across the physical sciences. However, exact prediction is
infeasible due to the exponential scaling of the Schrödinger equation---the
underlying quantum-mechanical description of materials. Decades of research
have yielded numerous first-principle \rev{methods} to describe electronic
structures approximately, each representing a compromise between computational
cost and prediction accuracy. Coupled cluster (CC) methods are typical examples
of high fidelity models which offer accuracy in exchange for considerable
computational cost. Specifically, CCSD(T)---the ``gold standard'' of
quantum-chemical prediction---suffers from a steep $O(N_{\rev{e}}^7)$ scaling
where $N_{\rev{e}}$ is the number of electrons in the target
system.~\cite{Harding2008} In contrast, density functional theory (DFT) methods
are popular baseline models since they provide adequate accuracy for most
systems while enjoying substantially lower cost (scaling only as
$O(N_{\rev{e}}^3)$).~\cite{dft_survey,dft_ladder,dft_dfa_benchmark,Teale2022}

Traditional workflows force users to settle on a single \rev{quantum chemical method} that is cheap enough to result in manageable cost but still sufficiently high fidelity to give meaningful predictions. However, the advent of modern data-driven modeling has introduced new tools which make it possible to decouple \rev{the step of estimating the properties of unseen molecular systems from the step of running expensive first-principles computations}. Statistical surrogate \rev{models} are trained with predictions from first-principle \rev{methods} and, once trained, can be substituted into workflows, making the prediction step considerably cheaper. In the context of building potential energy surfaces and interatomic potentials, these data-driven approaches have become the de facto standard.~\cite{Smith2019,Dral2020,Goodlett2023,Zaverkin2023} Simultaneously, large-scale data generation is facilitated by sophisticated  workflows for high-throughput calculations~\cite{Curtarolo2012,
Jain2011, Huber2020}  and robust DFT methods inspired by mathematical research.~\cite{Teale2022,cances2022numerical,Herbst2020,Herbst2022,Cances2023} \rev{Increasing amounts of data are openly available~\cite{Ruddigkeit2012,ramakrishnan2014quantum,ocp_dataset,Smith2017} and are, in principle, ready for incorporation into statistical surrogates,} but quality is generally not homogeneous as each dataset’s author has made different \rev{methodological} choices. Consequently, most surrogate models are single fidelity: all training data is generated using the same quantum-chemical method and the same numerical approach. Heterogeneity has thus been an obstacle to opportunistically exploiting the already available data. \rev{In} this
work, we \rev{construct} surrogates
which do not require a single choice between accurate (but expensive) and fast (but crude)
methods, but instead combine the advantages of both. Specifically, we consider
multitask methods which construct a relationship between multiple regression
problems without requiring precise information about the respective accuracy of
each source.~\cite{Bonilla2008,Leen2012}

    \begin{figure*}
        \centering
        \includegraphics[width=0.9\linewidth]{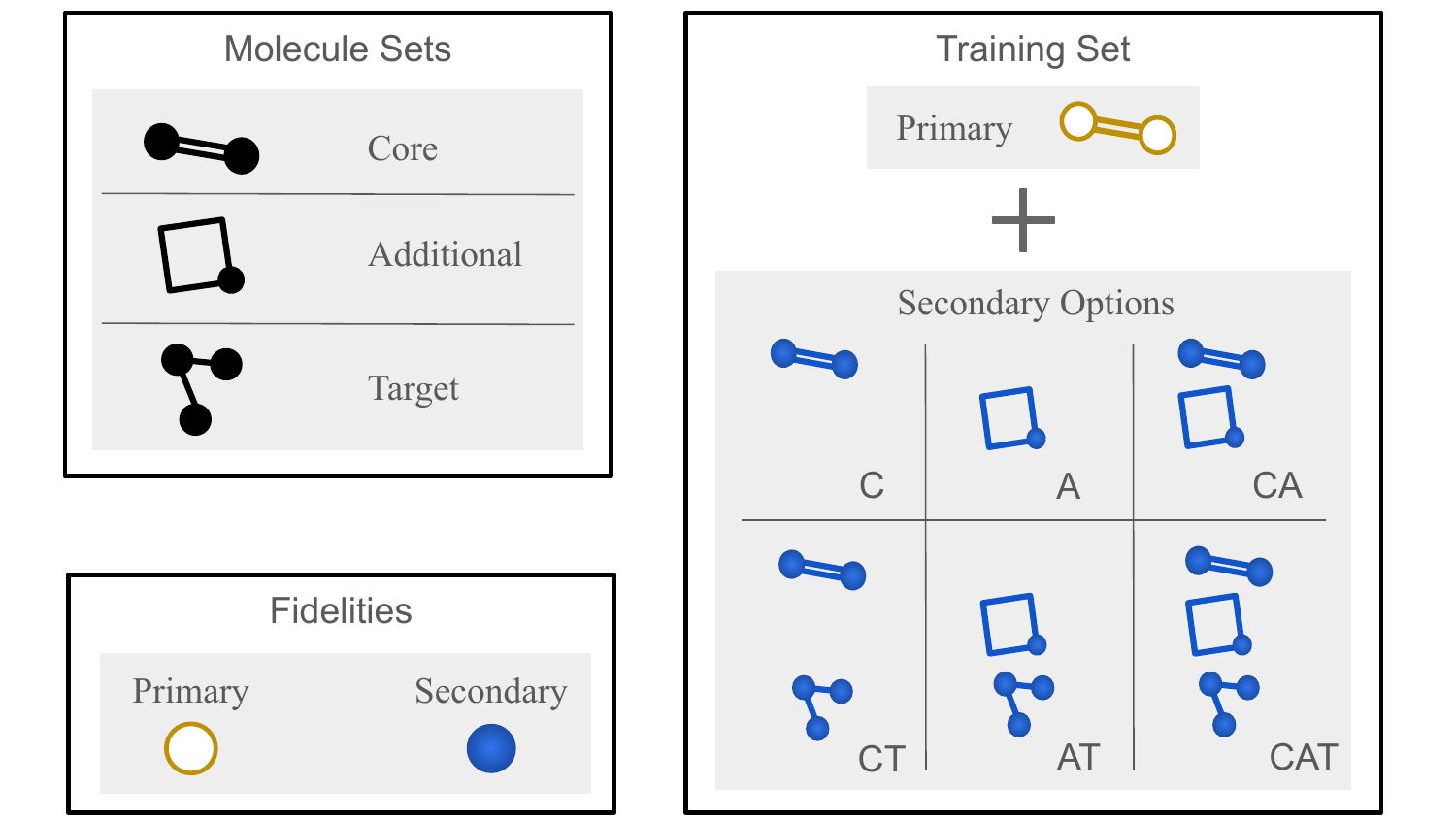}
        \caption{
        \label{fig:molecule_cartoons} Diagram of training data sets compatible with the methods considered in this work. Here, we define three sets of molecules---\textbf{core}, \textbf{additional}, and \textbf{target} systems---and two fidelities---\textbf{primary} and \textbf{secondary}. Our goal is to train a model that will predict our quantity of interest for the target molecules with the accuracy of the primary method. \rev{The molecular space covered by secondary training data is a design choice for which we present six options, labeled C, CT, A, AT, CA, and CAT.  Key terms are defined in Table \ref{tab:key_terms}.}}
    \end{figure*}

\rev{
    Our multitask framework can accommodate a range of heterogeneous training
    set structures. For example, suppose we aim to design a surrogate which
    predicts the three-body energy for any configuration of a water
    trimer with accuracy comparable to CCSD(T). We may only have access to
    $100$ CCSD(T) data points, all corresponding to configurations with HOH
    angles smaller $110$ degrees. Additional data may be provided by including
    a more exhaustive dataset based on DFT simulations with the PBE functional.
    Notably, even though the geometries of the molecular configurations in both
    datasets may not match, they can \textit{both} be fully employed to
    train a multitask surrogate. Moreover, if higher predictive accuracy is
    desired, further training data can be created using yet another
    quantum-chemical method. For example, we may employ BLYP-based calculations
    on all molecular configurations covered by CCSD(T) and PBE, plus some
    configurations which we know we would eventually target for prediction. 
    Remarkably, in this case, all available data can be used without imposing
    \textit{any assumption} about which of the DFT methods (here, BLYP or PBE)
    is more accurate for the considered systems. Internally, a multitask
    surrogate creates a statistical submodel for each of the categories of data
    we supply it with (here, different levels of theory, i.e., CCSD(T), PBE, BLYP).
    In the language used in the rest of the article we call the submodel
    associated to our main prediction objective---the three-body energy at CCSD(T)
    level---the \emph{primary task}.
    The other submodels---here associated to the DFT levels of theory---
    strive to improve the predictive accuracy of the primary task
    and are called \emph{secondary tasks}.
    For a concise overview of these and other terms employed in the article,
    see Table \ref{tab:key_terms}.
}

\rev{
In this work, we will investigate such multitask models and the impact of training set design (i.e., the distribution of training data between the primary and secondary tasks) on prediction accuracy. Fig.~\ref{fig:molecule_cartoons} visualizes the different training set structures which we consider. Secondary task training sets may share some molecular configurations with primary task training sets, include configurations unique to the secondary task, and may even include configurations from the testing data set. An example of the latter case is the use of a BLYP energy prediction at the same geometry where we later apply the surrogate to predict the CCSD(T) energy.} The impact of these choices on inference is under-explored in literature, and multitask \rev{models are} useful tool\rev{s} for investigating the benefits of different training set configurations. Crucially, the multitask framework enables training with a ``dataset of opportunity,'' amalgamated from existing data sets, as an alternative to expending computational time to generate entirely new data.

\rev{In this paper, we construct submodels for each task using }Gaussian process (GP) regression\rev{,} a popular tool for constructing surrogates informed by heterogeneous data. \rev{Several works}~\cite{Pilania2017,Batra2019, Patra2019} have explored the use of multifidelity GP models~\cite{Kennedy2000} trained using two fidelities, namely a high and a low fidelity DFT approach. Alternatively,  $\Delta$-learning may be employed to model the difference between two levels of theory, then by adding that difference to low level predictions of the target to produce high level predictions.~\cite{Ramakrishnan2015,dral2023learning,Batra2019} Though GP regression and the related technique of kernel ridge regression are common choices for fitting such models, in principle, any method of fitting a surrogate may be employed.~\cite{Batra2019,dral2023learning}. \rev{Hierarchical machine learning (hML) models extend the }$\Delta$-learning \rev{framework} to leverage more than two training data sets \rev{by learning and summing} multiple difference models.~\cite{Dral2020,dral2023learning} However, between each pair of training sets, data points must be in some way aligned. For example, in a model of the difference between DFT and CCSD(T), DFT predictions may only contribute to the training set if the corresponding CCSD(T) prediction is also available. Further, when either \rev{hML} or multifidelity GP regression learn from more than two heterogeneous training sets, these sets must be assigned some order---for example, following a hierarchy in accuracy. \rev{Recent work}~\cite{vinod2023optimized, Vinod2023} has proposed an optimized linear combination of submodels as an alternative, but all coefficients in this linear combination must be re-optimized whenever any submodel is altered, and opportunistic data set construction is not fully explored.

Progress towards multifidelity approaches has also been made in the context of neural network surrogates. For example, transfer learning methods train an initial neural network on a larger \rev{secondary} dataset, then re-optimise part of the parameters using a \rev{primary} dataset.~\cite{TOWELL1994119,Fu1989,Smith2019} For the prediction of atomistic properties, the \rev{secondary} dataset might contain millions of DFT calculations while the \rev{primary} might be a smaller set of CCSD(T) predictions. In prior studies, it has been observed that transfer learning can produce higher accuracy than training a network on only one of these datasets.~\cite{Dral2020,dral2023learning} While the framework we discuss in this work is not restricted to GP regression strategies, we will focus exclusively on GP methods. In comparison to neural network approaches, where a rigorous mathematical explanation for the low generalization error of well-trained models is still lacking~\cite{bartlett_montanari_rakhlin_2021,lotfi2022pacbayes}, GP approaches are more amenable to mathematical analysis.~\cite{gpml} This comparative transparency simplifies interpretation of the surrogate's predictive process in the diverse training data configurations we analyze. Additionally, GP methods have the advantage of yielding accurate predictions even for relatively small training set sizes. While the training cost of GP regression may become large as the size of the training set increases, many strategies for efficient scaling have been developed and can be routinely employed.\cite{Quinonero-Candela2005,Liu2018,Wilson2015,Cole2021}

\subsection{\label{summary} Advantages of the Multitask Method}

Our numerical experiments will demonstrate the following advantages of the multitask modeling approach. Key figures illustrating each claim are referenced in parentheses.

\begin{itemize}
    \item Multitask models can achieve target accuracy with less computational cost than Gaussian process regression (Fig.~\ref{fig:water_sets}). 
    \item Accuracy of multitask inference improves as more DFT data is added to the training set. Furthermore, increasing the amount of training data from different secondary methods for the same molecule improves accuracy (Fig.~\ref{fig:ip_levels}).
    \item Unlike the $\Delta$ and multifidelity approaches, multitask methods can model more than two observation data sets without imposing an arbitrary order on the sets (Figs.~\ref{fig:water_delta} and  \ref{fig:ip_levels})
    \item Multitask models have flexible training set requirements. The sets are constructed from multiple levels of theory, and the same molecules do not need to be included for each level (Fig.~\ref{fig:water_sets}).
\end{itemize}

Flexibility is a key advantage of the multitask method compared to a single task or $\Delta$-learning approach. The method is compatible with many training set structures. For instance, a multitask model can be trained with CCSD(T) data for one molecule set and DFT data for a different molecule set, with CCSD(T) and DFT data for the same molecule set, or with some in intermediary case. Models also have the option to train with DFT data for the molecules that we aim to make predictions for at the CCSD(T) level of accuracy. The flexible multitask framework makes it possible to explore the impact of different compositions of a training data set on inference as seen in Fig.~\ref{fig:water_sets}. Consequently, we can determine the balance of expensive high level and cost effective low level data which best suits the problem setting and available data.

In Section \ref{GPR}, we will review prediction with GP regression as well as the $\Delta$ and multifidelity methods before describing the multitask framework. Section \ref{design} provides detail on our approach to testing different training set configurations, and Section \ref{numerical} describes our numerical tests. We demonstrate the performance of the multitask method on two examples: prediction of three-body energy of water trimer configurations and prediction of ionization potential of small organic molecules.

\section{\label{GPR}Statistical Models}

We review the formulation of Gaussian process regression models. Subsections \ref{ss:multiple_observations} through \ref{ss:multitask} discuss variants of GP regression which leverage heterogeneous \rev{data} sets. See references \cite{gpml, Kennedy2000, Leen2012, Bonilla2008, Petter1997} for more detail. 

\subsection{\label{ss:gpr} Gaussian Process Regression}

Our goal is to predict values of some quantity of interest (QoI) $f$, given an input feature \rev{vector
$\bm{X}_{\rev{m}}$ describing the $m^{th}$ molecular configuration.} We train our predictive model using $\{ \rev{(}\bm{X}_{\rev{m}},\bm{Y}_{\rev{m}}\rev{)}\}_{{\rev{m}}=1}^M$ where each $\bm{Y}_{\rev{m}}$ is a noisy observation of $f(\bm{X}_{\rev{m}})$. In our setting, we interpret imperfect first-principle predictions of the QoI as noisy observations of the true quantity. Each $\bm{X}_{\rev{m}}$ is some representation of the electronic structure of a molecular system. For our numerical results, we build $\bm{X}_{\rev{m}}$ using the well established Smooth Overlap of Atomic Potentials (SOAP). \cite{Bartok2010, Bartok2013, Bartok2015} $\bm{Y}_{\rev{m}}$ is the CCSD(T) or DFT prediction of the quantity of interest for the ${\rev{m}}^{th}$ molecular configuration. 

We will assume that each observation is independently perturbed by additive noise which follows identical Gaussian distributions. The corresponding model is 
\begin{eqnarray}
    \label{eq:regression}
    \bm{Y}_{\rev{m}} \ &=& \ f(\bm{X}_{\rev{m}}) \ + \ \varepsilon_{\rev{m}}, \quad \rev{m=1,\dots,M,} \\
    \varepsilon_{\rev{m}} \ &\sim& \ N\left( 0, \ \sigma^2 \right). \nonumber
\end{eqnarray}
Before inference, we choose a mean function $\mu(\bm{X}_{\rev{m}})$ to describe the expected value of $f(\bm{X}_{\rev{m}})$ and a kernel function, $k(\cdot,\cdot)$, to encapsulate our assumptions about the relationship between $f(\rev{\bm{X}_{\rev{m}}})$ and $f(\rev{\bm{X}}_{\rev{m'}})$. Typically, $k(\cdot,\cdot)$ will have hyperparameters that are selected through optimization, then fixed during inference. 

We will employ a Gaussian process prior model for our regression function: 
\begin{eqnarray}
    \label{eq:gp_prior}
    f(\bm{X}) \ \sim \ \text{GP}\Big(\mu(\bm{X}), k(\bm{X},\bm{X}') \Big).
\end{eqnarray}
As a consequence of this prior assumption, any finite collection, $\{ f(\bm{X}_1), \dots, f(\bm{X}_{\rev{M}} \rev{)\}}$, will follow a multivariate normal distribution with mean $\bm{\mu}= \{ \mu(\bm{X}_1), \dots, \mu(\bm{X}_{\rev{M}}) \}$ and covariance matrix $K$ with entries at row $\rev{m}$ and column $\rev{m'}$ given by $k(\bm{X}_{\rev{m}},\bm{X}_{\rev{m'}})$. \rev{It is common practice to center the observation vector and set $\bm{\mu}=\bm{0}$. We retain a more general formulation here to emphasize that it is possible to construct a feature-dependent prior mean}. 

We are interested in inferring the values the \rev{QoI} takes on at target inputs, and we collect these values in the vector, $\bm{f}_*$.  Using this notation, we write the joint distribution of the observations and targets for inference:
\begin{eqnarray}
    \label{eq:joint_prior}
    \begin{bmatrix}
    \bm{Y} \\ \bm{f}_* 
    \end{bmatrix} \  \sim \ N\left( \begin{bmatrix}
        \bm{\mu} \\
        \bm{\mu}_*
    \end{bmatrix}, \ \begin{bmatrix}
    K_{pp}+\sigma^2 I & K_{p*} \\[3pt] K_{p*}^T & K_{**}
    \end{bmatrix}  \right).
\end{eqnarray}
where $I$ is the identity matrix. The specific subscripts of each covariance matrix represent the features that the matrix compares: $p$ corresponds to our primary training data and $*$ indicates the target set. We note that in this single \rev{task} setting, all training data is contained in the primary set, but this will no longer be the case when we consider surrogates informed by heterogeneous data.

The marginal distribution of $\bm{Y}$ is the model evidence, and conditioning $\bm{Y}$ on $\bm{f}_*$ yields the likelihood. To obtain the posterior, we employ the well known formula for conditioning part of a multivariate Gaussian random variable on the rest,~\cite{gpml} finding that 
\begin{eqnarray}
    \label{eq:posterior}
        \bm{f}_* \  \big| \bm{Y} \ \sim \ N\Big( \bm{\mu}_* \ + \ K_{p*}^T(K_{pp} + \sigma^2I)^{-1}\big(\bm{Y} - \bm{\mu}\big), \\ \quad K_{**} \ - \ K_{p*}^T(K_{pp} + \sigma^2I)^{-1}K_{p*} \Big). \nonumber
\end{eqnarray}
Eq.~(\ref{eq:posterior}) gives the probability distribution of our target values, conditioned on our observations. The mean of this distribution is a correction to our prior mean based on the deviations of the observations from that mean. The predicted covariance indicates how much our observations have reduced our uncertainty, provided our assumptions are reasonably aligned to reality.

\rev{\subsection{\label{ss:multiple_observations}Multiple Observation Sets}}

Suppose \rev{we have access to heterogeneous training data which includes predictions of our QoI by multiple first principles methods.} We thus seek an inference model that leverages \rev{all data $\bigcup_{n=1}^N\{ \bm{Y}_{mn} \}_{m\in\mathcal{M}_n}$, where each $n=1,\dots, N$ represents a task
and the corresponding set $\mathcal{M}_n$ contains indices of the molecular configurations available to train a model for task $n$.} A simple ``pooling'' approach \rev{could} put all observations into one vector and proceed with inference via Eq.~(\ref{eq:posterior}). Then, \rev{if $m\in\mathcal{M}_{n}\cap\mathcal{M}_{n'}$, we can interpret $\bm{Y}_{mn}$ and $\bm{Y}_{mn'}$} as independent observations of the \rev{same} QoI, $f(\bm{X}_{\rev{m}})$, for molecular configuration $\bm{X}_{\rev{m}}$. The distinct behavior of each quantum chemical method is treated as a perturbation of some mean value by a realization of the noise. 

\rev{Rather than doing this, w}e seek a model where the distribution of each $\bm{Y}_{\rev{mn}}$ is sensitive to $\rev{n}$. Ideally, we want to map differences in electronic structure methods to differences in their predictions. 
Method-specific features could be a solution, but construction of such features is challenging. One approach involves appending a vector which indicates the method category to each feature. This strategy puts a large burden on covariance hyperparameter optimization since one set of hyperparameters must encode differences between molecular configurations and between quantum chemical methods. Alternatively, to characterize the qualitative difference between methods within the features, we would have to solve the very problem for which we are creating the features in the first place: for instance, we would need to reliably predict CCSD(T) performance given CCSD(T) and DFT data from training molecules as well as DFT data for target molecules. 

Instead, we \rev{will extend the framework of GP regression to model relationships between multiple prediction tasks.}

\rev{\subsection{\label{ss:delta}$\Delta$-Learning}}

$\Delta$-learning performs Gaussian process regression to predict differences, $\Delta_{\rev{m}}$, between observations obtained with two \rev{levels of theory} for a given $\bm{X}_{\rev{m}}$. \rev{In this setting, the observation data available to us allows two interpretations: either as a single-task approach with the observations $\{ (\bm{Y}_{m2}- \bm{Y}_{m1}) \}_{m=1}^M$, or as a two-task approach with data $\bigcup_{n=1}^2  \{ Y_{mn} \}_{m\in\mathcal{M}_n}$. Unless stated otherwise, we will assume the latter.} If we retain our assumption that noise follows an independent, identically distributed Gaussian distribution, we have the model
\begin{eqnarray}
    \label{eq:delta}
    \Delta_{\rev{m}}\rev{(n,n')} \ &\coloneq& \ \bm{Y}_{\rev{mn}} - \bm{Y}_{\rev{mn'}}, \quad \rev{\forall\, m\in \mathcal{M}_n\cap \mathcal{M}_{n'}} \\ 
    \ &=& \ \rev{f_\Delta}\big( \bm{X}_{\rev{m}} \big) \ + \ \varepsilon_{\rev{m}} \nonumber \\
    \rev{f_\Delta} \ &\sim& \ \text{GP}\big( \bm{\mu}^{\rev{\Delta}}\rev{(\bm{X})}, \ k^{\rev{\Delta}}(\bm{X},\bm{X}')  \big).  \nonumber
\end{eqnarray}
where $\rev{n}$ and $\rev{n'}$ indicate two prediction methods. An application of Eq.~(\ref{eq:posterior}) yields a distribution on the regression function for $\Delta\rev{(n,n')}$. Suppose our ultimate interest is to predict $\bm{Y}$ at $\bm{X}_{*}$. Within the $\Delta$-learning framework, we can supply an observation from one method, say $\bm{Y}_{* \rev{n'}}$, to make a prediction for the value from the other method: 
\[  \bm{Y}_{* \rev{n}} \ \approx \ \Delta_{\rev{*}} \ + \ \bm{Y}_{* \rev{n'}}.  \]
Thus, $\Delta$-learning is suited to settings where calculations by method $\rev{n}$ are both more accurate and more time consuming than calculations that use $\rev{n'}$. 

Consider a case where we want to build a model from an existing data set which \rev{consists of CCSD(T) predictions for configurations in $\mathcal{M}_2$ and DFT predictions for configurations in $\mathcal{M}_1$. In the notation introduced by Table \ref{tab:key_terms}, $\mathcal{M}_2$ is the core set, $C$. $\mathcal{M}_1$ may contain any combination of molecules from the $C$, $A$, and $T$ sets. As \eqref{eq:delta} indicates, the training data for $\Delta$-learning will include only molecular configurations in $\mathcal{M}_1\cap\mathcal{M}_2 \subseteq C$. If $\mathcal{M}_1\cap\mathcal{M}_2 = C$, then $\Delta$-learning will be able to leverage more data than the conventional GP regression model described by \eqref{eq:joint_prior} trained on the CCSD(T) data. }Unfortunately, $\Delta$-learning is not-applicable in cases where predictions by different methods correspond to \textit{different} molecular configurations. \rev{When $\mathcal{M}_1\cap\mathcal{M}_2=\emptyset$, it is impossible to train a $\Delta$-learning model.} To \rev{include} a DFT prediction for a molecule $\bm{X}_{\rev{m}}$ \rev{in the training set for} a $\Delta$ model between CCSD(T) and DFT, we also require a CCSD(T) prediction for the same molecule. 

\rev{Hierarchical machine learning (hML)\cite{Dral2020,dral2023learning} extends the $\Delta$-learning approach to $N>2$ by representing the difference between $\bm{Y}_{mN}$ and $\bm{Y}_{m1}$ as}
\begin{eqnarray}
    \label{eq:hml}
    \rev{\Delta^{hML}_m(N,1)} \ \rev{=} \ \rev{\sum_{n=1}^{N-1} \Delta_m(n+1,n),}
\end{eqnarray}
\rev{where each $\Delta_m(n+1,n)$ is defined following \eqref{eq:delta} (with $f_\Delta$ potentially different for each $n$). As before, to ultimately predict, $\bm{Y}_{*N}$, we also require data for $\bm{Y}_{*1}$. Note that the summed difference models are independently trained, and their training sets need not be identical. Thus, an hML model designed to predict at the level of CCSD(T) would not necessarily require CCSD(T) data for every molecular configuration, $m$, in its training set. Rather, each training configuration must belong to \textit{some} intersection of two adjacent levels of theory, i.e., to
$\bigcup_{n=1}^{N-1} \left(\mathcal{M}_n\cap\mathcal{M}_{n+1}\right)$. }

In the following subsections, we consider alternative approaches which model disparities in data sources and have the ability to use all available observations.

\subsection{\label{ss:multifidelity}Multifidelity Method}

As a step toward considering statistical models which train on heterogeneous data, we describe the multifidelity data fusion model.~\cite{Kennedy2000}
The multifidelity approach defines two regression functions: $f_p$\rev{,} representing the predictions from some high fidelity model, and $f_s$, representing a low fidelity model.~\cite{Kennedy2000} We assume the relationships 
\begin{eqnarray}
    \bm{Y}_{\rev{m}p} \ &=& \ f_p\big( \bm{X}_{\rev{m}} \big) \ + \varepsilon_{\rev{m}p} \rev{, \qquad \forall\, m\in\mathcal{M}_p } \nonumber \\ 
    \label{eq:mf_p}
    &=& \  \rho f_s\big( \bm{X}_{\rev{m}} \big) \ + \ \delta_{ps}\big( \bm{X}_{\rev{m}} \big)  \ + \ \varepsilon_{\rev{m}p}, \\ 
    \label{eq:mf_s}
    \bm{Y}_{\rev{m}s} \ &=& \  f_s\big( \bm{X}_{\rev{m}} \big)  \ + \ \varepsilon_{\rev{m}s} \rev{, \qquad \forall\, m\in\mathcal{M}_s }. 
\end{eqnarray}
The term $\delta_{ps}$ captures the difference between the high fidelity model and the low fidelity model, scaled by the hyperparameter $\rho$. This disparity is endowed with a Gaussian process prior, as is the low fidelity regression function, $f_s$\rev{:}
\begin{eqnarray*}
     \rev{\delta_{ps}} \ &\rev{\sim}& \ \rev{\text{GP}\big( \bm{\mu}^{\delta}\rev{(\bm{X})}, \ k^{\delta}(\bm{X},\bm{X}')  \big),}\\
     \rev{f_s} \ &\rev{\sim}& \ \rev{\text{GP}\big( \bm{\mu}^s\rev{((\bm{X}))}, \ k^s(\bm{X},\bm{X}')  \big).} \\
\end{eqnarray*}
The correlation, $\rho$, will be optimized before inference. As before, we
assume that the additive noise terms, $\varepsilon_{\rev{m}p}$ and
$\varepsilon_{\rev{m}s}$, are independent and drawn from identical, centered Gaussian
distributions. Now, suppose we want to predict \rev{our QoI for the target
molecules, $\{ \bm{X}_m : m\in T\}$, at the primary fidelity. Let
$\bm{f}_*$ be the vector of all values in  $\{ f_p(\bm{X}_m) : m\in T\}$.}  We
can determine that the joint distribution of our high and low fidelity
observations and \rev{$\bm{f}_*$ is multivariate normal with mean and
covariance}
\begin{eqnarray}
    \label{eq:mf_cov}
    \rev{\mu}   \ &\rev{=}& \ \begin{bmatrix}
       \rev{\bm{\mu}^s \ + \ \bm{\mu}^{\delta}} \\
        \rev{\bm{\mu}^s} \\
        \rev{\bm{\mu}^s \ + \ \bm{\mu}^{\delta} } \\
    \end{bmatrix} \nonumber \\
    \Sigma \ &=& \ \begin{bmatrix}
        \rho^2K^s_{pp} + K^\delta_{pp} + \sigma^2I   &  \rho K^s_{ps}  &   \rho^2 K^s_{p*} + K^\delta_{p*} \\[5pt]
        \rho  K^s_{sp}                                   &  K^s_{ss} + \sigma^2I  & \rho K^s_{s*}  \\[5pt]
        \rho^2 K^s_{*p} + K^\delta_{*p}            &  \rho K^s_{*s}  &  \rho^2 K^s_{**} + K^\delta_{**}   \\[5pt]
    \end{bmatrix}. 
\end{eqnarray}
where $\sigma^2$ is the variance of the noise distribution. The superscript of each \rev{mean vector $\bm{\mu}$ and} covariance matrix, K, corresponds to the \rev{GP prior} used in its construction, and the subscripts indicate the pairs of features that are compared \rev{by the prior kernel function}. Bayesian inference yields an analytical posterior distribution, analogous to Eq.~(\ref{eq:posterior}).

\subsection{\label{ss:multitask}Multitask Method}

``Multitasking'' refers to a category of methods which consider several
regression tasks and assume some relationship between these tasks. As with
$\Delta$-learning or multifidelity fusion, multitasking can allow us to
incorporate a larger dataset into our inference problem without just pooling
the data into one observation vector. For instance, we can define a regression
problem trained on CCSD(T) data as well as one trained on each density-functional approximations~(DFA) we consider.
By modeling correlation between regression functions, we can use the DFA
regression tasks to support prediction in the CCSD(T) task. 

\rev{Symmetric} variants of multitask modeling treat all tasks equally. For instance, we may assume that all the data are modeled by a single Gaussian process whose covariance kernel is the Kronecker product of one kernel which relates the tasks and another which models relationships between data from the same task.\cite{Bonilla2008,Khatamsaz2023} In many cases, however, the goal is to perform regression for a primary task, and all other tasks \rev{are} considered secondary tasks which provide data that may be useful in learning the primary task. For our own setting, CCSD(T) is a natural choice of data for the primary regression task because we would like to make predictions which match this method in accuracy. Each set of DFT predictions made using a different DFA could inform a secondary regression task which supports the primary task. Thus, we consider an asymmetrical model~\cite{Leen2012} in which secondary tasks are related through the primary task.

\rev{Let $\mathcal{M}_p$ contain the molecular configurations available to train the primary task. Note that following the definitions provided in Table \ref{tab:key_terms}, $\mathcal{M}_p=C$. Likewise, we define $\mathcal{M}_{s_n}$ to be the set of training configurations for the $n^{th}$ } secondary task. As before, we assume that each task has its own regression function
\begin{eqnarray}
    \label{eq:mt_regress}
    \bm{Y}_{mp} \ = \ \ f_p(\bm{X}_m) \ + \ \varepsilon_{mp}, \  &\quad&  \rev{\forall\, m\, \in \mathcal{M}_p,}  \\
    \bm{Y}_{m s_n} \ = \ f_{s_n}(\bm{X}_m) \ + \ \varepsilon_{m s_n}, &\quad&  \rev{\forall\, m\, \in \mathcal{M}_{s_n}}, \\
    &\quad& \forall\, n=1,\dots,\rev{N-1}. \nonumber
\end{eqnarray}
where each noise term is drawn independently from a centered Gaussian distribution with variance $\sigma^2$. We now model a shared structure in the regression functions based on the primary function:
\begin{eqnarray}
    \label{eq:mt_relate}
    f_{s_{\rev{n}}}(\bm{X}_{\rev{m}}) \ &= \ \rho_{s_{\rev{n}}}f_{p}(\bm{X}_{\rev{m}}) \ + \ \delta_{s_{\rev{n}}}(\bm{X}_{\rev{m}}).
\end{eqnarray}
The task-specific component $\delta_{s_{\rev{n}}}$ of the secondary function serves to ``explain away'' behavior that is not captured by the shared component, $f_p$.~\cite{Leen2012} Both the correlation parameter $\rho_{s_{\rev{n}}}$ and the task-specific component aim to mitigate negative transfer---learning behavior from secondary tasks that is not representative of the primary task. We make the prior assumption that $f_p$ and $\delta_{s_{\rev{n}}}$ for $\rev{n}=1,\dots,\rev{N-1}$ follow Gaussian process distributions, each with a mean and kernel function which we will specify\rev{:}
\begin{eqnarray}
    \label{eq:multitask_prior1}
    \rev{f_p} \ &\rev{\sim}& \ \rev{\text{GP}\big( \bm{\mu}^p\rev{(\bm{X})}, \ k^p(\bm{X},\bm{X}')  \big),} \\ 
    \label{eq:multitask_prior2}
    \rev{\delta_{s_n}} \ &\rev{\sim}& \ \rev{\text{GP}\big( \bm{\mu}^{s_n}\rev{(\bm{X})}, \ k^{s_n}(\bm{X},\bm{X}')  \big).}
\end{eqnarray}

As was the case in subsections \ref{ss:gpr} through \ref{ss:multifidelity}, the
first step toward inference \rev{of $\bm{f}_*$} with the multitask method is to
determine the joint distribution of observations and targets. \rev{Since all
secondary tasks are modeled following \eqref{eq:mt_relate}, we can easily
represent this distribution for arbitrary $N$.} Note that the task-specific
component \rev{$\delta_{s_n}$} of each \rev{secondary} task is modeled as
independent of the others\rev{, $\delta_{s_{n'}}$ for $n\neq n'$}. Then, the
covariance between observations due to the tasks-specific components is a block
diagonal matrix\rev{, which we denote as $K^\delta$. The kernel function
$k^{s_n}$ is applied to construct the $s_n^{th}$  diagonal block. Since
the off-diagonal blocks correspond to the covariances of task-specific specific functions for
different tasks, they are all zero. We note that the number of rows
and columns of this matrix both equal $\sum_{n=1}^{N-1} |\mathcal{M}_{s_n}|$.}
Let $\bm{R}$ be a diagonal matrix with dimensions equal to the number of
\rev{secondary} observations, and let each diagonal element be the correlation
hyperparameter of the corresponding task. The \rev{mean and }covariance of the
\rev{multivariate normal} joint distribution is 
\begin{eqnarray}
    \label{eq:mt_cov}
    \rev{\mu}   \ &\rev{=}& \ \begin{bmatrix}
       \rev{\bm{\mu}^p} \\
        \rev{R\bm{\mu}^p + \bm{\mu}^s} \\
        \rev{\bm{\mu}^p } \\
    \end{bmatrix} \nonumber \\
    \Sigma \ &=& \ \begin{bmatrix}
        K^p_{pp} + \sigma^2I   &  K^p_{ps} \bm{R}  &   K^p_{p*}  \\[5pt]
        \bm{R} K^p_{sp}                  &  \bm{R} K^p_{ss} \bm{R} + K^\delta +\sigma^2I  &  \bm{R} K^p_{s*}  \\[5pt]
        K^p_{*p}                         &  K^p_{*s} \bm{R}  &  K^p_{**}  \\[5pt]
    \end{bmatrix}. 
\end{eqnarray}
\rev{S}uperscript\rev{s} indicate \rev{the GP prior structures used to create a vector or matrix, and}  subscripts indicate the features that are compared \rev{by a matrix. In both cases, $s$ indicates evaluations for \emph{all} secondary tasks, $n=1,\dots,N-1$.} Inference proceeds as described in Eq.~(\ref{eq:posterior}).

\subsection{\rev{Comparison of Methods for More than Two Fidelities}}

Asymmetric multitasking assumes a relationship between regression functions---given by Eq.~(\ref{eq:mt_relate})---which is structurally reminiscent of the multifidelity fusion model---Eq.~(\ref{eq:mf_p}). Both assume that the regression function of one dataset is related to the regression function of another by a scaling parameter, $\rho$, and an additive disparity function, $\delta$.

For only one secondary model, $s_1$, the relationships between the regression functions posited by both methods are mathematically equivalent. However, asymmetric multitasking diverges more significantly from the multifidelity method when more than two models are considered. In the multifidelity approach, model hierarchy is reflected in a nested relationship between the functions,
\begin{eqnarray*}
    f_{p}(\bm{X}) \ &=& \ \rho_{s_{1}} f_{s_{1}}(\bm{X}) \ + \ \delta_{s_{1}}(\bm{X}), \\
    f_{s_1}(\bm{X}) \ &=& \ \rho_{s_{2}} f_{s_{2}}(\bm{X}) \ + \ \delta_{s_{2}}(\bm{X}), \\
    &\vdots& \\
    f_{s_{\rev{N}-1}}(\bm{X}) \ &=& \ \rho_{s_{\rev{N}}} f_{s_{\rev{N}}}(\bm{X}) \ + \ \delta_{s_{\rev{N}}}(\bm{X}), 
\end{eqnarray*}
whereas in the asymmetric approach, all functions share the same relationship to the primary function
\begin{eqnarray*}
        f_{s_1}(\bm{X}) \ &=& \ \rho_{s_1} f_p(\bm{X}) \ + \ \delta_{s_1}(\bm{X}), \\
        &\vdots& \\
        f_{s_{\rev{N}}}(\bm{X}) \ &=& \ \rho_{s_{\rev{N}}} f_p(\bm{X}) \ + \ \delta_{s_{\rev{N}}}(\bm{X}). 
\end{eqnarray*}
Thus, compared to multifidelity fusion, the asymmetric multitasking approach more naturally accommodates a reasonably large number of regression tasks and relates each closely to the primary task of interest. 

Multitasking also holds a similar advantage over \rev{the hML extension to} $\Delta$-learning. \rev{ The indices $n=1,\dots,N$ suggest an order of the training sets, and the difference models in \eqref{eq:hml} correspond to consecutive pairs of training sets. } Such an order may be arbitrary, for example, if our data sets are generated by multiple DFAs with similar structure. In Section \ref{design}, we will describe different compositions of a training data set which the multitask framework allows us to test.

\section{\label{design}  Mapping Heterogeneous Data to Tasks}

        \begin{table*}
    \begin{center}
        \rev{\begin{tabular}{ lc|cl }
        \toprule
        $\mathcal{C}(t) \quad  t\in\{p,s_1,\dots,s_{N-1}\}$   &&& intersection between \emph{core} molecules and training molecules for task $t$ \\[5pt]
         $\mathcal{A}(t) \quad  t\in\{p,s_1,\dots,s_{N-1}\}$   &&& intersection between \emph{additional} molecules and training molecules for task $t$ \\[5pt]
        $\mathcal{T}(t) \quad  t\in\{p,s_1,\dots,s_{N-1}\}$   &&& intersection between \emph{target} molecules and training molecules for task $t$   \\ 
        \bottomrule
        \end{tabular}
        \caption{ Definition of functions which return sets of molecular configurations corresponding to primary task, p, and secondary tasks, $s_1, \dots, s_{N-1}$. Note that by definition, $\mathcal{C}(p)$ is the entire core set, $C$, and $\mathcal{A}(p)=\mathcal{T}(p)=\emptyset$.}}
     \label{tab:molecule_set}
    \end{center}
    \end{table*}

In our numerical examples, we train inference models using data for multiple molecular configurations computed with multiple quantum chemistry methods. The multitask method leaves many questions about the particulars of the training data set up to the researcher. \rev{In particular, it is necessary to define tasks and map molecules to the training set of each task.}

\subsection{\rev{Selecting Tasks}} 
\rev{The primary task is chosen according to our prediction goal. We aim}  to demonstrate that by training with data from multiple quantum chemistry methods, we can construct an inference model \rev{which has} prediction accuracy comparable to CCSD(T) \rev{for} less computational expense than if we trained a single task model on CCSD(T) alone. \rev{Thus, we will test our surrogates using a set of CCSD(T) predictions for some target molecules belonging to a set we call $T$. Our primary task must be informed by CCSD(T) predictions for a different set of molecules which we refer to as our core training set, $C$. See Table \ref{tab:key_terms} for an overview of our key definitions.}

\rev{ We consider two approaches to mapping between levels of theory and tasks.
In the first (subsections \ref{sec:num1} and \ref{sec:num3}), the training data
of our primary task will consist of direct computations of our QoI at CCSD(T) level,
and each secondary task will be trained by DFT predictions of the QoI obtained
with different DFT functionals.}

\rev{The second approach (Subsection \ref{sec:num2})
we call the  ``multitask $\Delta$'' option.
In that setting, the levels of theory include CCSD(T) and DFT computations performed with two
different DFAs. The \textit{difference} between CCSD(T) and the first DFA will be the
primary task while the \textit{difference} between the two DFAs informs the secondary
task.  Note that the $\Delta$-learning method could be interpreted as the
single task counterpart to the multitask $\Delta$ option.}

\subsection{\rev{Assigning Molecular Configurations to Tasks}}

\rev{ We are specifically interested in the flexibility which the multitask
framework provides towards assigning molecular configurations to secondary
training sets. Our pool of available molecules is divided into Core ($C$),
Additional ($A$), and Target ($T$) subsets to systematically explore the impact
of the molecular space covered by the secondary training data and its
relationship to our primary training and testing data. The $A$ set contains all
molecular configurations that are not part of the primary training or testing
set. In our setting, these are configurations which are not our targets for inference and for which high level CCSD(T)
predictions are unavailable. A major difference between the multitask framework and $\Delta$-learning is the
former's ability to leverage this $A$ set.}

\rev{Fig.~\ref{fig:molecule_cartoons} shows the six main secondary training set
structures considered in this work. We name the training set structures using
combinations of the letters $C$, $A$, and $T$. Inclusion of a letter in a
structure name indicates that at least one molecular configuration from the
corresponding subset is included in the secondary training data set. For
example, a CT training structure in our setting would contain CCSD(T)
predictions for the C set of molecules, DFT predictions for some subset of the
C set, and DFT predictions for some subset of the T set. The data requirement
of implementing a multitask method with this CT training set structure is
comparable to the data demands of a $\Delta$-learning model. However, a crucial
distinction is that the multitask approach does \emph{not} require a DFT
prediction for \emph{every} molecular configuration in the C and T
sets. Furthermore, the multitask model can make use of an additional $A$
training set for which CCSD(T) predictions are unavailable.
$\Delta$-learning models cannot be
implemented for such a training set structure. }

\subsection{\rev{Multiple Secondary Tasks}}

\rev{The multitask framework can accommodate an arbitrary number of secondary
tasks and allows the observational data used to train each task to correspond
to non-identical sets of inputs. We will use the same definition for the $C$,
$A$, and $T$ molecular sets regardless of the total number of tasks, $N$. In
this context, when a training structure is labeled CAT, this name indicates
that at least one configuration from each of these sets is present in the
training set for at least one secondary task. To the best of our knowledge,
most work which explores training set design for multifidelity,
$\Delta$-learning, or hML models focuses on optimizing the size of the training
set for each source. Questions of whether different secondary sources of
information should explore the same or new areas of molecular space are
underexplored in literature. For example, if our secondary training data
includes DFT predictions at the PBE level for CO$_2$ configurations and at the
BLYP level for ONH$_4$ configurations, we aim to determine how much
our model will improve if we fill in the blanks (i.e., add PBE level
predictions for ONH$_4$ and BLYP level predictions for CO$_2$
versus adding data on previously unseen molecules). Such questions
are investigated in Subsection~\ref{sec:num3}.
}

\begin{figure}
    \begin{tabular}{c}
        \includegraphics[width=0.9\linewidth]{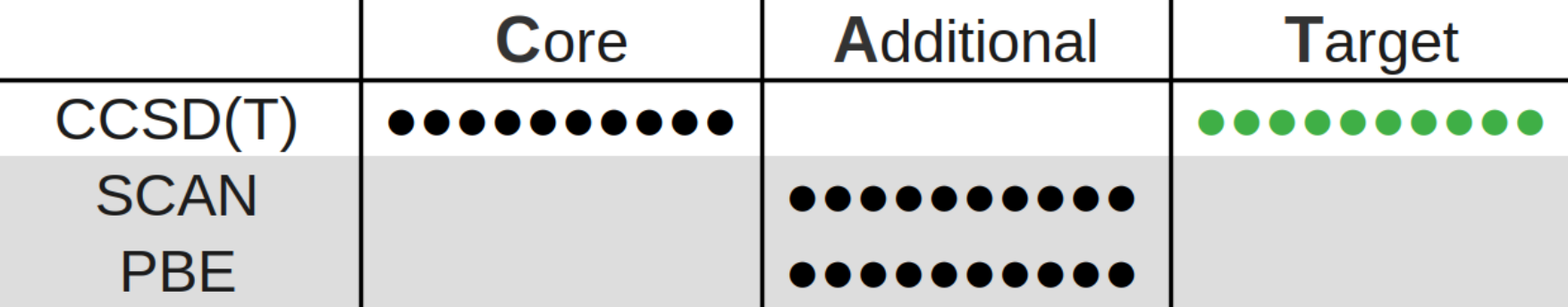} \\[6pt]
        \includegraphics[width=0.9\linewidth]{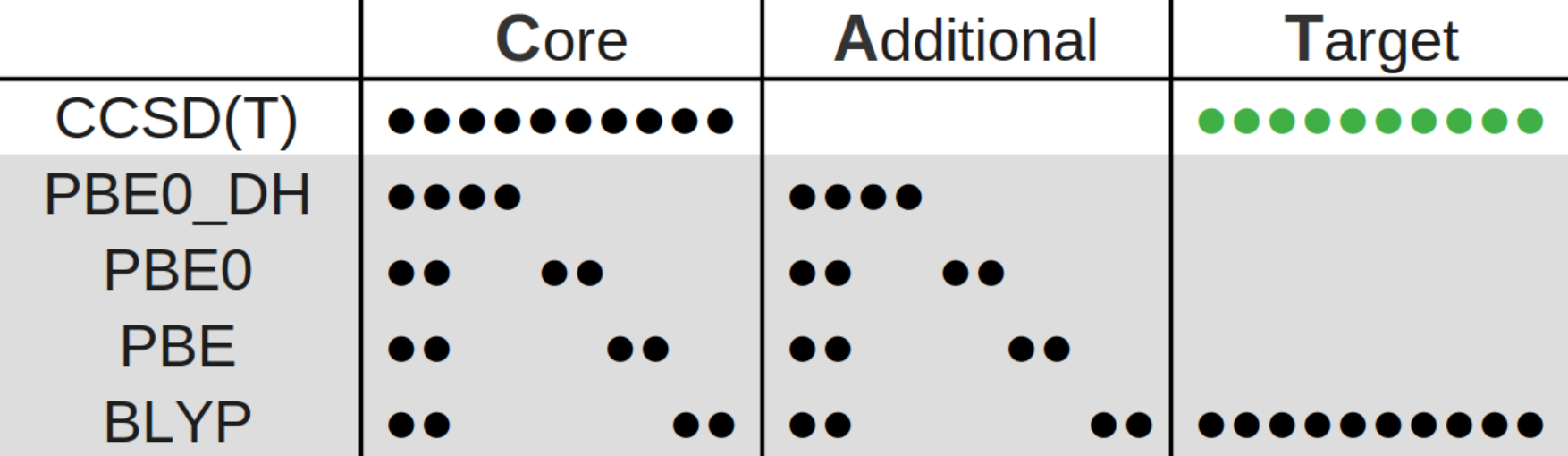} \\[6pt]
    \end{tabular}
    \caption{\label{fig:5level} Example data set structures. $C$, $A$, and $T$ are sets of molecular configurations. $C$ contains all configurations that the primary task, CCSD(T), is trained on. $T$ contains the Target configurations; our goal is to estimate their properties with the accuracy of CCSD(T). The testing set is colored green. All secondary tasks are highlighted in gray. Configurations in set A can only be used for training secondary tasks. The vertical alignment of dots indicates that multiple tasks have access to the same molecular configurations.}
\end{figure}

\rev{
To make this discussion more tractable, we introduce notation which allows us to compare the sets of molecules used to train each task. Table \ref{tab:molecule_set} defines three set-valued functions: $\mathcal{C}$, $\mathcal{A}$, and $\mathcal{T}$. To illustrate this notation further, 
Figure~\ref{fig:5level}  provides a visual representation of two possible training set structures.
Each line in the figure corresponds to a different task,
here defined by the level of theory employed during data generation.
CCSD(T) informs the primary task, and different DFT functionals correspond to each secondary task (shaded in grey).
Black dots represent
training data, and their location indicates whether molecules from a given set
are available for a particular task. Vertically aligned dots indicate the
presence of the same molecular configuration in the training set for multiple
tasks. For example, consider the top plot of Figure~\ref{fig:5level}: SCAN and PBE are the secondary tasks, $s_1$ and $s_2$, and we have:} 
\begin{eqnarray*}
    \rev{\bigl |\mathcal{C}(s_n)\bigr |} \ &\rev{=}& \ \rev{0,} \quad  \rev{ n\in{1,2}} \\
    \rev{\frac{\bigl|\{\mathcal{A}(s_1)\cap \mathcal{A}(s_{2})\}\bigr |}{|A|}} \ &\rev{=}& \ \rev{1,}    
\end{eqnarray*}
\rev{ where $|\cdot|$ returns the number of elements in a given set, and $A$ is the entire `additional' set. The training data for both secondary tasks only includes molecules from the A set, so this plot represents an A test case. The lower plot represents a CAT data structure where $\forall\, 1\leq n<n' \leq 4$ and}
\begin{eqnarray*}
     \rev{\frac{\bigl|\{\mathcal{C}(s_n)\cap \mathcal{C}(s_{n'})\}\bigr|}{|C|}} \ \rev{=} \ \rev{\frac{\bigl|\{\mathcal{A}(s_n)\cap \mathcal{A}(s_{n'})\}\bigr|}{|A|}} \ \rev{=} \ \rev{\frac{2}{10}.}    
\end{eqnarray*}

\subsection{\label{ss:target0}\rev{Training with Target Molecules}}

\rev{We stress that whenever $\mathcal{T}(s_n)\neq\emptyset$, the secondary training data includes low level predictions of at least some molecules which we will use to test the multitask model. In Fig.~\ref{fig:5level}, the testing molecules---target molecules at the primary fidelity---are represented in green. The black dots in the third column of the lower plot indicate that BLYP predictions for the test molecules are included in the training set. }

\rev{
All lower level predictions for target molecular configurations which are
available at training time can be used to obtain a multitask model. If
new lower level T predictions become available after this initial training,
a retraining is needed to use this data. Note that a $\Delta$-learning
model can leverage such additional T data without any retraining. However,
each target data point will only contribute to a single prediction of the
$\Delta$-learning framework. In contrast, retraining the
multitask model buys the benefit of this additional data point for \emph{all future}
predictions. Furthermore, sparse update techniques such as
the Sherman-Morrison formula offer efficient alternatives
to a full retraining of the multitask model.
}

\section{\label{numerical}Numerical Examples}

\subsection{Numerical Setup}

We consider two case studies. In the first, the quantity of interest is the three-body (3-b) interaction energy of randomly selected water trimer configurations. The 3-b interaction term is isolated by subtraction of all 1-b and 2-b energies from the full energy of the trimer. We selected water because it is a simple molecule with unique properties which make it crucial for life and a popular subject for research.~\cite{Chen2017,Dasgupta2021,Gillan2016} The second case study considers the prediction of the ionization potential of small organic molecules. Ionization potential is known to be challenging for DFT to capture accurately, so it is insightful to test the ability of inference models to predict this quantity.

\subsubsection{Data Generation}

\rev{Multitask models can be trained on heterogeneous data sets constructed
opportunistically from multiple sources without the need to specify a predefined
accuracy hierarchy (apart from singling out the primary task).
Our test cases were selected to explore this flexibility by tapping mostly
on previously published data sets supplemented with only few additional computations.}

For our first test case---inference of 3-b interaction energy---we consider 5986 water
trimer configurations and corresponding CCSD(T) calculations.
\rev{These} were randomly
selected from 45,332 configurations available from
\href{https://github.com/jmbowma/q-AQUA}{https://github.com/jmbowma/q-AQUA}~\cite{Yu2022}
\rev{to obtain training data sizes approximately matching our second test case.}
\rev{To illustrate that even such truncated datasets can be employed
to yield reasonable predictive qualities with multitask models,
we enrich this data set again by running a small number
of cheap density-functional theory simulations:}
\rev{on 2992 of these configurations we employ Psi4~\cite{Smith2020}
to compute the 3-b interaction energies
at PBE\cite{PBE} and SCAN\cite{SCAN} level.}
The former was selected for its popularity
and the latter because its description of
intermediate range dispersion has been found to contribute to accurate
prediction of water energy differences~\cite{Chen2017, Dasgupta2021} Following
recent work on dispersion corrections,~\cite{Price2021} we paired the PBE
approximation with a D3 correction using finite damping and the SCAN
approximation with a D3 correction using zero damping. Both DFT and CCSD(T)
calculations employed a standard counterpoise correction to alleviate basis set
superposition error.~\cite{Huang2006}

The second test case is informed by the ionization potential of 3,165 molecular configurations, combinations of 479 different molecules and 7 configuration options.~\cite{Duan2023} Configurations are subselected~\cite{Duan2020} from the ANI-1 dataset~\cite{Smith2017} and contain only elements from the set $\{H,C,N,O\}$.
We compute~\cite{Duan2021} ionization potential for these configurations using CCSD(T) as well as secondary tasks corresponding to data from four density functionals: PBE0\_DH,~\cite{PBE0_DH} PBE0,~\cite{PBE0} PBE,~\cite{PBE} and BLYP.~\cite{BLYP} 

Quantification of cost in our numerical results refers to the cost of data set generation. The cost of inference is negligible next to quantum chemistry calculations. We base our cost model for each method on the average runtime of Psi4 computations for ten representative systems. For more details about the cost model, see Appendix \ref{cost}. \rev{We make the code required to reproduce our results and figures openly available as described in Statement \ref{DA}. Similarly, the first principles data as well as surrogate prediction data are openly accessible along with the cost of training and applying surrogate models.}

\subsubsection{Features}

We require features which can distinguish between molecular systems to serve as inputs, $\bm{X}_i$, to our Gaussian process models. Kernel functions use features to model the relationships between quantities of interest. As noted in Section \ref{ss:gpr}, in our experiments we consider Smooth Overlap of Atomic Positions (SOAP) features, which have been successfully used to fit interatomic potentials in a Gaussian process based approach.~ \cite{Bartok2010, Bartok2013, Bartok2015} SOAP features are constructed based on the neighborhood of each atom in a system, so challenges arise when two systems contain different numbers of constituent atoms of elements. Multiple strategies have been proposed for constructing ``global'' features to capture entire systems.~\cite{De_2016} The simplest approach is to take an average over the SOAP features for each atom in the system. Since this approach is computationally efficient and works well in practice, we apply it in our test cases. We chose SOAP parameters following experimentation and the advice of literature. ~\cite{Deringer2021, Musil2021} More details are available in Appendices \ref{SOAP} and \ref{features}.

\subsubsection{\label{ss:h_opt}Hyperparameter Optimization}

Squared exponential kernels are used within all Gaussian process priors in our experiments\rev{:}~\cite{Petter1997} 
\begin{eqnarray*}
    \rev{k(\bm{X},\bm{X}')}  \ \rev{=} \ \rev{v \ \exp\left( \frac{-(\bm{X}-\bm{X}')^T(\bm{X}-\bm{X}')}{2\ell} \right) }
\end{eqnarray*}
These kernels introduce mean, variance ($v$), and length scale ($\ell$) hyperparameters in addition to the multitask correlation ($\rho$). We use the mean of our training data for each task as that task's prior mean, calculated at training time. To estimate the remaining parameters, we construct an optimization data set containing $150$ randomly selected molecular configurations which we subsequently exclude from our training and testing data sets. We assume here that we have access to predictions of our quantities of interest from each quantum chemistry method for each of these $150$ molecular configurations, but when this assumption does not hold, a conventional GP regression model may be used to fill in missing data. 

We estimate correlation parameters, $\rho$, using Pearson's correlation coefficient between each secondary task optimization data set and the corresponding primary task data. Popular belief in literature holds that correlation parameters require a detailed optimization procedure,~\cite{dral2023learning} but we find that these easy to obtain estimates work reasonably well in practice. A more detailed optimization is left for future work. We estimate $\ell\in\mathbb{R}$ by maximizing its log-marginal likelihood. This approach is easily extended to multifidelity and multitask cases by treating $Y_{ij} - \rho_j Y_{i1}$ as the $i^{th}$ observation.~\cite{Forrester2007} For the organic molecules example, we also use a maximum likelihood estimate for $v$ while for the water trimer example, we find better results from using an estimate of $v$ which minimizes mean absolute prediction error. \rev{See Appendix \ref{opt} for a comparison of MLE and minimum MAE estimates as well as a more detailed discussion of the resources required for hyperparameter optimization.}

\subsubsection{\label{ss:target}\rev{Isolating the Impact of Target Training Data}}

\rev{ Consider a molecular configuration, $\bm{X}_m$. We are interested in contrasting the accuracy of a multitask model trained using a CA set and a model trained on a CAT set which is identical except for the inclusion of a DFT prediction for $\bm{X}_m$. The Target set may contain a large number of molecular configurations, and we cannot make a fair comparison if we simultaneously include DFT predictions for all Target molecules in the training set. To facilitate comparison, whenever $|\mathcal{T}(s_n)|>0$ for $n=1,\dots,N-1$ we train a new multitask model for each element of $\mathcal{T}(s_n)$. Each model will train on only one configuration from the $T$ set, and we will test that model by making predictions for the same configuration. We report the mean absolute error (MAE) over all these models and compare this value to the MAE of the predictions of a CA trained model for the same set of targets.} 

\rev{We implement this retraining procedure to investigate the impact of
training set design. Retraining is \emph{not} a prerequisite for leveraging all
molecules in $\mathcal{T}(s_n)$. All lower level data for target molecules
which is available at training time can be simultaneously incorporated into one
multitask model. See Subsection \ref{ss:target0} for further discussion. }

\subsection{Three-Body Interaction Energy of Water Trimer}

\subsubsection{\label{sec:num1}Multitask Performance}

    \begin{figure}
        \begin{tabular}{cc}
        \includegraphics[width=0.9\linewidth]{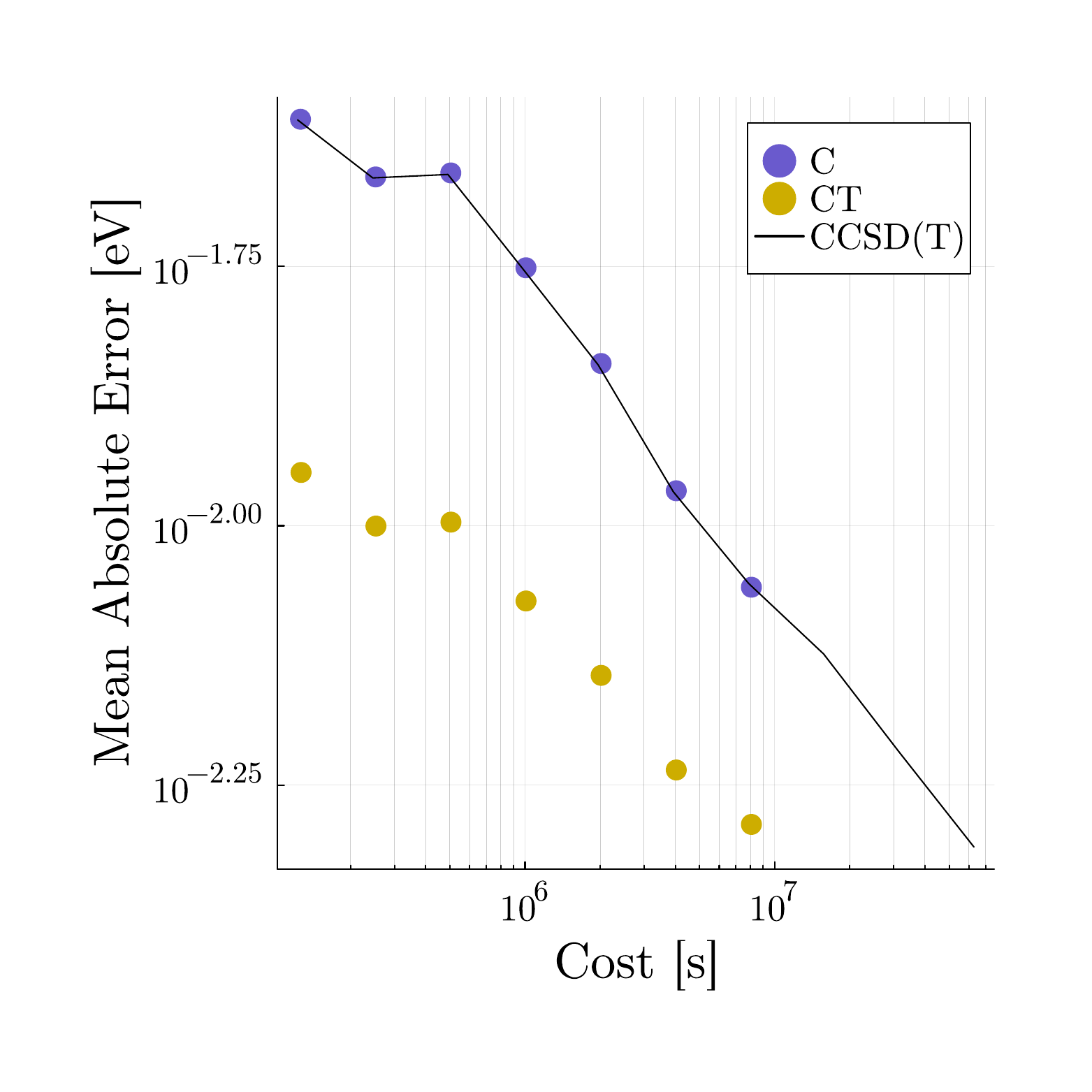}  \\[-25pt]
        \includegraphics[width=0.9\linewidth]{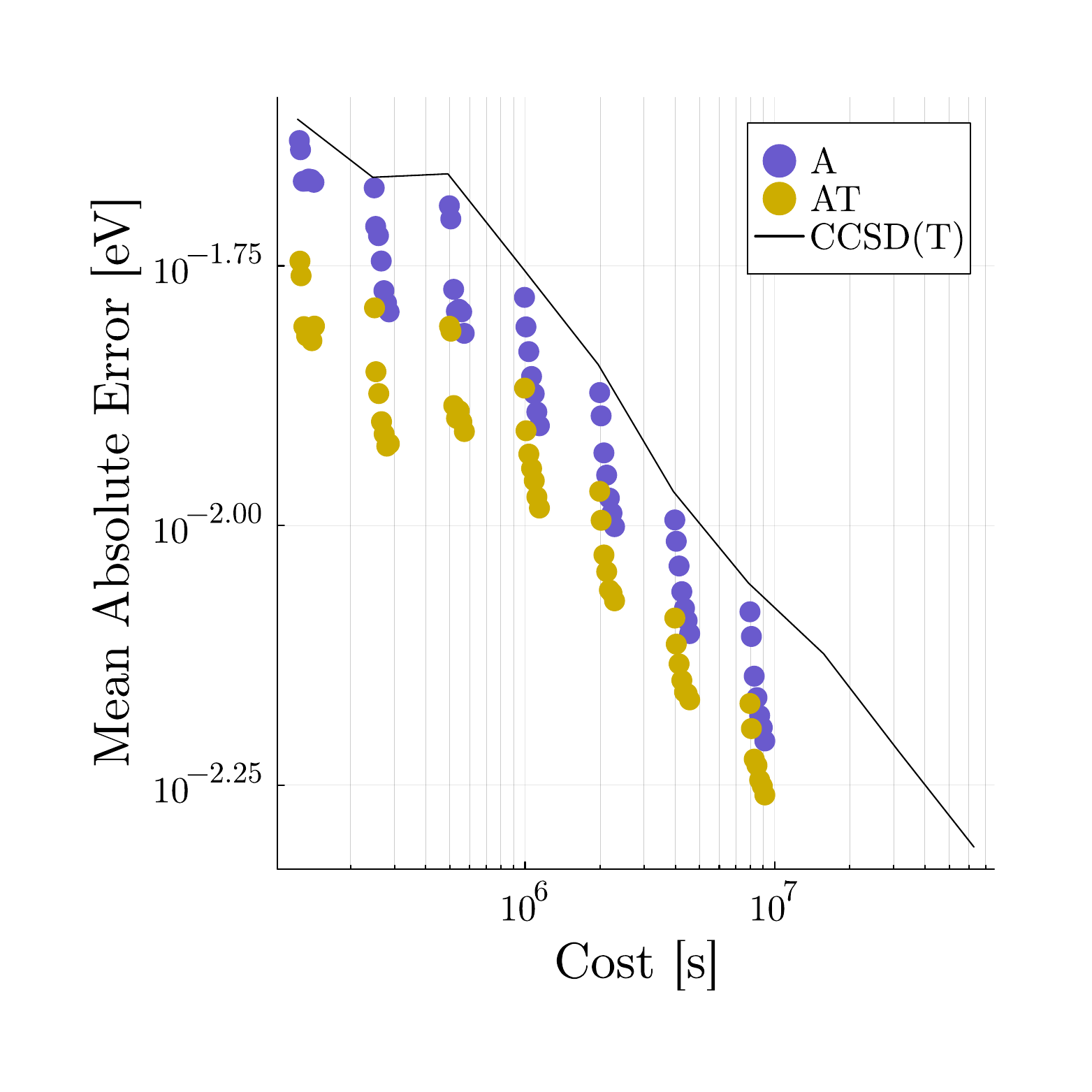} \\[-25pt]
        \includegraphics[width=0.9\linewidth]{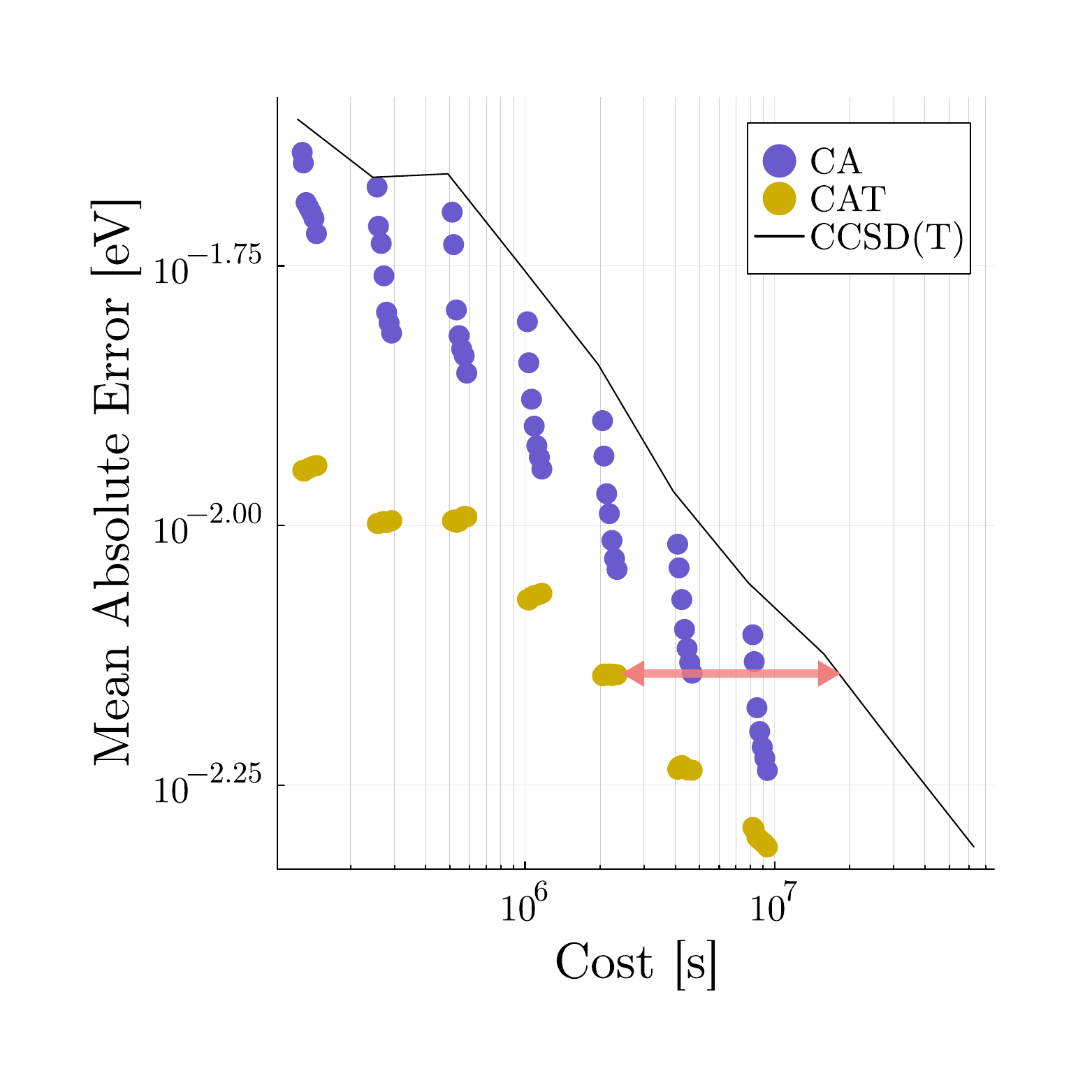} \\[-25pt]
        \end{tabular}
        \caption{\label{fig:water_sets} \textbf{Water Trimer Case.}  Comparison of the MAE of the multitask method for a range of training set constructions (scatter points) to the MAE of a GP regression model trained only with CCSD(T) data (black line). Predictions by the multitask method are informed by CCSD(T) and SCAN. The legends of the plots indicate the structure of the secondary training set according to the system presented by Fig.~\ref{fig:molecule_cartoons}. \rev{The red arrows emphasizes the order of magnitude improvement in data generation cost achieved by CAT multitask models, quantified in Table \ref{tab:magnitude}.}}
    \end{figure}

         We first compare the performance of multitask models with only one secondary task \rev{ ($N=2$)} to the performance of a traditional GP model \rev{ ($N=1$)} . Fig.~\ref{fig:water_sets} shows the error incurred by several implementations of each model when predicting the three-body (3-b) energy of water trimer configurations.  We used the SCAN functional to generate the secondary task data\rev{.} Pearson's correlation coefficient between the primary and secondary task \rev{training data} is $0.997$, indicating strong positive correlation. The three subplots of Fig.~\ref{fig:water_sets} cover six secondary training set configurations (C, CT, A, AT, CA, and CAT), labeled within the legends.

         All models were tested on a target set of 320 molecular configurations. \rev{For multitask models, we consider seven possible sizes of the core set: $|C|\in \{5,10,20,40,80,160,320\}$. We use the same seven core sets as well as sets with $|C|\in \{640,1280,2560\}$ to train the corresponding single task \mbox{($N=1$)} models. Performance of this single task reference is marked with a black line on each subplot. Larger $|C|$ is considered in the single task case to demonstrate the cost required for a these models to reach the accuracy of the best performing multitask models. We choose the additional training sets for multitask models so that
        $|A|=r|C|$ where $r$ is allowed to take on each value in $\{0.5,1,2,3,4,5,6\}$. Whenever core data is included in the secondary training set, we only consider $\mathcal{C}(s_1)=C$. Thus, we consider a total of $7$ different training set sizes for the C and CT structures and $49$ different training set sizes for the A, AT, CA, and CAT structures. } Each point on \rev{each subplot of Fig.~\ref{fig:water_sets}} represents a model with a different training set size. The x axis indicates the cost of each model's training set\rev{; details of our cost model are provided in Appendix \ref{cost}. T}he y axis gives the average \rev{value of} mean absolute error \rev{obtained from six models trained on a given training set size and structure. For each of these six implementations, available molecular configurations are randomly assigned} to the C, A, and T sets \rev{without replacement}. 

        For every \rev{secondary training set structure} except C, the multitask method outperforms the single task reference\rev{.} Including T data from DFT in the training set leads to better results than corresponding cases without T in every example. In particular, training with secondary T data can enable the multitask method to exhibit an accuracy comparable to a GP model that is an order of magnitude more expensive.  \rev{Table \ref{tab:magnitude} provides representative examples. The data points corresponding to the last line in the table are marked with arrows on Fig.~\ref{fig:water_sets}. This analysis of MAE demonstrates that each statistical model can achieve chemical accuracy. For insight into prediction for individual molecules, we report that the best performing multitask models produce predictions with $\geq 95\%$ correlation to the CCSD(T) target data.}

    \begin{table}
    \begin{center}
        \rev{\begin{tabular}{ cc|cccc }
        \toprule
         Accuracy [meV] &&&  & Cost [millions of seconds] &   \\[3pt]
         MAE      &&& Single Task & CA & CAT  \\
         \midrule
         $14$  &&& $2.0$  & $0.58$ &  $0.13$  \\
         $7$  &&& $16.0$  & $4.5$  &  $2.0$   \\
        \bottomrule
        \end{tabular}
         \label{tab:magnitude}
        \caption{ \textbf{Water Trimer.} Data generation cost to achieve the threshold of accuracy given in the left column.}}
    \end{center}
    \end{table}

    \begin{figure*}
        \begin{tabular}{cc}
        \includegraphics[width=0.4\linewidth]{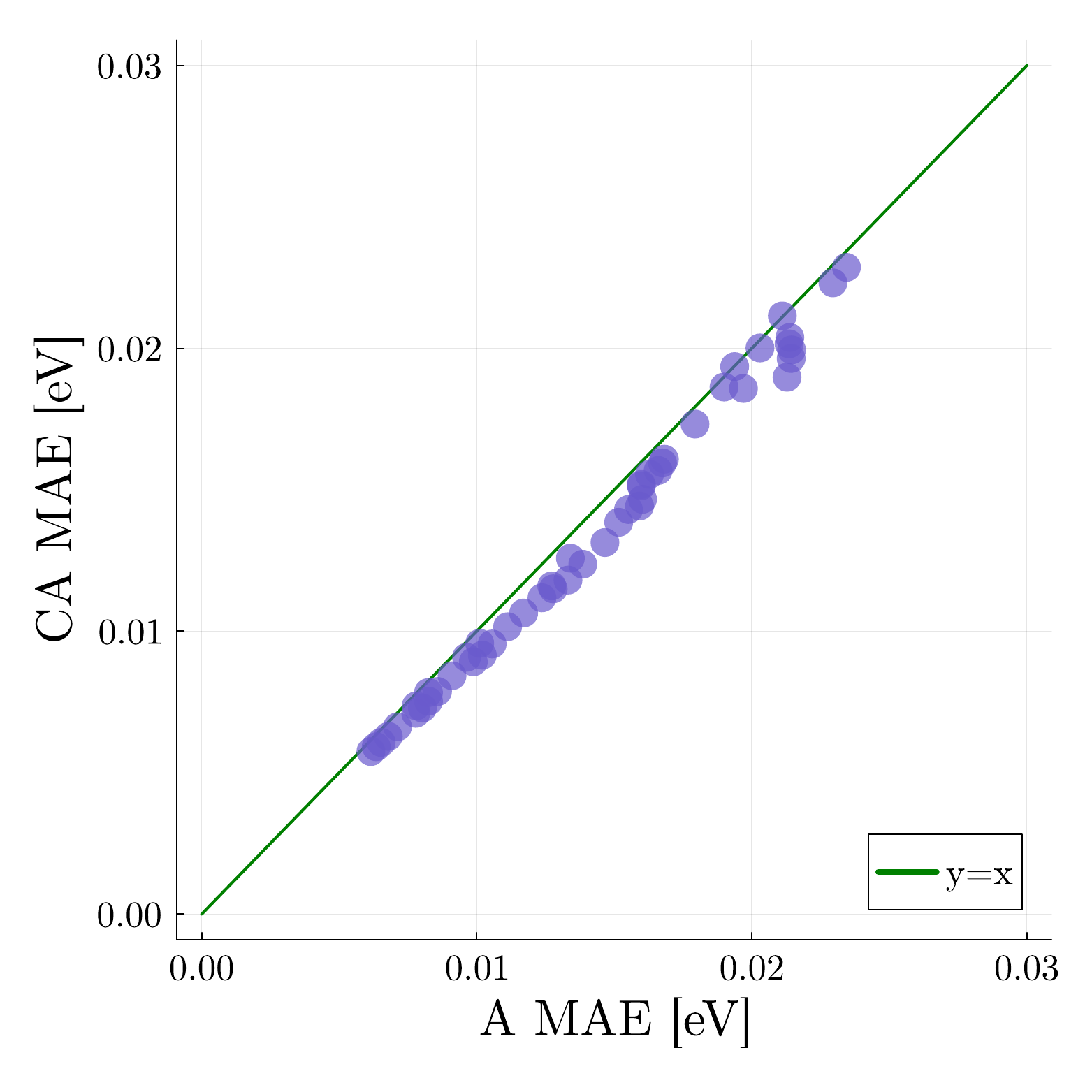}  &
        \includegraphics[width=0.4\linewidth]{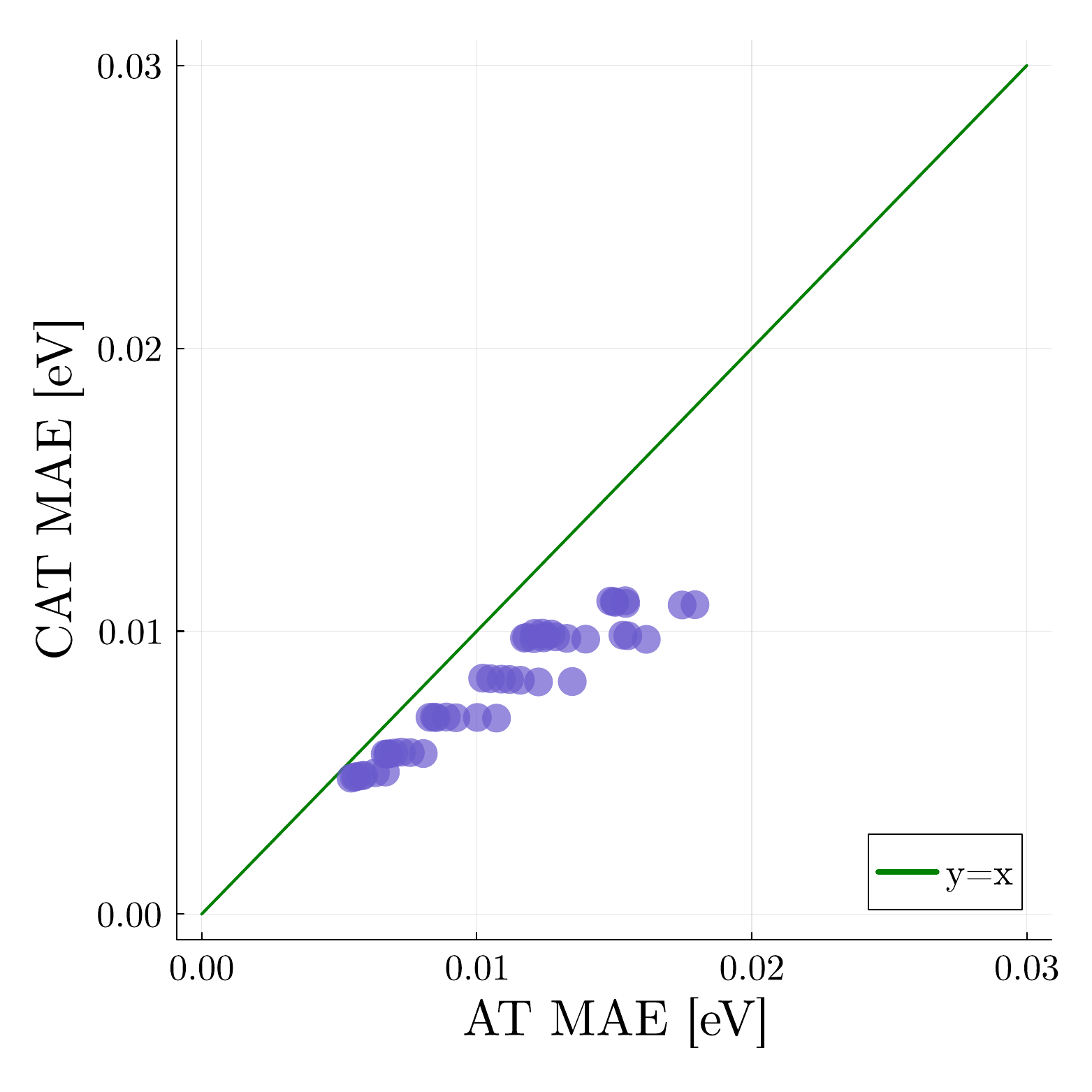} \\
        (a) & (b) \\
        \end{tabular}
        \caption{\label{fig:flexibility} \textbf{Water Trimer Case.} Effect of including Core and Additional data \rev{(defined in Table \ref{tab:key_terms})} in secondary training sets. Each indigo point represents a comparison between two multitask models where one model uses a training set which is a subset of the other. In (a), this comparison is between A and CA training configurations, and in (b), we compare AT and CAT configurations. For each pair of models compared, both the primary training data and the secondary data from the A set are identical. The $y=x$ line is plotted in green for reference. }
    \end{figure*}

    While the C results demonstrate no significant improvement on the single task model, both A and CA results show steady improvement as the size of the A set increases. \rev{We can distinguish the impact of increasing $|A|$ from the impact of increasing $|C|$ in Fig.~\ref{fig:water_sets} because CCSD(T) is significantly more expensive than DFT. Thus, every jump in $|C|$ leads to a significant increase in cost while jumps in $|A|$ for fixed $|C|$ produce a comparatively slight increase to cost. The positive correlation between $|A|$ and performance whenever $\mathcal{T}(s_1)=\emptyset$} suggests that for the multitask method to be beneficial, secondary training data must cover molecular space that is not included in the CCSD(T) training data. We mention that it is possible that the behavior of C models would improve if we consider task dependent features or different relationships between task regression functions. We do not investigate these alternatives further here; instead, we focus on the usual setting where secondary DFT data for additional systems is cheap to obtain.
    
    CT and CAT models are the best-performing multitask models, demonstrating very similar MAE per cost to each other. Furthermore, unlike the AT model, the CAT model does not show improvement as the size of the A set increases. It seems that for our setting, the combination of secondary C and T training data is so valuable to the model that any benefit from \rev{the A set} is dwarfed. We can hypothesize that when the model has access to \rev{training }data for the Core molecules from \rev{both} the primary and secondary tasks, it implicitly learns information about the difference between the tasks. Given secondary data on the Target, it can leverage the captured relationship between tasks to make an accurate prediction for the Target on the primary fidelity.

    Fig.~\ref{fig:flexibility} provides more insight to the relative roles of the training data sets and to the inherent flexibility of the multitask framework. While Fig.~\ref{fig:water_sets} demonstrates that multitask models \rev{outperform single task models for several training set configurations,} it does not directly compare the \rev{efficacy} of the A and CA \rev{training structures}. Such a comparison will illuminate the resilience of the multitask model to the loss of points in the training set which are common to primary and secondary tasks. We emphasize that this resilience is an advantage of the multitask method in comparison with $\Delta$-learning---where no training is feasible without access to secondary C data. Each point on subplot (a) of Fig.~\ref{fig:flexibility} compares the average MAE of two multitask models: one with a CA secondary training set, the other with an identical A set but no secondary C data. As before, we are inferring the 3-b energy of water trimer configurations with only one secondary task. The closeness of the points to the reference $y=x$ line shows that there is minimal effect on accuracy when the secondary C data is excluded from the training set. In fact, the median improvement in accuracy from using the CA training model instead of an A training model model is $0.0007$ eV. By comparison, the median improvement in accuracy from combining DFT data from the A set with CCSD(T) training data---that is, switching from a single task approach to a multitask method trained on an A set---is $0.004$ eV.
    
    Subplot (b) of Fig.~\ref{fig:flexibility} also compares models trained with and without C secondary data in the case where \rev{$\mathcal{T}(s_1)\neq\emptyset$ for all models}. While the AT models perform well, there is a penalty to losing access to secondary C training data when secondary Target data is also available. The median improvement in accuracy when switching from an AT model to a CAT model is $0.002$ eV, a third of the $0.006$ eV median improvement incurred from switching from a single task approach to an AT multitask model with the same CCSD(T) training cost. As we remarked previously, when the model trains on secondary C data, it likely gains information about the difference between the primary and secondary tasks. This information is most useful to the model when it also has access to secondary predictions for the Target molecules.

    \subsubsection{\label{sec:num2}Comparison with $\Delta$-learning}

    \begin{table}[h!]
    \begin{center}
        \begin{tabular}{ cccc }
        \toprule
         & CCSD(T) [eV] $\ $ & SCAN [eV]  $\ $ &  PBE [eV] \\
        \midrule
        absolute mean & $0.0205$ & $0.0223$ & $0.0225$  \\
        absolute median & $0.0126$ & $0.0144$ & $0.0155$ \\
        standard deviation & $0.0300$ & $0.0316$ & $0.0332$ \\
        \bottomrule
        \end{tabular}
        \end{center}
        \begin{center}
        \begin{tabular}{ ccc }
        \toprule 
         & CCSD(T)$-$SCAN [eV] $\ \ $& CCSD(T)$-$PBE [eV] \\
        \midrule
        absolute mean &  $0.0021$  & $0.0037$ \\
        absolute median &  $0.0013$ & $0.0018$ \\
        standard deviation & $0.0033$ & $0.0057$ \\
        \bottomrule
        \end{tabular}
        \caption{ \textbf{Water Trimer case.} Summary statistics for \rev{training data.}}
     \label{tab:diff_stats}
    \end{center}
    \end{table}

    \begin{figure}
        \begin{tabular}{cc}
        \includegraphics[width=0.9\linewidth]{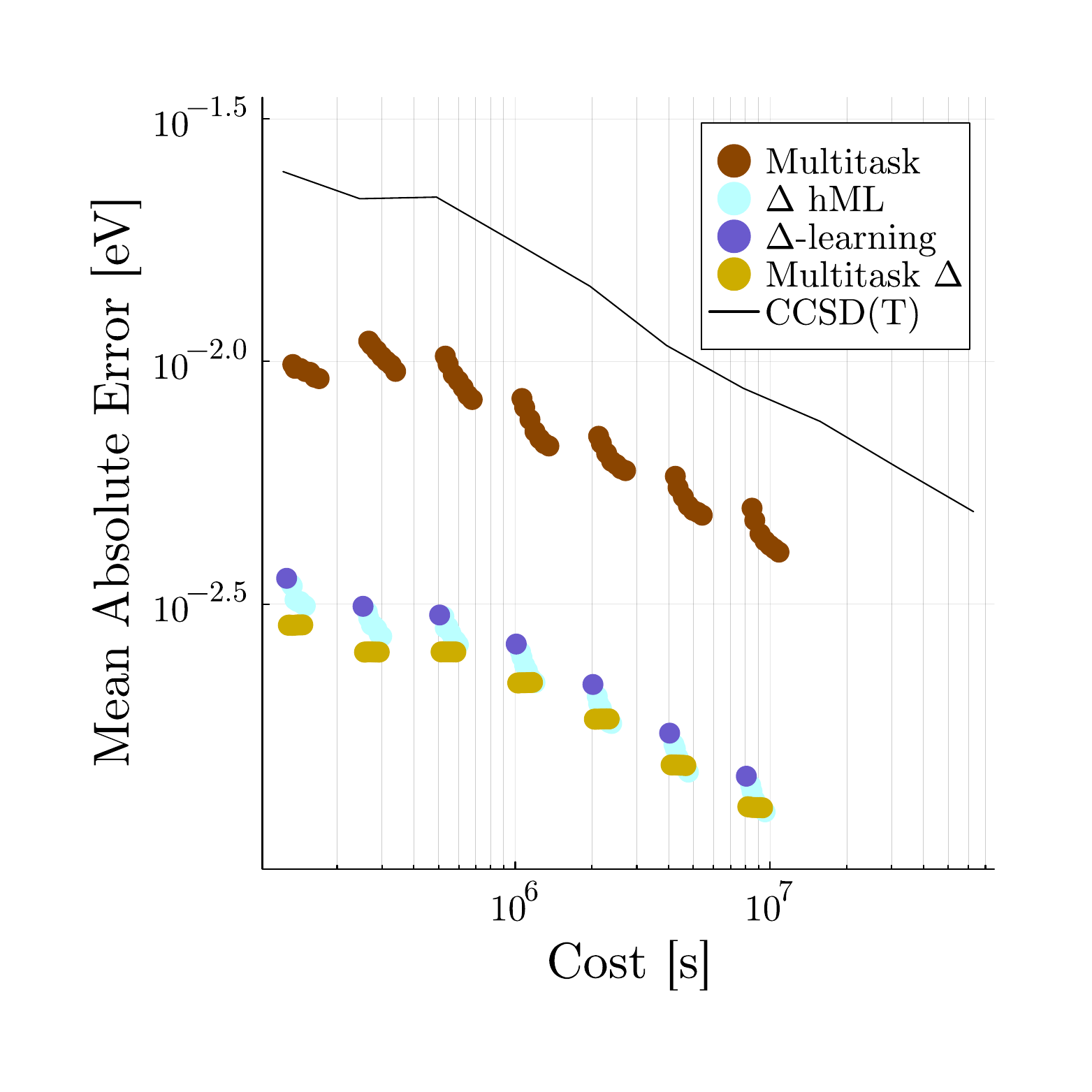}  \\[-15pt]
        (a) Estimation of CCSD(T)-PBE\\
        \includegraphics[width=0.9\linewidth]{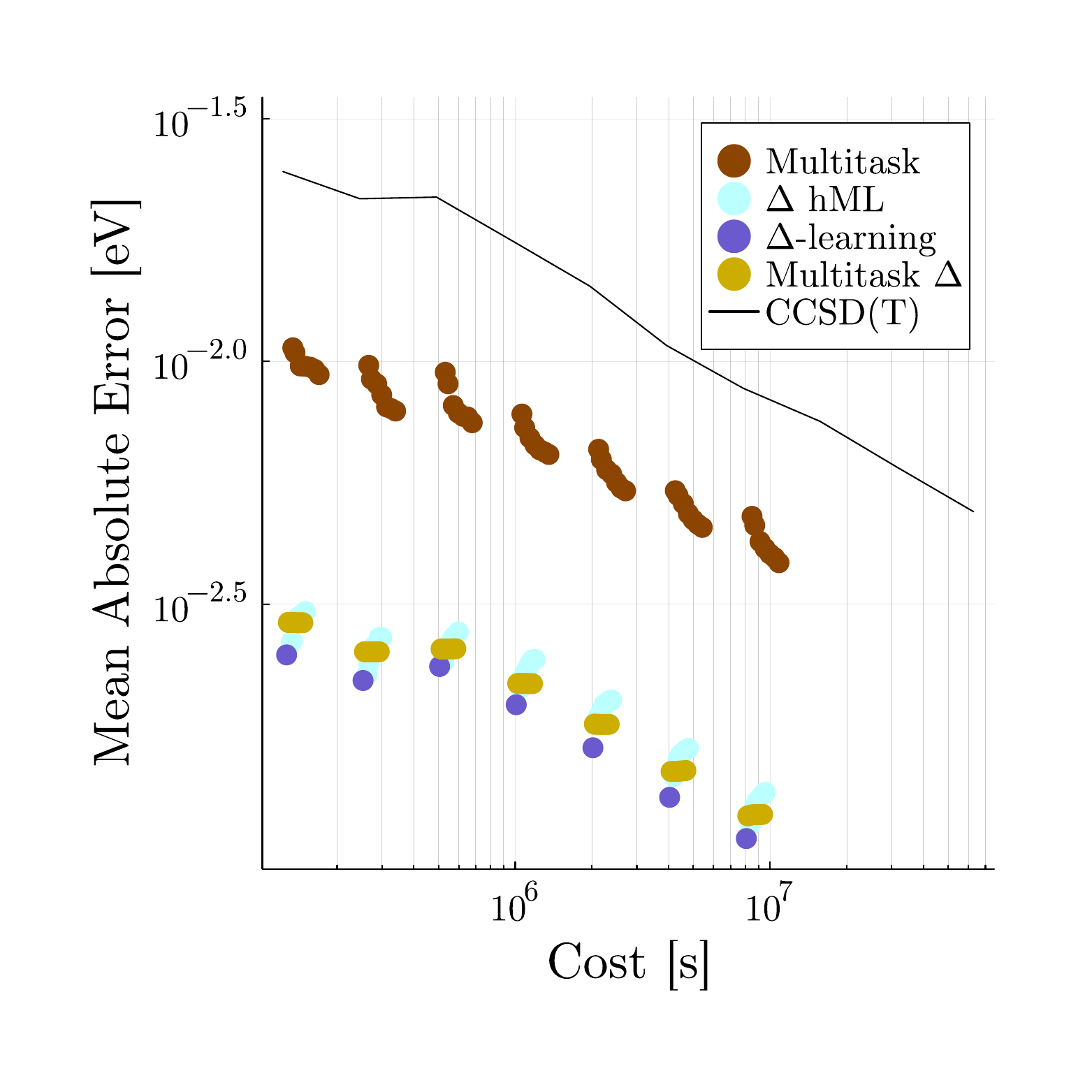} \\[-15pt]
        (b) Estimation of CCSD(T)-SCAN\\
        \end{tabular}
        \caption{\label{fig:water_delta} \textbf{Water Trimer Case.} The MAE of different inference methods: GP regression (line), multitask regression trained with a CAT secondary set (orange), two level $\Delta$-learning (indigo), three level $\Delta$-learning (light blue), and multitask inference on $\Delta$s (gold). Reported performance is the average from six draws of each training set.}
    \end{figure}

$\Delta$-learning is trained on differences between predictions of our quantity
of interest while the single and multitask methods are trained on absolute
predictions of the quantity of interest. Table \rev{
\ref{tab:diff_stats} reports} summary statistics for these data sets. In this
case, \rev{\emph{both}} the mean and \rev{standard deviation} of the differences between
quantum chemistry methods are smaller by at least an order of magnitude than
\rev{than the corresponding statistics for} any set of quantum chemistry
predictions. \rev{Greater standard deviation in the distribution of target data tends to
result in greater standard deviation in the prediction error of a statistical surrogate
which in turn leads to larger MAE.}
Since $\Delta$-learning is trained
on differences to predict differences, it is \rev{thus} expected 
that this method will produce smaller mean absolute error
than a method trained on absolutes to predict absolutes.

By modifying our choice of tasks, we can design a multitask model which trains on and predicts differences. \rev{This reinterpretation of tasks will be referred to as} the multitask $\Delta$ option. Fig.~\ref{fig:water_delta} compares this option to \rev{$\Delta$-learning as well as our original implementation of the multitask model, trained on absolute predictions of the QoI.} For each inference method, \rev{we report the }average \rev{MAE} obtained \rev{over} six random assignments of the training data \rev{to the C,  A, and T sets. We consider} the same \rev{molecule} set sizes which were considered in our previous experiments. The CAT secondary training set structure is used for the multitask absolute \rev{QoI} models as well as the multitask \rev{$\Delta$ option} models. As expected, the multitask absolute model is not as successful as those which train on and learn differences.
 
In subplot (a) of Fig.~\ref{fig:water_delta}, the object of estimation is the difference between CCSD(T) and PBE computations for each molecular system. For the multitask $\Delta$ option, we let the primary task be exactly this difference. The secondary task is the difference between SCAN and PBE computations which has correlation coefficient $0.782$ with the primary task data. Results of the multitask $\Delta$ option are plotted in gold. $\Delta$-learning \rev{predictions for CCSD(T)-PBE are} plotted in indigo\rev{.}  Note that these \rev{$\Delta$-learning} models are the single task \rev{counterpart to} the multitask $\Delta$ approach---they are \rev{a conventional GP model} trained only on \rev{its} primary task. Therefore, Subplot (a) provides another example where the multitask method outperforms its single task counterpart, in this case with \rev{a smaller} correlation parameter \rev{than we considered in Subsection \ref{sec:num1}:} $\rho=0.782$. 

The turquoise points represent a \rev{hierarchical machine learning (hML) extension to the} $\Delta$-learning model\rev{,} trained with three levels of theory: CCSD(T), SCAN, and PBE. \rev{In this method, the difference between CCSD(T) and PBE is represented as }
\begin{eqnarray*}
    &&\rev{\Delta^{hML}(\text{CCSD(T)}, \ \text{PBE})} \\ 
    \rev{\coloneqq}&& \  \rev{\Delta(\text{CCSD(T)}, \ \text{SCAN}) \ + \ \Delta(\text{SCAN}, \ \text{PBE})},
\end{eqnarray*}
\rev{where the submodels $\Delta$(CCSD(T),  \text{SCAN}) and $\Delta$(SCAN,  PBE) are $\Delta$-learning models as defined in \eqref{eq:delta}. Note that while $\Delta$(CCSD(T), SCAN) can only be trained with molecular configurations in $C$, $\Delta$(SCAN, PBE) can be trained with $C\cup A$. }  Given sufficiently large training sets for \rev{SCAN and PBE}, this method is comparable with the multitask $\Delta$ option. The multitask $\Delta$ option outperforms most other models, particularly when either \rev{$|C|$} or \rev{$|A|$} is relatively small. 

Subplot (b) of Fig.~\ref{fig:water_delta} shows the performance of the same set of methods when the primary and secondary tasks \rev{do not have significant correlation}. In this case, the extra data from the secondary task does not contribute to improved inference of the primary task. Here, we estimate the difference between CCSD(T) and SCAN predictions. For all descriptions accompanying Subplot (a), we reverse the roles of SCAN and PBE for Subplot (b). The major difference between the cases is the correlation coefficient between data for CCSD(T)-SCAN and PBE-SCAN: $\rho=-0.271$. The multitask models shown in Subplot (b) demonstrate the limitations of choosing Pearson’s coefficient as the correlation parameter. The multitask $\Delta$ option and \rev{hML extension to} $\Delta$-learning do not perform as well as \rev{standard } $\Delta$-learning in this setting. \rev{Since standard $\Delta$-learning  is the single task counterpart to the multitask $\Delta$ option, t}his behavior suggests that the multitask method requires a stronger correlation between tasks to outperform the single task method. Further, the subplot shows that the multitask $\Delta$ option outperforms \rev{hML} for models trained on larger amounts of DFT data. Thus,  \rev{hML} seems to suffer more from problematic data than the Multitask $\Delta$ option. Importantly, $\Delta$-learning \rev{and its hML extension have} no correlation parameter\rev{s} and cannot adjust the contribution of each \rev{level of theory}. Though we have left a detailed optimization of the Multitask $\Delta$ option’s correlation parameter for future work, we observe that $\rho=0$ can always be chosen. Therefore, the fully optimized multitask $\Delta$ option will always perform at least as well as \rev{standard} $\Delta$-learning. Consequently, the multitask framework makes it possible to benefit from useful secondary datasets which \rev{standard} $\Delta$-learning cannot accommodate while offering more protection against uncorrelated data than \rev{hML} models.

\rev{We conclude our comparison between hML and the multitask $\Delta$ option with a remark about scaling when training data is available for more levels of theory}. Suppose that alongside PBE and CCSD(T) we want to include \rev{$L$} \emph{additional} DFT functionals $y_1,\dots,y_{\rev{L}}$. Then, \rev{an hML model which uses all levels of theory would require that these levels be assigned} some order---with CCSD(T) first and PBE last and \rev{$y_1,\dots,y_L$} in between. The difference between neighboring pairs of methods is then learned. However, hierarchies in DFA accuracy can be difficult to predict and differ between problem settings. This necessary ordering can thus become rather arbitrary. By contrast, the multitask $\Delta$ option would incorporate the $\rev{L}$ new secondary tasks based on the differences to PBE, i.e., $y_{\rev{\ell}}-\text{PBE}$. Each of these secondary tasks can be directly related to the primary task through a correlation parameter---which thus eliminates the need to impose an arbitrary order.

\subsection{\label{sec:num3}Ionization Potential of Small Organic Molecules}

    \begin{table}[h!]
    \begin{center}
        \begin{tabular}{ ccccc }
        \toprule
         & PBE0\_DH  & PBE0  &  PBE & BLYP \\
        \midrule
        $\ \rho \ $ & $0.970$ & $0.967$ & $0.954$ & $0.955$ \\
        \bottomrule
        \end{tabular}
        \caption{ \textbf{Organic Molecules Case.} Pearson's correlation coefficients between each secondary task and the primary task data set. The column headers give the DFA used to generate the secondary set.}
     \label{tab:correlation_ip}
    \end{center}
    \end{table}

    \begin{figure*}
        \begin{tabular}{ccc}
        &\Large \rev{$k=0$} & \\[-2pt]
        \includegraphics[width=0.4\linewidth]{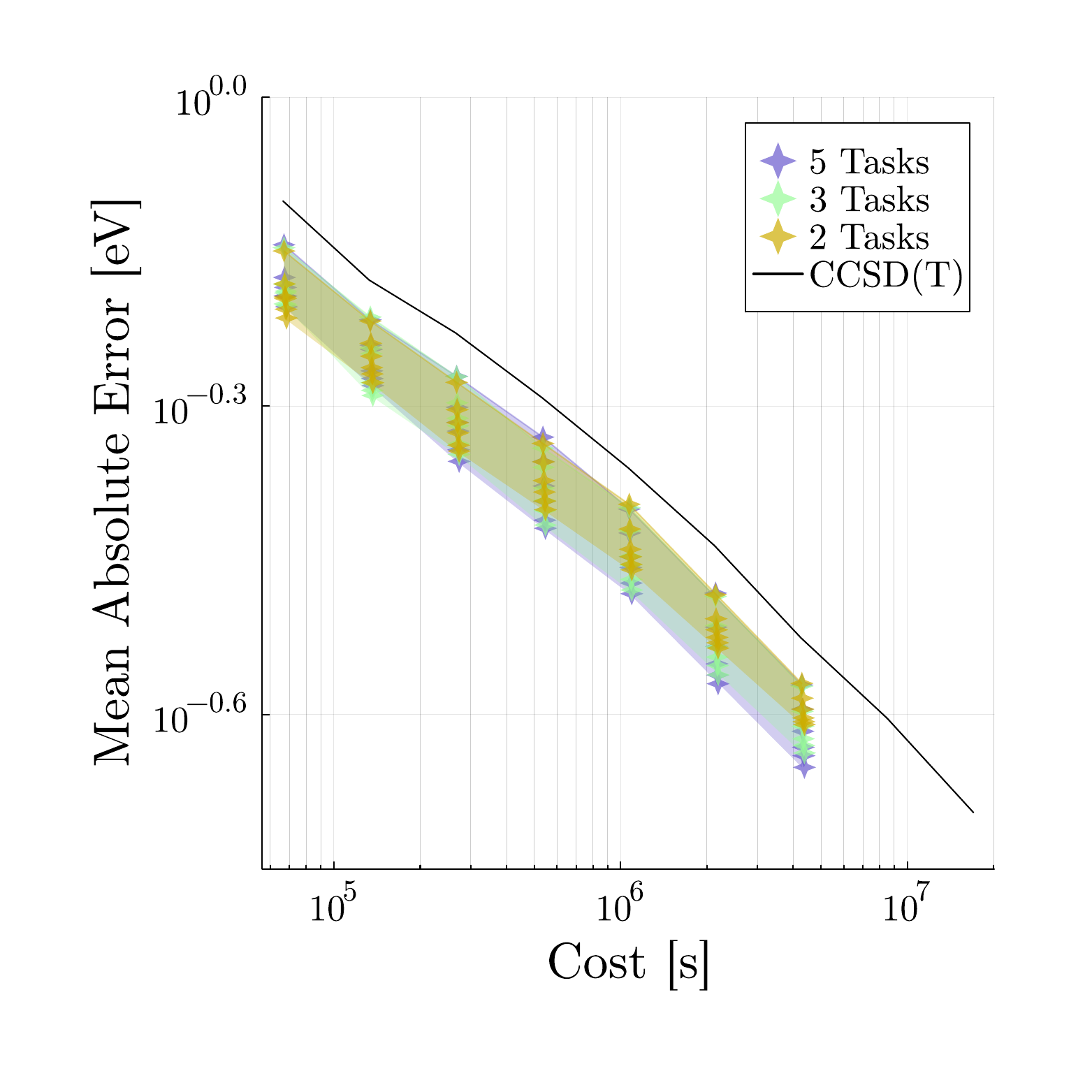} &&   \includegraphics[width=0.4\linewidth]{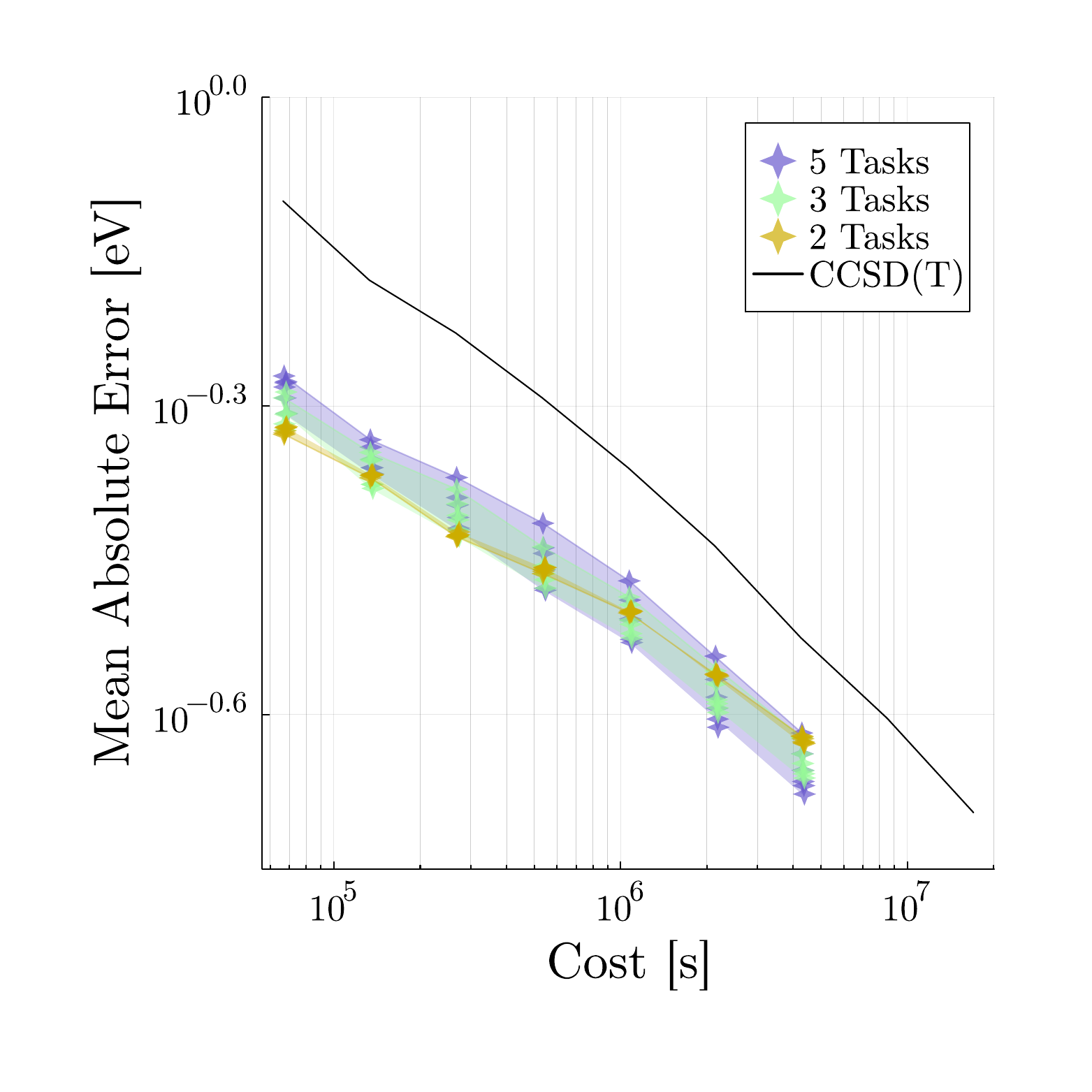} \\[-5pt]
        \midrule
        &\Large \rev{$k=0.5$} & \\[-2pt]
        \includegraphics[width=0.4\linewidth]{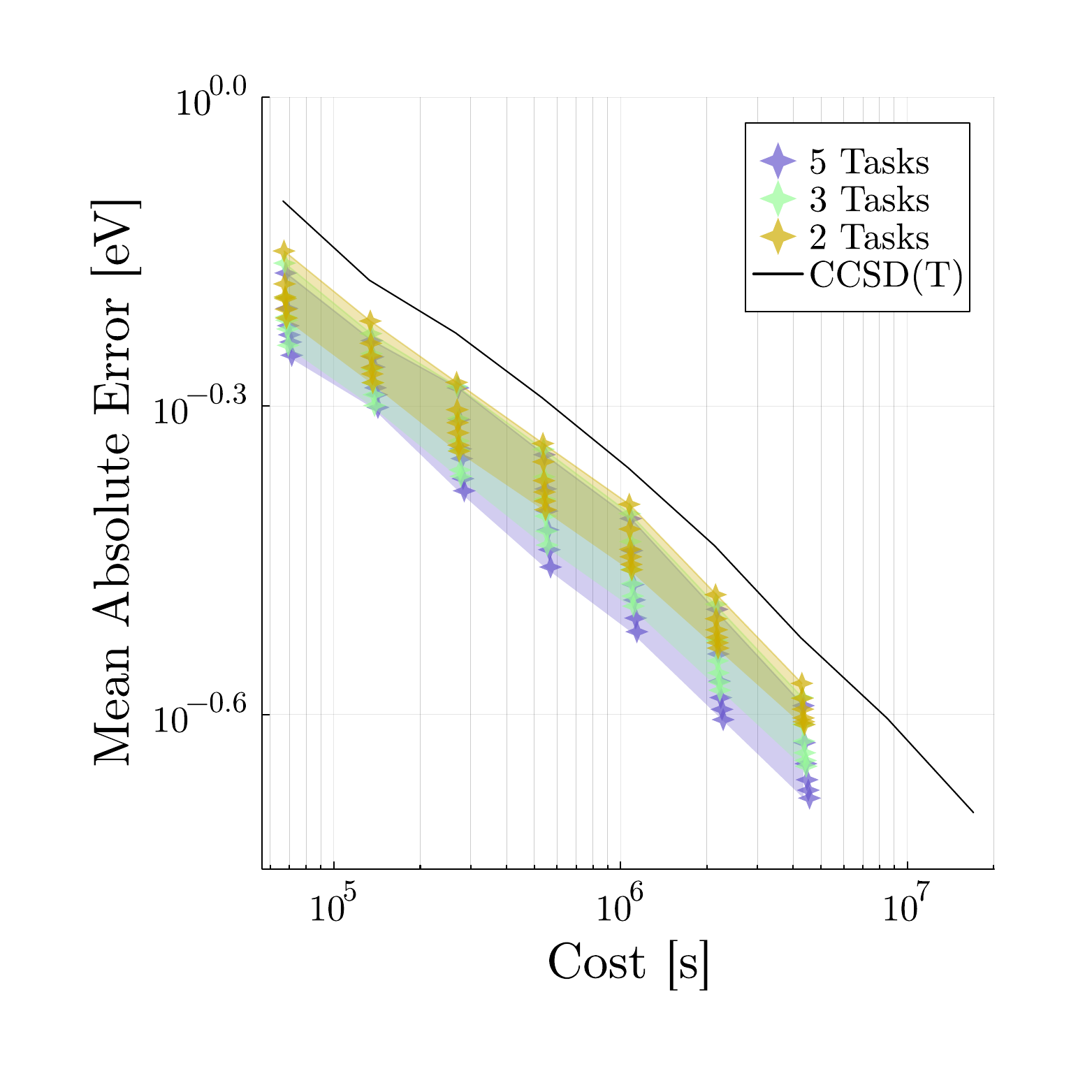} &&   \includegraphics[width=0.4\linewidth]{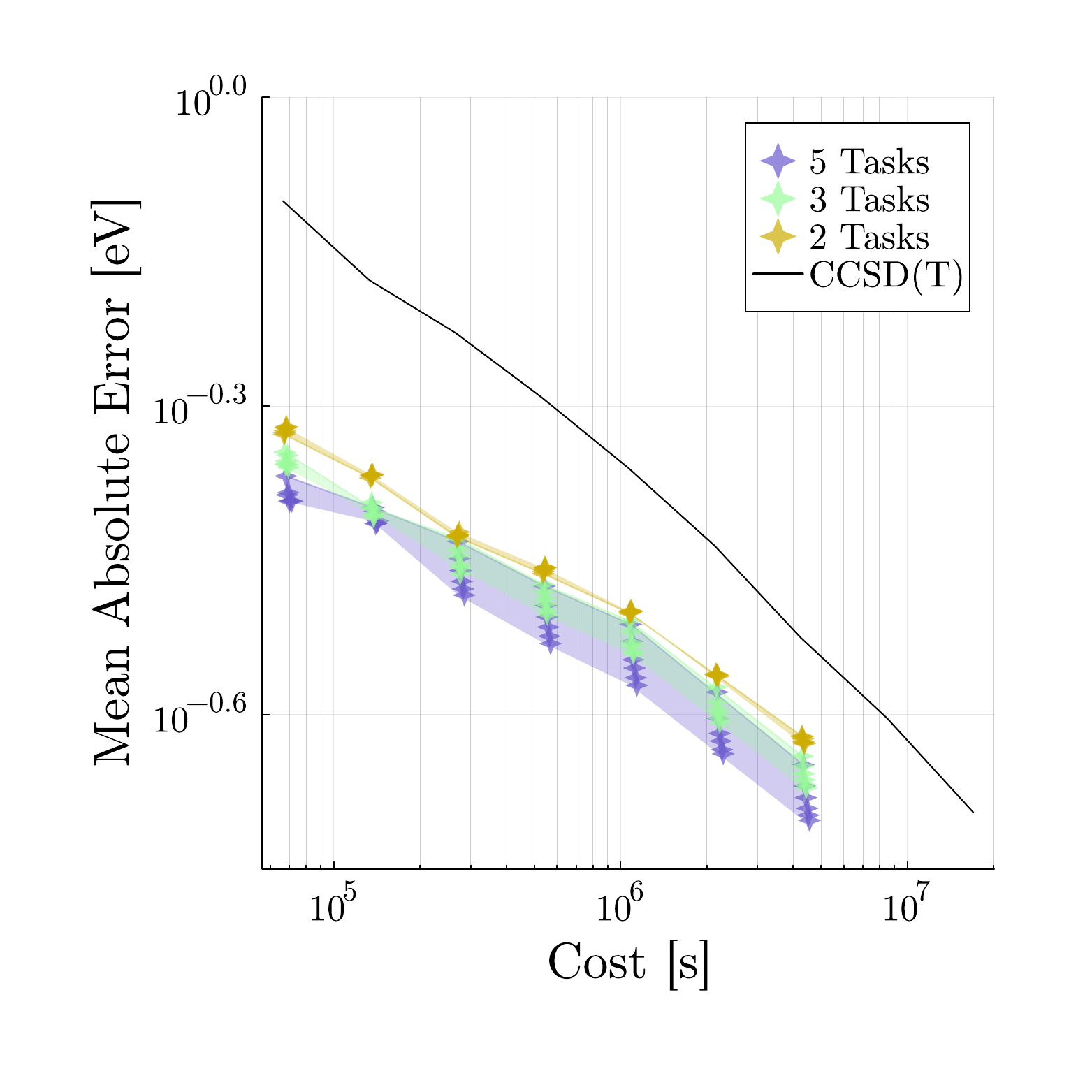} \\[-5pt]
        \midrule
        &\Large \rev{$k=1$} & \\[-2pt]
        \includegraphics[width=0.4\linewidth]{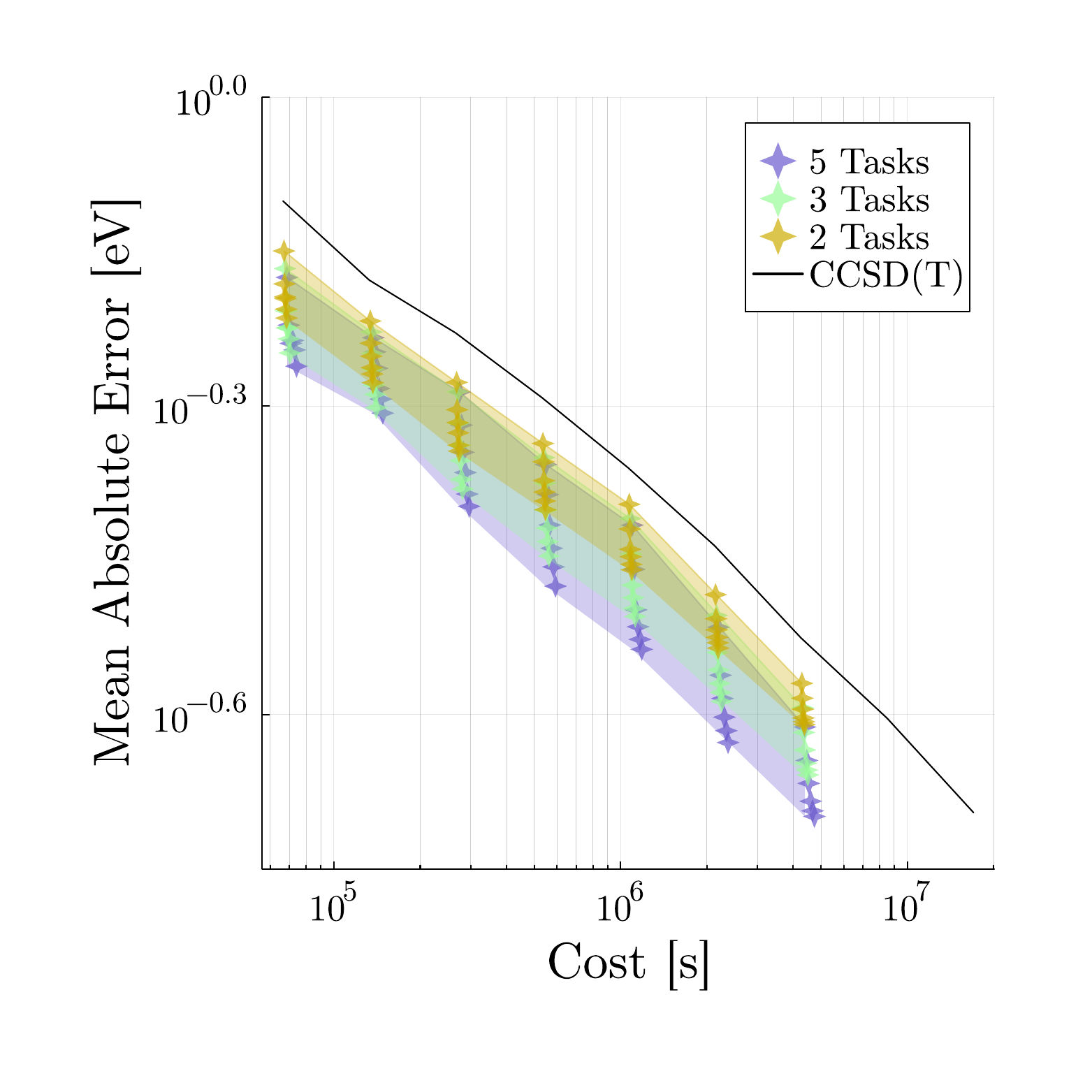} &&   \includegraphics[width=0.4\linewidth]{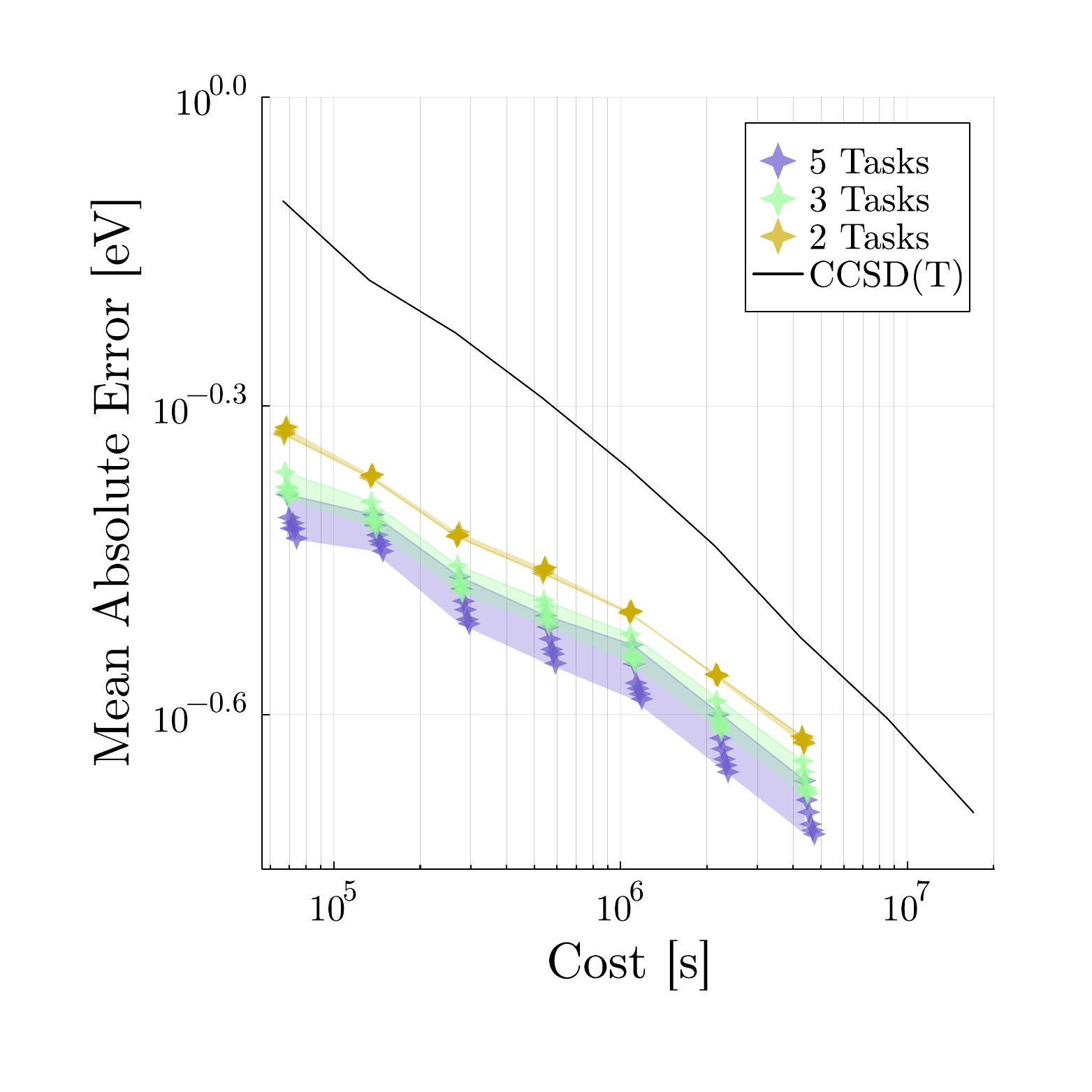} \\[-10pt]
             (a) Trained without DFT data from Target Set && (b) Trained with DFT data from Target Set \\[6pt]
        \end{tabular}
        \caption{\label{fig:ip_levels} \textbf{Organic Molecules Case.} For different numbers of tasks (indicated by color), plots show MAE versus cost. Note that $10^{-0.3} \textrm{eV}$ is approximately half an electronvolt. For each row \rev{of plots, $k$ determines the fraction of training data which is shared between any pair of secondary sets}. Models represented in column (a) were trained without DFT data from the T set while models in column (b) had access to this data. }
    \end{figure*}

    We now turn our attention to the prediction of the ionization potential (IP) of small organic molecules made up of elements in $\{$C,N,O,H$\}$. Our secondary data for this example includes up to four tasks, generated with the PBE, PBE0, PBE\_DH, and BLYP functionals, respectively. Table \ref{tab:correlation_ip} shows the correlation parameter used to relate each of these secondary tasks to the primary task data, generated with CCSD(T). As in the water trimer example, we consider Core sets of sizes $\{5,10,20,40,80,160,320\}$. \rev{We provide results for $|A|=r|C|$ with $r=1,2,\dots,6$ and choose a }Target set \rev{which} contains 320 molecular configurations. As Section \ref{design} visualizes, once the C and A sets have been fixed, there is still a design choice in allotting the molecular configurations to secondary tasks\rev{: we must determine how to construct each $\mathcal{C}(s_n)$ and $\mathcal{A}(s_n)$. In this numerical example, we enforce the constraint that for $1\leq n < n' \leq N-1$,} 
    \begin{eqnarray*}
        \rev{\frac{\bigl|\{\mathcal{C}(s_n)\cap \mathcal{C}(s_{n'})\}\bigr|}{|C|}} \ \rev{=} \ \rev{\frac{\bigl|\{\mathcal{A}(s_n)\cap \mathcal{A}(s_{n'})\}\bigr|}{|A|}} \ \rev{=}  \ \rev{k} .
    \end{eqnarray*}
    \rev{We provide results for $k\in\{0,\frac{1}{2},1\}$.} These cases correspond---in order, from the top down---to the three rows of Fig.~\ref{fig:ip_levels}.

    In Fig.~\ref{fig:ip_levels}, as for previous results, the reported mean absolute error is the average \rev{result of} six tests produced by randomly assigning molecules to the C, A, and T sets. The left column of subplots compares results for multitask models trained \rev{using the CA structure to the single task reference} while the right column considers multitask models trained \rev{with the CAT structure}. Translucent stripes encompass the results for models with \rev{a given number} of tasks, $N$. For the two task case, the secondary task data is generated with PBE, and for the three task case the secondary data comes from PBE and PBE0. In all cases, secondary Target data is only generated with PBE. Recall \rev{from Subsection \ref{ss:target} that our training and testing procedure is designed to produce fair comparisons between models trained on CA and CAT training structures. Though we test the predictive performance of each training set structure on $320$ different target molecular configurations, we do not simultaneously include all of these configurations in $\mathcal{T}(\text{PBE})$. Rather, we iterate over the target configurations and train a multitask model for each configuration such that $\mathcal{T}(\text{PBE})=1$ in each case.}  We remark that in practice it would not be necessary to train a new model for each molecule in the Target set.

    As we observed in the water trimer case, the multitask models consistently perform better than single task models of comparable cost. Especially for relatively small training sets, the multitask method shows an improvement in performance when secondary Target data is included in the training set. \rev{ Further, we report that for larger secondary training sets, correlation between multitask predictions and the CCSD(T) target data exceeds $95\%$, but we defer a full evaluation of prediction quality at the level of individual molecular configurations to future work.  }

    Fig.~\ref{fig:ip_levels} also informs us about the utility of including more than one secondary task in a model. Note that \rev{we consider the same combination of sizes for the C, A, and T sets regardless of the total number of tasks, $N$.} In the top row, \rev{$k=0$ implies that molecular configurations} in the C and A sets are each assigned to exactly one secondary task, regardless of the total number of secondary tasks. Therefore, in this row, \rev{we can compare models implemented with different $N$ but nearly identical training budgets.} The results show that for sufficiently large training sets, \rev{given a fixed budget, there is a benefit to dividing the budget across multiple secondary tasks rather than employing only one secondary task.} For smaller datasets, \rev{choosing $N=2$} may lead to performance as good or better than a larger number of tasks. In particular, the top subplot in column (b) suggests that models \rev{where $N=2$} perform particularly well when given a relatively inexpensive training set of the CAT structure. Note that, in each model, only one \rev{of the} secondary task\rev{s} is \rev{trained using} T data\rev{: only $\mathcal{T}(\text{PBE})$ is nonempty}. Thus, in this setting, the decrease in performance when new tasks are added likely comes from siphoning off part of the computational budget to train tasks which do not have access to T data. 
    
    We can also consider whether it is useful to \rev{include the same molecular configuration in the training set of multiple secondary tasks. In the final row of Fig.~\ref{fig:ip_levels}, $k=1$ implies that $\mathcal{C}(s_n)\cup \mathcal{A}(s_n)$ is the same set for all $n=1,\dots,N-1$. By contrast, in the middle row, to enforce $k=\frac{1}{2}$, we assign $50\%$ of the configurations in $C$ and $A$ to every secondary task while the remaining configurations in these sets are distributed between the secondary tasks so that each is assigned to exactly one secondary task.  The improved performance of the $N=3$ and $N=5$ cases in the last row compared with the middle row suggest} that there is a benefit to training our model on multiple secondary predictions for the same molecule.


\section{Conclusion}
The present work has demonstrated that inference models see a general benefit from leveraging all available data. Furthermore, use of the multitask framework can reduce the cost of generating entirely new training data by an order of magnitude and can also facilitate the opportunistic incorporation of existing datasets of heterogeneous quality into the training set. \rev{To our knowledge, o}ur work is the first to explore a\rev{n asymmetric} multitask GP regression model in a materials science setting. Significantly, the multitask framework is not restricted to the choice of GP inference. \rev{F}uture work can \rev{preserve the same relationship between primary and secondary task submodels and} investigate training \rev{each sub}model with neural networks \rev{or} other tools for regression.

Unlike other machine learning methods for constructing surrogates from ab initio data, the multitask model does not require all data sources be ordered in a hierarchy of accuracy. In this work, the data which informed our primary task---generated with CCSD(T)---was the most accurate in our training set, but we were also able to leverage an arbitrary number of secondary data sets without any information on their accuracy relative to each other. This feature of the multitask method is particularly useful when employing datasets constructed from different density functional approximations. We found that multitask model performance improves as we increase the number of secondary sets in the training data which include predictions for a molecule of interest. The influence of each secondary dataset is controlled by a correlation parameter, and we found that in most settings an estimate of Pearson’s correlation coefficient from training data works well. Use of this estimate simplifies implementation of the method. The multitask method also supports training on differences, and with appropriate choice of correlation parameter, multitask difference models can perform at least as well as and often better than $\Delta$ models. To explore the framework’s flexibility, we demonstrated the success of the multitask method on diverse training data set configurations.

\begin{acknowledgments}
The authors would like to thank Yeongsu Cho, Chenru Duan, Dallas Foster, and Heather Kulik for insightful discussions. MIT SuperCloud and Lincoln Laboratory Supercomputing Center are acknowledged for providing the HPC resources that have contributed to the research results reported within this paper.~\cite{reuther2018interactive} This material is based upon work supported by the Department of Energy, National Nuclear Security Administration PSAAP-III program, under Award Number DE-NA0003965 as well as work supported by the National Science Foundation Graduate Research Fellowship under Grant No. 1745302.
This research was supported by the NCCR MARVEL, a National Centre of Competence in Research, funded by the Swiss National Science Foundation (grant number 205602).
\end{acknowledgments}

\section*{Author Contributions Statement}

\textbf{Katharine Fisher:} Conceptualization (equal); Data Curation (equal); Methodology (equal); Software (lead); Writing – original draft (lead); Writing – review and editing (equal); Visualization (equal). \textbf{Michael Herbst:} Conceptualization (equal);  Data Curation (equal); Methodology (equal); Supervision (equal); Writing – review and editing (equal); Visualization (equal). \textbf{Youssef Marzouk:} Conceptualization (equal);  Methodology (equal); Supervision (equal); Writing – review and editing (equal); Visualization (equal).

\section*{Conflict of Interest Statement}

The authors have no conflicts to disclose.

\section*{\label{DA} Data Availability Statement}

All data supporting this work are openly available. All code to reproduce figures as well as all DFT predictions for the three-body energy of water trimers can be found at \href{https://doi.org/10.5281/zenodo.10387647}{https://doi.org/10.5281/zenodo.10387647} or \href{https://github.com/kefisher98/Multitask_GP/tree/Multitask}{https://github.com/kefisher98/Multitask\_GP/tree/Multitask}. Water trimer configurations and CCSD(T) predictions for three-body energy are available at \href{https://github.com/jmbowma/q-AQUA}{ https://github.com/jmbowma/q-AQUA}. The data that support the findings of
this study for the small organic molecules example are at \href{https://doi.org/10.5281/zenodo.10215421}{https://doi.org/10.5281/zenodo.10215421}. 

\nocite{*}
\bibliography{references}

\providecommand{\noopsort}[1]{}\providecommand{\singleletter}[1]{#1}%
\begin{thebibliography}{66}%
\makeatletter
\providecommand \@ifxundefined [1]{%
 \@ifx{#1\undefined}
}%
\providecommand \@ifnum [1]{%
 \ifnum #1\expandafter \@firstoftwo
 \else \expandafter \@secondoftwo
 \fi
}%
\providecommand \@ifx [1]{%
 \ifx #1\expandafter \@firstoftwo
 \else \expandafter \@secondoftwo
 \fi
}%
\providecommand \natexlab [1]{#1}%
\providecommand \enquote  [1]{``#1''}%
\providecommand \bibnamefont  [1]{#1}%
\providecommand \bibfnamefont [1]{#1}%
\providecommand \citenamefont [1]{#1}%
\providecommand \href@noop [0]{\@secondoftwo}%
\providecommand \href [0]{\begingroup \@sanitize@url \@href}%
\providecommand \@href[1]{\@@startlink{#1}\@@href}%
\providecommand \@@href[1]{\endgroup#1\@@endlink}%
\providecommand \@sanitize@url [0]{\catcode `\\12\catcode `\$12\catcode `\&12\catcode `\#12\catcode `\^12\catcode `\_12\catcode `\%12\relax}%
\providecommand \@@startlink[1]{}%
\providecommand \@@endlink[0]{}%
\providecommand \url  [0]{\begingroup\@sanitize@url \@url }%
\providecommand \@url [1]{\endgroup\@href {#1}{\urlprefix }}%
\providecommand \urlprefix  [0]{URL }%
\providecommand \Eprint [0]{\href }%
\providecommand \doibase [0]{http://dx.doi.org/}%
\providecommand \selectlanguage [0]{\@gobble}%
\providecommand \bibinfo  [0]{\@secondoftwo}%
\providecommand \bibfield  [0]{\@secondoftwo}%
\providecommand \translation [1]{[#1]}%
\providecommand \BibitemOpen [0]{}%
\providecommand \bibitemStop [0]{}%
\providecommand \bibitemNoStop [0]{.\EOS\space}%
\providecommand \EOS [0]{\spacefactor3000\relax}%
\providecommand \BibitemShut  [1]{\csname bibitem#1\endcsname}%
\let\auto@bib@innerbib\@empty
\bibitem [{\citenamefont {Harding}\ \emph {et~al.}(2008)\citenamefont {Harding}, \citenamefont {Metzroth}, \citenamefont {Gauss},\ and\ \citenamefont {Auer}}]{Harding2008}%
  \BibitemOpen
  \bibfield  {author} {\bibinfo {author} {\bibfnamefont {M.~E.}\ \bibnamefont {Harding}}, \bibinfo {author} {\bibfnamefont {T.}~\bibnamefont {Metzroth}}, \bibinfo {author} {\bibfnamefont {J.}~\bibnamefont {Gauss}}, \ and\ \bibinfo {author} {\bibfnamefont {A.~A.}\ \bibnamefont {Auer}},\ }\bibfield  {title} {\enquote {\bibinfo {title} {Parallel calculation of ccsd and ccsd(t) analytic first and second derivatives},}\ }\href {\doibase 10.1021/ct700152c} {\bibfield  {journal} {\bibinfo  {journal} {Journal of Chemical Theory and Computation}\ }\textbf {\bibinfo {volume} {4}},\ \bibinfo {pages} {64--74} (\bibinfo {year} {2008})}\BibitemShut {NoStop}%
\bibitem [{\citenamefont {Ying}, \citenamefont {Yu},\ and\ \citenamefont {Ying}(2019)}]{dft_survey}%
  \BibitemOpen
  \bibfield  {author} {\bibinfo {author} {\bibfnamefont {L.}~\bibnamefont {Ying}}, \bibinfo {author} {\bibfnamefont {J.}~\bibnamefont {Yu}}, \ and\ \bibinfo {author} {\bibfnamefont {L.}~\bibnamefont {Ying}},\ }\bibfield  {title} {\enquote {\bibinfo {title} {Numerical methods for kohn–sham density functional theory},}\ }\href {\doibase https://doi.org/10.1017/S0962492919000047} {\bibfield  {journal} {\bibinfo  {journal} {Acta Numerica}\ }\textbf {\bibinfo {volume} {28}},\ \bibinfo {pages} {405--539} (\bibinfo {year} {2019})}\BibitemShut {NoStop}%
\bibitem [{\citenamefont {Perdew}\ and\ \citenamefont {Schmidt}(2001)}]{dft_ladder}%
  \BibitemOpen
  \bibfield  {author} {\bibinfo {author} {\bibfnamefont {J.}~\bibnamefont {Perdew}}\ and\ \bibinfo {author} {\bibfnamefont {K.}~\bibnamefont {Schmidt}},\ }\bibfield  {title} {\enquote {\bibinfo {title} {Jacob’s ladder of density functional approximations for the exchange-correlation energy},}\ }in\ \href {\doibase https://doi.org/10.1063/1.1390175} {\emph {\bibinfo {booktitle} {AIP Conference Proceedings}}}\ (\bibinfo {year} {2001})\ \bibinfo {note} {presented at AIP Conference Proceedings 577}\BibitemShut {NoStop}%
\bibitem [{\citenamefont {Goerigkab}\ and\ \citenamefont {Grimme}(2011)}]{dft_dfa_benchmark}%
  \BibitemOpen
  \bibfield  {author} {\bibinfo {author} {\bibfnamefont {L.}~\bibnamefont {Goerigkab}}\ and\ \bibinfo {author} {\bibfnamefont {S.}~\bibnamefont {Grimme}},\ }\bibfield  {title} {\enquote {\bibinfo {title} {A thorough benchmark of density functional methods for general main group thermochemistry, kinetics, and noncovalent interactions},}\ }\href {\doibase https://doi.org/10.1039/C0CP02984J} {\bibfield  {journal} {\bibinfo  {journal} {Physical Chemistry Chemical Physics}\ }\textbf {\bibinfo {volume} {13}},\ \bibinfo {pages} {6670--6688} (\bibinfo {year} {2011})}\BibitemShut {NoStop}%
\bibitem [{\citenamefont {Teale}\ \emph {et~al.}(2022)\citenamefont {Teale}, \citenamefont {Helgaker}, \citenamefont {Savin}, \citenamefont {Adano}, \citenamefont {Aradi}, \citenamefont {Arbuznikov}, \citenamefont {Ayers}, \citenamefont {Baerends}, \citenamefont {Barone}, \citenamefont {Calaminici}, \citenamefont {Cances}, \citenamefont {Carter}, \citenamefont {Chattaraj}, \citenamefont {Chermette}, \citenamefont {Ciofini}, \citenamefont {Crawford}, \citenamefont {De~Proft}, \citenamefont {Dobson}, \citenamefont {Draxl}, \citenamefont {Frauenheim}, \citenamefont {Fromager}, \citenamefont {Fuentealba}, \citenamefont {Gagliardi}, \citenamefont {Galli}, \citenamefont {Gao}, \citenamefont {Geerlings}, \citenamefont {Gidopoulos}, \citenamefont {Gill}, \citenamefont {Gori-Giorgi}, \citenamefont {G\"{o}rling}, \citenamefont {Gould}, \citenamefont {Grimme}, \citenamefont {Gritsenko}, \citenamefont {Jensen}, \citenamefont {Johnson}, \citenamefont {Jones}, \citenamefont {Kaupp}, \citenamefont {Koster}, \citenamefont
  {Kronik}, \citenamefont {Krylov}, \citenamefont {Kvaal}, \citenamefont {Laestadius}, \citenamefont {Levy}, \citenamefont {Lewin}, \citenamefont {Liu}, \citenamefont {Loos}, \citenamefont {Maitra}, \citenamefont {Neese}, \citenamefont {Perdew}, \citenamefont {Pernal}, \citenamefont {Pernot}, \citenamefont {Piecuch}, \citenamefont {Rebolini}, \citenamefont {Reining}, \citenamefont {Romaniello}, \citenamefont {Ruzsinszky}, \citenamefont {Salahub}, \citenamefont {Scheffler}, \citenamefont {Schwerdtfeger}, \citenamefont {Staroverov}, \citenamefont {Sun}, \citenamefont {Tellgren}, \citenamefont {Tozer}, \citenamefont {Trickey}, \citenamefont {Ullrich}, \citenamefont {Vela}, \citenamefont {Vignale}, \citenamefont {Wesolowski}, \citenamefont {Xu},\ and\ \citenamefont {Yang}}]{Teale2022}%
  \BibitemOpen
  \bibfield  {author} {\bibinfo {author} {\bibfnamefont {A.~M.}\ \bibnamefont {Teale}}, \bibinfo {author} {\bibfnamefont {T.}~\bibnamefont {Helgaker}}, \bibinfo {author} {\bibfnamefont {A.}~\bibnamefont {Savin}}, \bibinfo {author} {\bibfnamefont {C.}~\bibnamefont {Adano}}, \bibinfo {author} {\bibfnamefont {B.}~\bibnamefont {Aradi}}, \bibinfo {author} {\bibfnamefont {A.~V.}\ \bibnamefont {Arbuznikov}}, \bibinfo {author} {\bibfnamefont {P.}~\bibnamefont {Ayers}}, \bibinfo {author} {\bibfnamefont {E.~J.}\ \bibnamefont {Baerends}}, \bibinfo {author} {\bibfnamefont {V.}~\bibnamefont {Barone}}, \bibinfo {author} {\bibfnamefont {P.}~\bibnamefont {Calaminici}}, \bibinfo {author} {\bibfnamefont {E.}~\bibnamefont {Cances}}, \bibinfo {author} {\bibfnamefont {E.~A.}\ \bibnamefont {Carter}}, \bibinfo {author} {\bibfnamefont {P.~K.}\ \bibnamefont {Chattaraj}}, \bibinfo {author} {\bibfnamefont {H.}~\bibnamefont {Chermette}}, \bibinfo {author} {\bibfnamefont {I.}~\bibnamefont {Ciofini}}, \bibinfo {author} {\bibfnamefont
  {T.~D.}\ \bibnamefont {Crawford}}, \bibinfo {author} {\bibfnamefont {F.}~\bibnamefont {De~Proft}}, \bibinfo {author} {\bibfnamefont {J.}~\bibnamefont {Dobson}}, \bibinfo {author} {\bibfnamefont {C.}~\bibnamefont {Draxl}}, \bibinfo {author} {\bibfnamefont {T.}~\bibnamefont {Frauenheim}}, \bibinfo {author} {\bibfnamefont {E.}~\bibnamefont {Fromager}}, \bibinfo {author} {\bibfnamefont {P.}~\bibnamefont {Fuentealba}}, \bibinfo {author} {\bibfnamefont {L.}~\bibnamefont {Gagliardi}}, \bibinfo {author} {\bibfnamefont {G.}~\bibnamefont {Galli}}, \bibinfo {author} {\bibfnamefont {J.}~\bibnamefont {Gao}}, \bibinfo {author} {\bibfnamefont {P.}~\bibnamefont {Geerlings}}, \bibinfo {author} {\bibfnamefont {N.}~\bibnamefont {Gidopoulos}}, \bibinfo {author} {\bibfnamefont {P.~M.~W.}\ \bibnamefont {Gill}}, \bibinfo {author} {\bibfnamefont {P.}~\bibnamefont {Gori-Giorgi}}, \bibinfo {author} {\bibfnamefont {A.}~\bibnamefont {G\"{o}rling}}, \bibinfo {author} {\bibfnamefont {T.}~\bibnamefont {Gould}}, \bibinfo {author}
  {\bibfnamefont {S.}~\bibnamefont {Grimme}}, \bibinfo {author} {\bibfnamefont {O.}~\bibnamefont {Gritsenko}}, \bibinfo {author} {\bibfnamefont {H.~J.~A.}\ \bibnamefont {Jensen}}, \bibinfo {author} {\bibfnamefont {E.~R.}\ \bibnamefont {Johnson}}, \bibinfo {author} {\bibfnamefont {R.~O.}\ \bibnamefont {Jones}}, \bibinfo {author} {\bibfnamefont {M.}~\bibnamefont {Kaupp}}, \bibinfo {author} {\bibfnamefont {A.}~\bibnamefont {Koster}}, \bibinfo {author} {\bibfnamefont {L.}~\bibnamefont {Kronik}}, \bibinfo {author} {\bibfnamefont {A.~I.}\ \bibnamefont {Krylov}}, \bibinfo {author} {\bibfnamefont {S.}~\bibnamefont {Kvaal}}, \bibinfo {author} {\bibfnamefont {A.}~\bibnamefont {Laestadius}}, \bibinfo {author} {\bibfnamefont {M.~P.}\ \bibnamefont {Levy}}, \bibinfo {author} {\bibfnamefont {M.}~\bibnamefont {Lewin}}, \bibinfo {author} {\bibfnamefont {S.}~\bibnamefont {Liu}}, \bibinfo {author} {\bibfnamefont {P.-F.}\ \bibnamefont {Loos}}, \bibinfo {author} {\bibfnamefont {N.~T.}\ \bibnamefont {Maitra}}, \bibinfo {author}
  {\bibfnamefont {F.}~\bibnamefont {Neese}}, \bibinfo {author} {\bibfnamefont {J.}~\bibnamefont {Perdew}}, \bibinfo {author} {\bibfnamefont {K.}~\bibnamefont {Pernal}}, \bibinfo {author} {\bibfnamefont {P.}~\bibnamefont {Pernot}}, \bibinfo {author} {\bibfnamefont {P.}~\bibnamefont {Piecuch}}, \bibinfo {author} {\bibfnamefont {E.}~\bibnamefont {Rebolini}}, \bibinfo {author} {\bibfnamefont {L.}~\bibnamefont {Reining}}, \bibinfo {author} {\bibfnamefont {P.}~\bibnamefont {Romaniello}}, \bibinfo {author} {\bibfnamefont {A.}~\bibnamefont {Ruzsinszky}}, \bibinfo {author} {\bibfnamefont {D.}~\bibnamefont {Salahub}}, \bibinfo {author} {\bibfnamefont {M.}~\bibnamefont {Scheffler}}, \bibinfo {author} {\bibfnamefont {P.}~\bibnamefont {Schwerdtfeger}}, \bibinfo {author} {\bibfnamefont {V.~N.}\ \bibnamefont {Staroverov}}, \bibinfo {author} {\bibfnamefont {J.}~\bibnamefont {Sun}}, \bibinfo {author} {\bibfnamefont {E.}~\bibnamefont {Tellgren}}, \bibinfo {author} {\bibfnamefont {D.~J.}\ \bibnamefont {Tozer}}, \bibinfo
  {author} {\bibfnamefont {S.}~\bibnamefont {Trickey}}, \bibinfo {author} {\bibfnamefont {C.~A.}\ \bibnamefont {Ullrich}}, \bibinfo {author} {\bibfnamefont {A.}~\bibnamefont {Vela}}, \bibinfo {author} {\bibfnamefont {G.}~\bibnamefont {Vignale}}, \bibinfo {author} {\bibfnamefont {T.~A.}\ \bibnamefont {Wesolowski}}, \bibinfo {author} {\bibfnamefont {X.}~\bibnamefont {Xu}}, \ and\ \bibinfo {author} {\bibfnamefont {W.}~\bibnamefont {Yang}},\ }\bibfield  {title} {\enquote {\bibinfo {title} {Dft exchange: Sharing perspectives on the workhorse of quantum chemistry and materials science},}\ }\href {\doibase 10.1039/d2cp02827a} {\bibfield  {journal} {\bibinfo  {journal} {Physical Chemistry Chemical Physics}\ } (\bibinfo {year} {2022}),\ 10.1039/d2cp02827a}\BibitemShut {NoStop}%
\bibitem [{\citenamefont {Smith}\ \emph {et~al.}(2019)\citenamefont {Smith}, \citenamefont {Nebgen}, \citenamefont {Zubatyuk}, \citenamefont {Lubbers}, \citenamefont {Devereux}, \citenamefont {Barros}, \citenamefont {Tretiak}, \citenamefont {Isayev},\ and\ \citenamefont {Roitberg}}]{Smith2019}%
  \BibitemOpen
  \bibfield  {author} {\bibinfo {author} {\bibfnamefont {J.~S.}\ \bibnamefont {Smith}}, \bibinfo {author} {\bibfnamefont {B.~T.}\ \bibnamefont {Nebgen}}, \bibinfo {author} {\bibfnamefont {R.}~\bibnamefont {Zubatyuk}}, \bibinfo {author} {\bibfnamefont {N.}~\bibnamefont {Lubbers}}, \bibinfo {author} {\bibfnamefont {C.}~\bibnamefont {Devereux}}, \bibinfo {author} {\bibfnamefont {K.}~\bibnamefont {Barros}}, \bibinfo {author} {\bibfnamefont {S.}~\bibnamefont {Tretiak}}, \bibinfo {author} {\bibfnamefont {O.}~\bibnamefont {Isayev}}, \ and\ \bibinfo {author} {\bibfnamefont {A.~E.}\ \bibnamefont {Roitberg}},\ }\bibfield  {title} {\enquote {\bibinfo {title} {Approaching coupled cluster accuracy with a general-purpose neural network potential through transfer learning},}\ }\href {\doibase 10.1038/s41467-019-10827-4} {\bibfield  {journal} {\bibinfo  {journal} {Nature Communications}\ }\textbf {\bibinfo {volume} {10}} (\bibinfo {year} {2019}),\ 10.1038/s41467-019-10827-4}\BibitemShut {NoStop}%
\bibitem [{\citenamefont {Dral}\ \emph {et~al.}(2020)\citenamefont {Dral}, \citenamefont {Owens}, \citenamefont {Dral},\ and\ \citenamefont {Cs\'{a}nyi}}]{Dral2020}%
  \BibitemOpen
  \bibfield  {author} {\bibinfo {author} {\bibfnamefont {P.~O.}\ \bibnamefont {Dral}}, \bibinfo {author} {\bibfnamefont {A.}~\bibnamefont {Owens}}, \bibinfo {author} {\bibfnamefont {A.}~\bibnamefont {Dral}}, \ and\ \bibinfo {author} {\bibfnamefont {G.}~\bibnamefont {Cs\'{a}nyi}},\ }\bibfield  {title} {\enquote {\bibinfo {title} {Hierarchical machine learning of potential energy surfaces},}\ }\href {\doibase 10.1063/5.0006498} {\bibfield  {journal} {\bibinfo  {journal} {The Journal of Chemical Physics}\ }\textbf {\bibinfo {volume} {152}} (\bibinfo {year} {2020}),\ 10.1063/5.0006498}\BibitemShut {NoStop}%
\bibitem [{\citenamefont {Goodlett}, \citenamefont {Turney},\ and\ \citenamefont {Schaefer}(2023)}]{Goodlett2023}%
  \BibitemOpen
  \bibfield  {author} {\bibinfo {author} {\bibfnamefont {S.~M.}\ \bibnamefont {Goodlett}}, \bibinfo {author} {\bibfnamefont {J.~M.}\ \bibnamefont {Turney}}, \ and\ \bibinfo {author} {\bibfnamefont {H.~F.}\ \bibnamefont {Schaefer}},\ }\bibfield  {title} {\enquote {\bibinfo {title} {Comparison of multifidelity machine learning models for potential energy surfaces},}\ }\href {\doibase 10.1063/5.0158919} {\bibfield  {journal} {\bibinfo  {journal} {The Journal of Chemical Physics}\ }\textbf {\bibinfo {volume} {159}} (\bibinfo {year} {2023}),\ 10.1063/5.0158919}\BibitemShut {NoStop}%
\bibitem [{\citenamefont {Zaverkin}\ \emph {et~al.}(2023)\citenamefont {Zaverkin}, \citenamefont {Holzm\"{u}ller}, \citenamefont {Bonfirraro},\ and\ \citenamefont {K\"{a}stner}}]{Zaverkin2023}%
  \BibitemOpen
  \bibfield  {author} {\bibinfo {author} {\bibfnamefont {V.}~\bibnamefont {Zaverkin}}, \bibinfo {author} {\bibfnamefont {D.}~\bibnamefont {Holzm\"{u}ller}}, \bibinfo {author} {\bibfnamefont {L.}~\bibnamefont {Bonfirraro}}, \ and\ \bibinfo {author} {\bibfnamefont {J.}~\bibnamefont {K\"{a}stner}},\ }\bibfield  {title} {\enquote {\bibinfo {title} {Transfer learning for chemically accurate interatomic neural network potentials},}\ }\href {\doibase 10.1039/d2cp05793j} {\bibfield  {journal} {\bibinfo  {journal} {Physical Chemistry Chemical Physics}\ }\textbf {\bibinfo {volume} {25}},\ \bibinfo {pages} {5383--5396} (\bibinfo {year} {2023})}\BibitemShut {NoStop}%
\bibitem [{\citenamefont {Curtarolo}\ \emph {et~al.}(2012)\citenamefont {Curtarolo}, \citenamefont {Setyawan}, \citenamefont {Hart}, \citenamefont {Jahnatek}, \citenamefont {Chepulskii}, \citenamefont {Taylor}, \citenamefont {Wang}, \citenamefont {Xue}, \citenamefont {Yang}, \citenamefont {Levy}, \citenamefont {Mehl}, \citenamefont {Stokes}, \citenamefont {Demchenko},\ and\ \citenamefont {Morgan}}]{Curtarolo2012}%
  \BibitemOpen
  \bibfield  {author} {\bibinfo {author} {\bibfnamefont {S.}~\bibnamefont {Curtarolo}}, \bibinfo {author} {\bibfnamefont {W.}~\bibnamefont {Setyawan}}, \bibinfo {author} {\bibfnamefont {G.~L.}\ \bibnamefont {Hart}}, \bibinfo {author} {\bibfnamefont {M.}~\bibnamefont {Jahnatek}}, \bibinfo {author} {\bibfnamefont {R.~V.}\ \bibnamefont {Chepulskii}}, \bibinfo {author} {\bibfnamefont {R.~H.}\ \bibnamefont {Taylor}}, \bibinfo {author} {\bibfnamefont {S.}~\bibnamefont {Wang}}, \bibinfo {author} {\bibfnamefont {J.}~\bibnamefont {Xue}}, \bibinfo {author} {\bibfnamefont {K.}~\bibnamefont {Yang}}, \bibinfo {author} {\bibfnamefont {O.}~\bibnamefont {Levy}}, \bibinfo {author} {\bibfnamefont {M.~J.}\ \bibnamefont {Mehl}}, \bibinfo {author} {\bibfnamefont {H.~T.}\ \bibnamefont {Stokes}}, \bibinfo {author} {\bibfnamefont {D.~O.}\ \bibnamefont {Demchenko}}, \ and\ \bibinfo {author} {\bibfnamefont {D.}~\bibnamefont {Morgan}},\ }\bibfield  {title} {\enquote {\bibinfo {title} {Aflow: An automatic framework for high-throughput
  materials discovery},}\ }\href {\doibase 10.1016/j.commatsci.2012.02.005} {\bibfield  {journal} {\bibinfo  {journal} {Computational Materials Science}\ }\textbf {\bibinfo {volume} {58}},\ \bibinfo {pages} {218--226} (\bibinfo {year} {2012})}\BibitemShut {NoStop}%
\bibitem [{\citenamefont {Jain}\ \emph {et~al.}(2011)\citenamefont {Jain}, \citenamefont {Hautier}, \citenamefont {Moore}, \citenamefont {Ping~Ong}, \citenamefont {Fischer}, \citenamefont {Mueller}, \citenamefont {Persson},\ and\ \citenamefont {Ceder}}]{Jain2011}%
  \BibitemOpen
  \bibfield  {author} {\bibinfo {author} {\bibfnamefont {A.}~\bibnamefont {Jain}}, \bibinfo {author} {\bibfnamefont {G.}~\bibnamefont {Hautier}}, \bibinfo {author} {\bibfnamefont {C.~J.}\ \bibnamefont {Moore}}, \bibinfo {author} {\bibfnamefont {S.}~\bibnamefont {Ping~Ong}}, \bibinfo {author} {\bibfnamefont {C.~C.}\ \bibnamefont {Fischer}}, \bibinfo {author} {\bibfnamefont {T.}~\bibnamefont {Mueller}}, \bibinfo {author} {\bibfnamefont {K.~A.}\ \bibnamefont {Persson}}, \ and\ \bibinfo {author} {\bibfnamefont {G.}~\bibnamefont {Ceder}},\ }\bibfield  {title} {\enquote {\bibinfo {title} {A high-throughput infrastructure for density functional theory calculations},}\ }\href {\doibase 10.1016/j.commatsci.2011.02.023} {\bibfield  {journal} {\bibinfo  {journal} {Computational Materials Science}\ }\textbf {\bibinfo {volume} {50}},\ \bibinfo {pages} {2295--2310} (\bibinfo {year} {2011})}\BibitemShut {NoStop}%
\bibitem [{\citenamefont {Huber}\ \emph {et~al.}(2020)\citenamefont {Huber}, \citenamefont {Zoupanos}, \citenamefont {Uhrin}, \citenamefont {Talirz}, \citenamefont {Kahle}, \citenamefont {H\"{a}uselmann}, \citenamefont {Gresch}, \citenamefont {M\"{u}ller}, \citenamefont {Yakutovich}, \citenamefont {Andersen}, \citenamefont {Ramirez}, \citenamefont {Adorf}, \citenamefont {Gargiulo}, \citenamefont {Kumbhar}, \citenamefont {Passaro}, \citenamefont {Johnston}, \citenamefont {Merkys}, \citenamefont {Cepellotti}, \citenamefont {Mounet}, \citenamefont {Marzari}, \citenamefont {Kozinsky},\ and\ \citenamefont {Pizzi}}]{Huber2020}%
  \BibitemOpen
  \bibfield  {author} {\bibinfo {author} {\bibfnamefont {S.~P.}\ \bibnamefont {Huber}}, \bibinfo {author} {\bibfnamefont {S.}~\bibnamefont {Zoupanos}}, \bibinfo {author} {\bibfnamefont {M.}~\bibnamefont {Uhrin}}, \bibinfo {author} {\bibfnamefont {L.}~\bibnamefont {Talirz}}, \bibinfo {author} {\bibfnamefont {L.}~\bibnamefont {Kahle}}, \bibinfo {author} {\bibfnamefont {R.}~\bibnamefont {H\"{a}uselmann}}, \bibinfo {author} {\bibfnamefont {D.}~\bibnamefont {Gresch}}, \bibinfo {author} {\bibfnamefont {T.}~\bibnamefont {M\"{u}ller}}, \bibinfo {author} {\bibfnamefont {A.~V.}\ \bibnamefont {Yakutovich}}, \bibinfo {author} {\bibfnamefont {C.~W.}\ \bibnamefont {Andersen}}, \bibinfo {author} {\bibfnamefont {F.~F.}\ \bibnamefont {Ramirez}}, \bibinfo {author} {\bibfnamefont {C.~S.}\ \bibnamefont {Adorf}}, \bibinfo {author} {\bibfnamefont {F.}~\bibnamefont {Gargiulo}}, \bibinfo {author} {\bibfnamefont {S.}~\bibnamefont {Kumbhar}}, \bibinfo {author} {\bibfnamefont {E.}~\bibnamefont {Passaro}}, \bibinfo {author}
  {\bibfnamefont {C.}~\bibnamefont {Johnston}}, \bibinfo {author} {\bibfnamefont {A.}~\bibnamefont {Merkys}}, \bibinfo {author} {\bibfnamefont {A.}~\bibnamefont {Cepellotti}}, \bibinfo {author} {\bibfnamefont {N.}~\bibnamefont {Mounet}}, \bibinfo {author} {\bibfnamefont {N.}~\bibnamefont {Marzari}}, \bibinfo {author} {\bibfnamefont {B.}~\bibnamefont {Kozinsky}}, \ and\ \bibinfo {author} {\bibfnamefont {G.}~\bibnamefont {Pizzi}},\ }\bibfield  {title} {\enquote {\bibinfo {title} {Aiida 1.0, a scalable computational infrastructure for automated reproducible workflows and data provenance},}\ }\href {\doibase 10.1038/s41597-020-00638-4} {\bibfield  {journal} {\bibinfo  {journal} {Scientific Data}\ }\textbf {\bibinfo {volume} {7}} (\bibinfo {year} {2020}),\ 10.1038/s41597-020-00638-4}\BibitemShut {NoStop}%
\bibitem [{\citenamefont {Canc{\`e}s}\ \emph {et~al.}(2022)\citenamefont {Canc{\`e}s}, \citenamefont {Levitt}, \citenamefont {Maday},\ and\ \citenamefont {Yang}}]{cances2022numerical}%
  \BibitemOpen
  \bibfield  {author} {\bibinfo {author} {\bibfnamefont {E.}~\bibnamefont {Canc{\`e}s}}, \bibinfo {author} {\bibfnamefont {A.}~\bibnamefont {Levitt}}, \bibinfo {author} {\bibfnamefont {Y.}~\bibnamefont {Maday}}, \ and\ \bibinfo {author} {\bibfnamefont {C.}~\bibnamefont {Yang}},\ }\bibfield  {title} {\enquote {\bibinfo {title} {Numerical methods for kohn--sham models: Discretization, algorithms, and error analysis},}\ }in\ \href@noop {} {\emph {\bibinfo {booktitle} {Density Functional Theory: Modeling, Mathematical Analysis, Computational Methods, and Applications}}}\ (\bibinfo  {publisher} {Springer},\ \bibinfo {year} {2022})\ pp.\ \bibinfo {pages} {333--400}\BibitemShut {NoStop}%
\bibitem [{\citenamefont {Herbst}\ and\ \citenamefont {Levitt}(2020)}]{Herbst2020}%
  \BibitemOpen
  \bibfield  {author} {\bibinfo {author} {\bibfnamefont {M.~F.}\ \bibnamefont {Herbst}}\ and\ \bibinfo {author} {\bibfnamefont {A.}~\bibnamefont {Levitt}},\ }\bibfield  {title} {\enquote {\bibinfo {title} {Black-box inhomogeneous preconditioning for self-consistent field iterations in density functional theory},}\ }\href {\doibase 10.1088/1361-648x/abcbdb} {\bibfield  {journal} {\bibinfo  {journal} {Journal of Physics: Condensed Matter}\ } (\bibinfo {year} {2020}),\ 10.1088/1361-648x/abcbdb}\BibitemShut {NoStop}%
\bibitem [{\citenamefont {Herbst}\ and\ \citenamefont {Levitt}(2022)}]{Herbst2022}%
  \BibitemOpen
  \bibfield  {author} {\bibinfo {author} {\bibfnamefont {M.~F.}\ \bibnamefont {Herbst}}\ and\ \bibinfo {author} {\bibfnamefont {A.}~\bibnamefont {Levitt}},\ }\bibfield  {title} {\enquote {\bibinfo {title} {A robust and efficient line search for self-consistent field iterations},}\ }\href {\doibase 10.1016/j.jcp.2022.111127} {\bibfield  {journal} {\bibinfo  {journal} {Journal of Computational Physics}\ }\textbf {\bibinfo {volume} {459}},\ \bibinfo {pages} {111127} (\bibinfo {year} {2022})}\BibitemShut {NoStop}%
\bibitem [{\citenamefont {Canc\`{e}s}\ \emph {et~al.}(2023)\citenamefont {Canc\`{e}s}, \citenamefont {Herbst}, \citenamefont {Kemlin}, \citenamefont {Levitt},\ and\ \citenamefont {Stamm}}]{Cances2023}%
  \BibitemOpen
  \bibfield  {author} {\bibinfo {author} {\bibfnamefont {E.}~\bibnamefont {Canc\`{e}s}}, \bibinfo {author} {\bibfnamefont {M.~F.}\ \bibnamefont {Herbst}}, \bibinfo {author} {\bibfnamefont {G.}~\bibnamefont {Kemlin}}, \bibinfo {author} {\bibfnamefont {A.}~\bibnamefont {Levitt}}, \ and\ \bibinfo {author} {\bibfnamefont {B.}~\bibnamefont {Stamm}},\ }\bibfield  {title} {\enquote {\bibinfo {title} {Numerical stability and efficiency of response property calculations in density functional theory},}\ }\href {\doibase 10.1007/s11005-023-01645-3} {\bibfield  {journal} {\bibinfo  {journal} {Letters in Mathematical Physics}\ }\textbf {\bibinfo {volume} {113}} (\bibinfo {year} {2023}),\ 10.1007/s11005-023-01645-3}\BibitemShut {NoStop}%
\bibitem [{\citenamefont {Ruddigkeit}\ \emph {et~al.}(2012)\citenamefont {Ruddigkeit}, \citenamefont {van Deursen}, \citenamefont {Blum},\ and\ \citenamefont {Reymond}}]{Ruddigkeit2012}%
  \BibitemOpen
  \bibfield  {author} {\bibinfo {author} {\bibfnamefont {L.}~\bibnamefont {Ruddigkeit}}, \bibinfo {author} {\bibfnamefont {R.}~\bibnamefont {van Deursen}}, \bibinfo {author} {\bibfnamefont {L.~C.}\ \bibnamefont {Blum}}, \ and\ \bibinfo {author} {\bibfnamefont {J.-L.}\ \bibnamefont {Reymond}},\ }\bibfield  {title} {\enquote {\bibinfo {title} {Enumeration of 166 billion organic small molecules in the chemical universe database gdb-17},}\ }\href {\doibase 10.1021/ci300415d} {\bibfield  {journal} {\bibinfo  {journal} {Journal of Chemical Information and Modeling}\ }\textbf {\bibinfo {volume} {52}},\ \bibinfo {pages} {2864--2875} (\bibinfo {year} {2012})},\ \bibinfo {note} {pMID: 23088335},\ \Eprint {http://arxiv.org/abs/https://doi.org/10.1021/ci300415d} {https://doi.org/10.1021/ci300415d} \BibitemShut {NoStop}%
\bibitem [{\citenamefont {Ramakrishnan}\ \emph {et~al.}(2014)\citenamefont {Ramakrishnan}, \citenamefont {Dral}, \citenamefont {Rupp},\ and\ \citenamefont {von Lilienfeld}}]{ramakrishnan2014quantum}%
  \BibitemOpen
  \bibfield  {author} {\bibinfo {author} {\bibfnamefont {R.}~\bibnamefont {Ramakrishnan}}, \bibinfo {author} {\bibfnamefont {P.~O.}\ \bibnamefont {Dral}}, \bibinfo {author} {\bibfnamefont {M.}~\bibnamefont {Rupp}}, \ and\ \bibinfo {author} {\bibfnamefont {O.~A.}\ \bibnamefont {von Lilienfeld}},\ }\bibfield  {title} {\enquote {\bibinfo {title} {Quantum chemistry structures and properties of 134 kilo molecules},}\ }\href@noop {} {\bibfield  {journal} {\bibinfo  {journal} {Scientific Data}\ }\textbf {\bibinfo {volume} {1}} (\bibinfo {year} {2014})}\BibitemShut {NoStop}%
\bibitem [{\citenamefont {Chanussot*}\ \emph {et~al.}(2021)\citenamefont {Chanussot*}, \citenamefont {Das*}, \citenamefont {Goyal*}, \citenamefont {Lavril*}, \citenamefont {Shuaibi*}, \citenamefont {Riviere}, \citenamefont {Tran}, \citenamefont {Heras-Domingo}, \citenamefont {Ho}, \citenamefont {Hu}, \citenamefont {Palizhati}, \citenamefont {Sriram}, \citenamefont {Wood}, \citenamefont {Yoon}, \citenamefont {Parikh}, \citenamefont {Zitnick},\ and\ \citenamefont {Ulissi}}]{ocp_dataset}%
  \BibitemOpen
  \bibfield  {author} {\bibinfo {author} {\bibfnamefont {L.}~\bibnamefont {Chanussot*}}, \bibinfo {author} {\bibfnamefont {A.}~\bibnamefont {Das*}}, \bibinfo {author} {\bibfnamefont {S.}~\bibnamefont {Goyal*}}, \bibinfo {author} {\bibfnamefont {T.}~\bibnamefont {Lavril*}}, \bibinfo {author} {\bibfnamefont {M.}~\bibnamefont {Shuaibi*}}, \bibinfo {author} {\bibfnamefont {M.}~\bibnamefont {Riviere}}, \bibinfo {author} {\bibfnamefont {K.}~\bibnamefont {Tran}}, \bibinfo {author} {\bibfnamefont {J.}~\bibnamefont {Heras-Domingo}}, \bibinfo {author} {\bibfnamefont {C.}~\bibnamefont {Ho}}, \bibinfo {author} {\bibfnamefont {W.}~\bibnamefont {Hu}}, \bibinfo {author} {\bibfnamefont {A.}~\bibnamefont {Palizhati}}, \bibinfo {author} {\bibfnamefont {A.}~\bibnamefont {Sriram}}, \bibinfo {author} {\bibfnamefont {B.}~\bibnamefont {Wood}}, \bibinfo {author} {\bibfnamefont {J.}~\bibnamefont {Yoon}}, \bibinfo {author} {\bibfnamefont {D.}~\bibnamefont {Parikh}}, \bibinfo {author} {\bibfnamefont {C.~L.}\ \bibnamefont {Zitnick}}, \
  and\ \bibinfo {author} {\bibfnamefont {Z.}~\bibnamefont {Ulissi}},\ }\bibfield  {title} {\enquote {\bibinfo {title} {Open catalyst 2020 (oc20) dataset and community challenges},}\ }\href {\doibase 10.1021/acscatal.0c04525} {\bibfield  {journal} {\bibinfo  {journal} {ACS Catalysis}\ } (\bibinfo {year} {2021}),\ 10.1021/acscatal.0c04525}\BibitemShut {NoStop}%
\bibitem [{\citenamefont {Smith}, \citenamefont {Isayev},\ and\ \citenamefont {Roitberg}(2017)}]{Smith2017}%
  \BibitemOpen
  \bibfield  {author} {\bibinfo {author} {\bibfnamefont {J.}~\bibnamefont {Smith}}, \bibinfo {author} {\bibfnamefont {O.}~\bibnamefont {Isayev}}, \ and\ \bibinfo {author} {\bibfnamefont {A.}~\bibnamefont {Roitberg}},\ }\bibfield  {title} {\enquote {\bibinfo {title} {A data set of 20 million calculated off-equilibrium conformations for organic molecules},}\ }\href@noop {} {\bibfield  {journal} {\bibinfo  {journal} {Scientific Data}\ }\textbf {\bibinfo {volume} {4}} (\bibinfo {year} {2017})}\BibitemShut {NoStop}%
\bibitem [{\citenamefont {Bonilla}, \citenamefont {Chai},\ and\ \citenamefont {Williams}(2008)}]{Bonilla2008}%
  \BibitemOpen
  \bibfield  {author} {\bibinfo {author} {\bibfnamefont {E.}~\bibnamefont {Bonilla}}, \bibinfo {author} {\bibfnamefont {K.}~\bibnamefont {Chai}}, \ and\ \bibinfo {author} {\bibfnamefont {C.}~\bibnamefont {Williams}},\ }\enquote {\bibinfo {title} {Multi-task gaussian process prediction},}\ in\ \href@noop {} {\emph {\bibinfo {booktitle} {Advances in neural information processing systems}}},\ \bibinfo {editor} {edited by\ \bibinfo {editor} {\bibfnamefont {J.}~\bibnamefont {Platt}}, \bibinfo {editor} {\bibfnamefont {D.}~\bibnamefont {Koller}}, \bibinfo {editor} {\bibfnamefont {Y.}~\bibnamefont {Singer}}, \ and\ \bibinfo {editor} {\bibfnamefont {S.}~\bibnamefont {Roweis}}}\ (\bibinfo  {publisher} {MIT Press},\ \bibinfo {address} {Cambridge, Massachusetts},\ \bibinfo {year} {2008})\ pp.\ \bibinfo {pages} {153--160}\BibitemShut {NoStop}%
\bibitem [{\citenamefont {Leen}, \citenamefont {Peltonen},\ and\ \citenamefont {Kaski}(2012)}]{Leen2012}%
  \BibitemOpen
  \bibfield  {author} {\bibinfo {author} {\bibfnamefont {G.}~\bibnamefont {Leen}}, \bibinfo {author} {\bibfnamefont {J.}~\bibnamefont {Peltonen}}, \ and\ \bibinfo {author} {\bibfnamefont {S.}~\bibnamefont {Kaski}},\ }\bibfield  {title} {\enquote {\bibinfo {title} {Focused multi-task learning in a gaussian process framework},}\ }\href {\doibase 10.1007/s10994-012-5302-y} {\bibfield  {journal} {\bibinfo  {journal} {Machine Learning}\ }\textbf {\bibinfo {volume} {1-2}},\ \bibinfo {pages} {157--182} (\bibinfo {year} {2012})}\BibitemShut {NoStop}%
\bibitem [{\citenamefont {Pilania}, \citenamefont {Gubernatis},\ and\ \citenamefont {Lookman}(2017)}]{Pilania2017}%
  \BibitemOpen
  \bibfield  {author} {\bibinfo {author} {\bibfnamefont {G.}~\bibnamefont {Pilania}}, \bibinfo {author} {\bibfnamefont {J.}~\bibnamefont {Gubernatis}}, \ and\ \bibinfo {author} {\bibfnamefont {T.}~\bibnamefont {Lookman}},\ }\bibfield  {title} {\enquote {\bibinfo {title} {Multi-fidelity machine learning models for accurate bandgap predictions of solids},}\ }\href {\doibase 10.1016/j.commatsci.2016.12.004} {\bibfield  {journal} {\bibinfo  {journal} {Computational Materials Science}\ }\textbf {\bibinfo {volume} {129}},\ \bibinfo {pages} {156--162} (\bibinfo {year} {2017})}\BibitemShut {NoStop}%
\bibitem [{\citenamefont {Batra}\ \emph {et~al.}(2019)\citenamefont {Batra}, \citenamefont {Pilania}, \citenamefont {Uberuaga},\ and\ \citenamefont {Ramprasad}}]{Batra2019}%
  \BibitemOpen
  \bibfield  {author} {\bibinfo {author} {\bibfnamefont {R.}~\bibnamefont {Batra}}, \bibinfo {author} {\bibfnamefont {G.}~\bibnamefont {Pilania}}, \bibinfo {author} {\bibfnamefont {B.}~\bibnamefont {Uberuaga}}, \ and\ \bibinfo {author} {\bibfnamefont {R.}~\bibnamefont {Ramprasad}},\ }\bibfield  {title} {\enquote {\bibinfo {title} {Multifidelity information fusion with machine learning: A case study of dopant formation energies in hafnia},}\ }\href {\doibase 10.1021/acsami.9b02174} {\bibfield  {journal} {\bibinfo  {journal} {ACS Applied Materials \& Inference}\ }\textbf {\bibinfo {volume} {11}},\ \bibinfo {pages} {24906--24918} (\bibinfo {year} {2019})}\BibitemShut {NoStop}%
\bibitem [{\citenamefont {Patra}\ \emph {et~al.}(2020)\citenamefont {Patra}, \citenamefont {Batra}, \citenamefont {Chandrasekaran}, \citenamefont {Kim}, \citenamefont {Huan},\ and\ \citenamefont {Ramprasad}}]{Patra2019}%
  \BibitemOpen
  \bibfield  {author} {\bibinfo {author} {\bibfnamefont {A.}~\bibnamefont {Patra}}, \bibinfo {author} {\bibfnamefont {R.}~\bibnamefont {Batra}}, \bibinfo {author} {\bibfnamefont {A.}~\bibnamefont {Chandrasekaran}}, \bibinfo {author} {\bibfnamefont {C.}~\bibnamefont {Kim}}, \bibinfo {author} {\bibfnamefont {T.~D.}\ \bibnamefont {Huan}}, \ and\ \bibinfo {author} {\bibfnamefont {R.}~\bibnamefont {Ramprasad}},\ }\bibfield  {title} {\enquote {\bibinfo {title} {A multi-fidelity information-fusion approach to machine learn and predict polymer bandgap},}\ }\href {\doibase https://doi.org/10.1016/j.commatsci.2019.109286} {\bibfield  {journal} {\bibinfo  {journal} {Computational Materials Science}\ }\textbf {\bibinfo {volume} {172}},\ \bibinfo {pages} {109286} (\bibinfo {year} {2020})}\BibitemShut {NoStop}%
\bibitem [{\citenamefont {Kennedy}\ and\ \citenamefont {O'Hagan}(2000)}]{Kennedy2000}%
  \BibitemOpen
  \bibfield  {author} {\bibinfo {author} {\bibfnamefont {M.}~\bibnamefont {Kennedy}}\ and\ \bibinfo {author} {\bibfnamefont {A.}~\bibnamefont {O'Hagan}},\ }\bibfield  {title} {\enquote {\bibinfo {title} {Predicting the output from a complex computer code when fast approximations are available},}\ }\href {\doibase 10.1093/biomet/87.1.1} {\bibfield  {journal} {\bibinfo  {journal} {Biometrika}\ }\textbf {\bibinfo {volume} {87}},\ \bibinfo {pages} {1--13} (\bibinfo {year} {2000})}\BibitemShut {NoStop}%
\bibitem [{\citenamefont {Ramakrishnan}\ \emph {et~al.}(2015)\citenamefont {Ramakrishnan}, \citenamefont {Dral}, \citenamefont {Rupp},\ and\ \citenamefont {von Lilienfeld}}]{Ramakrishnan2015}%
  \BibitemOpen
  \bibfield  {author} {\bibinfo {author} {\bibfnamefont {R.}~\bibnamefont {Ramakrishnan}}, \bibinfo {author} {\bibfnamefont {P.~O.}\ \bibnamefont {Dral}}, \bibinfo {author} {\bibfnamefont {M.}~\bibnamefont {Rupp}}, \ and\ \bibinfo {author} {\bibfnamefont {O.~A.}\ \bibnamefont {von Lilienfeld}},\ }\bibfield  {title} {\enquote {\bibinfo {title} {Big data meets quantum chemistry approximations: The \ensuremath{\Delta}-machine learning approach},}\ }\href {\doibase 10.1021/acs.jctc.5b00099} {\bibfield  {journal} {\bibinfo  {journal} {Journal of Chemical Theory and Computation}\ }\textbf {\bibinfo {volume} {11}},\ \bibinfo {pages} {2087--2096} (\bibinfo {year} {2015})}\BibitemShut {NoStop}%
\bibitem [{\citenamefont {Dral}, \citenamefont {Zubatiuk},\ and\ \citenamefont {Xue}(2023)}]{dral2023learning}%
  \BibitemOpen
  \bibfield  {author} {\bibinfo {author} {\bibfnamefont {P.~O.}\ \bibnamefont {Dral}}, \bibinfo {author} {\bibfnamefont {T.}~\bibnamefont {Zubatiuk}}, \ and\ \bibinfo {author} {\bibfnamefont {B.-X.}\ \bibnamefont {Xue}},\ }\bibfield  {title} {\enquote {\bibinfo {title} {Learning from multiple quantum chemical methods: $\delta$-learning, transfer learning, co-kriging, and beyond},}\ }in\ \href@noop {} {\emph {\bibinfo {booktitle} {Quantum Chemistry in the Age of Machine Learning}}}\ (\bibinfo  {publisher} {Elsevier},\ \bibinfo {year} {2023})\ pp.\ \bibinfo {pages} {491--507}\BibitemShut {NoStop}%
\bibitem [{\citenamefont {Vinod}, \citenamefont {Kleinekathöfer},\ and\ \citenamefont {Zaspel}(2023)}]{vinod2023optimized}%
  \BibitemOpen
  \bibfield  {author} {\bibinfo {author} {\bibfnamefont {V.}~\bibnamefont {Vinod}}, \bibinfo {author} {\bibfnamefont {U.}~\bibnamefont {Kleinekathöfer}}, \ and\ \bibinfo {author} {\bibfnamefont {P.}~\bibnamefont {Zaspel}},\ }\href@noop {} {\enquote {\bibinfo {title} {Optimized multifidelity machine learning for quantum chemistry},}\ } (\bibinfo {year} {2023}),\ \Eprint {http://arxiv.org/abs/2312.05661} {arXiv:2312.05661 [physics.chem-ph]} \BibitemShut {NoStop}%
\bibitem [{\citenamefont {Vinod}\ \emph {et~al.}(2023)\citenamefont {Vinod}, \citenamefont {Maity}, \citenamefont {Zaspel},\ and\ \citenamefont {Kleinekathöfer}}]{Vinod2023}%
  \BibitemOpen
  \bibfield  {author} {\bibinfo {author} {\bibfnamefont {V.}~\bibnamefont {Vinod}}, \bibinfo {author} {\bibfnamefont {S.}~\bibnamefont {Maity}}, \bibinfo {author} {\bibfnamefont {P.}~\bibnamefont {Zaspel}}, \ and\ \bibinfo {author} {\bibfnamefont {U.}~\bibnamefont {Kleinekathöfer}},\ }\bibfield  {title} {\enquote {\bibinfo {title} {Multifidelity machine learning for molecular excitation energies},}\ }\href {\doibase 10.1021/acs.jctc.3c00882} {\bibfield  {journal} {\bibinfo  {journal} {Journal of Chemical Theory and Computation}\ }\textbf {\bibinfo {volume} {19}},\ \bibinfo {pages} {7658--7670} (\bibinfo {year} {2023})},\ \bibinfo {note} {pMID: 37862054}\BibitemShut {NoStop}%
\bibitem [{\citenamefont {Towell}\ and\ \citenamefont {Shavlik}(1994)}]{TOWELL1994119}%
  \BibitemOpen
  \bibfield  {author} {\bibinfo {author} {\bibfnamefont {G.~G.}\ \bibnamefont {Towell}}\ and\ \bibinfo {author} {\bibfnamefont {J.~W.}\ \bibnamefont {Shavlik}},\ }\bibfield  {title} {\enquote {\bibinfo {title} {Knowledge-based artificial neural networks},}\ }\href {\doibase https://doi.org/10.1016/0004-3702(94)90105-8} {\bibfield  {journal} {\bibinfo  {journal} {Artificial Intelligence}\ }\textbf {\bibinfo {volume} {70}},\ \bibinfo {pages} {119--165} (\bibinfo {year} {1994})}\BibitemShut {NoStop}%
\bibitem [{\citenamefont {Fu}(1989)}]{Fu1989}%
  \BibitemOpen
  \bibfield  {author} {\bibinfo {author} {\bibfnamefont {L.-M.}\ \bibnamefont {Fu}},\ }\bibfield  {title} {\enquote {\bibinfo {title} {Integration of neural heuristics into knowledge-based inference},}\ }in\ \href {\doibase 10.1109/IJCNN.1989.118425} {\emph {\bibinfo {booktitle} {International 1989 Joint Conference on Neural Networks}}}\ (\bibinfo {year} {1989})\ pp.\ \bibinfo {pages} {606 vol.2--}\BibitemShut {NoStop}%
\bibitem [{\citenamefont {Bartlett}, \citenamefont {Montanari},\ and\ \citenamefont {Rakhlin}(2021)}]{bartlett_montanari_rakhlin_2021}%
  \BibitemOpen
  \bibfield  {author} {\bibinfo {author} {\bibfnamefont {P.~L.}\ \bibnamefont {Bartlett}}, \bibinfo {author} {\bibfnamefont {A.}~\bibnamefont {Montanari}}, \ and\ \bibinfo {author} {\bibfnamefont {A.}~\bibnamefont {Rakhlin}},\ }\bibfield  {title} {\enquote {\bibinfo {title} {Deep learning: a statistical viewpoint},}\ }\href {\doibase 10.1017/S0962492921000027} {\bibfield  {journal} {\bibinfo  {journal} {Acta Numerica}\ }\textbf {\bibinfo {volume} {30}},\ \bibinfo {pages} {87–201} (\bibinfo {year} {2021})}\BibitemShut {NoStop}%
\bibitem [{\citenamefont {Lotfi}\ \emph {et~al.}(2022)\citenamefont {Lotfi}, \citenamefont {Finzi}, \citenamefont {Kapoor}, \citenamefont {Potapczynski}, \citenamefont {Goldblum},\ and\ \citenamefont {Wilson}}]{lotfi2022pacbayes}%
  \BibitemOpen
  \bibfield  {author} {\bibinfo {author} {\bibfnamefont {S.}~\bibnamefont {Lotfi}}, \bibinfo {author} {\bibfnamefont {M.}~\bibnamefont {Finzi}}, \bibinfo {author} {\bibfnamefont {S.}~\bibnamefont {Kapoor}}, \bibinfo {author} {\bibfnamefont {A.}~\bibnamefont {Potapczynski}}, \bibinfo {author} {\bibfnamefont {M.}~\bibnamefont {Goldblum}}, \ and\ \bibinfo {author} {\bibfnamefont {A.~G.}\ \bibnamefont {Wilson}},\ }\href@noop {} {\enquote {\bibinfo {title} {Pac-bayes compression bounds so tight that they can explain generalization},}\ } (\bibinfo {year} {2022}),\ \Eprint {http://arxiv.org/abs/2211.13609} {arXiv:2211.13609 [cs.LG]} \BibitemShut {NoStop}%
\bibitem [{\citenamefont {Rasmussen}\ and\ \citenamefont {Williams}(2006)}]{gpml}%
  \BibitemOpen
  \bibfield  {author} {\bibinfo {author} {\bibfnamefont {C.~E.}\ \bibnamefont {Rasmussen}}\ and\ \bibinfo {author} {\bibfnamefont {C.~K.~I.}\ \bibnamefont {Williams}},\ }\href {www.GaussianProcess.org/gpml} {\emph {\bibinfo {title} {Gaussian Processes for Machine Learning}}}\ (\bibinfo  {publisher} {the MIT Press},\ \bibinfo {year} {2006})\BibitemShut {NoStop}%
\bibitem [{\citenamefont {Quinonero-Candela}\ and\ \citenamefont {Rasmussen}(2005)}]{Quinonero-Candela2005}%
  \BibitemOpen
  \bibfield  {author} {\bibinfo {author} {\bibfnamefont {J.}~\bibnamefont {Quinonero-Candela}}\ and\ \bibinfo {author} {\bibfnamefont {C.}~\bibnamefont {Rasmussen}},\ }\bibfield  {title} {\enquote {\bibinfo {title} {A unifying view of sparse approximate gaussian process regression},}\ }\href {https://www.jmlr.org/papers/volume6/quinonero-candela05a/quinonero-candela05a.pdf} {\bibfield  {journal} {\bibinfo  {journal} {Journal of Machine Learning Research}\ }\textbf {\bibinfo {volume} {6}},\ \bibinfo {pages} {1939–1959} (\bibinfo {year} {2005})}\BibitemShut {NoStop}%
\bibitem [{\citenamefont {Liu}\ \emph {et~al.}(2020)\citenamefont {Liu}, \citenamefont {Ong}, \citenamefont {Shen},\ and\ \citenamefont {Cai}}]{Liu2018}%
  \BibitemOpen
  \bibfield  {author} {\bibinfo {author} {\bibfnamefont {H.}~\bibnamefont {Liu}}, \bibinfo {author} {\bibfnamefont {Y.-S.}\ \bibnamefont {Ong}}, \bibinfo {author} {\bibfnamefont {X.}~\bibnamefont {Shen}}, \ and\ \bibinfo {author} {\bibfnamefont {J.}~\bibnamefont {Cai}},\ }\bibfield  {title} {\enquote {\bibinfo {title} {When gaussian process meets big data: A review of scalable gps},}\ }\href {\doibase 10.1109/tnnls.2019.2957109} {\bibfield  {journal} {\bibinfo  {journal} {IEEE Transactions on Neural Networks and Learning Systems}\ }\textbf {\bibinfo {volume} {31}},\ \bibinfo {pages} {4405--4423} (\bibinfo {year} {2020})}\BibitemShut {NoStop}%
\bibitem [{\citenamefont {Wilson}, \citenamefont {Dann},\ and\ \citenamefont {Nickisch}(2015)}]{Wilson2015}%
  \BibitemOpen
  \bibfield  {author} {\bibinfo {author} {\bibfnamefont {A.~G.}\ \bibnamefont {Wilson}}, \bibinfo {author} {\bibfnamefont {C.}~\bibnamefont {Dann}}, \ and\ \bibinfo {author} {\bibfnamefont {H.}~\bibnamefont {Nickisch}},\ }\bibfield  {title} {\enquote {\bibinfo {title} {Thoughts on massively scalable gaussian processes},}\ }\href {http://arxiv.org/abs/1511.01870} {\bibfield  {journal} {\bibinfo  {journal} {CoRR}\ }\textbf {\bibinfo {volume} {abs/1511.01870}} (\bibinfo {year} {2015})},\ \Eprint {http://arxiv.org/abs/1511.01870} {1511.01870} \BibitemShut {NoStop}%
\bibitem [{\citenamefont {Cole}, \citenamefont {Christianson},\ and\ \citenamefont {Gramacy}(2021)}]{Cole2021}%
  \BibitemOpen
  \bibfield  {author} {\bibinfo {author} {\bibfnamefont {D.~A.}\ \bibnamefont {Cole}}, \bibinfo {author} {\bibfnamefont {R.~B.}\ \bibnamefont {Christianson}}, \ and\ \bibinfo {author} {\bibfnamefont {R.~B.}\ \bibnamefont {Gramacy}},\ }\bibfield  {title} {\enquote {\bibinfo {title} {Locally induced gaussian processes for large-scale simulation experiments},}\ }\href {\doibase https://doi.org/10.1007/s11222-021-10007-9} {\bibfield  {journal} {\bibinfo  {journal} {Statistics and Computing}\ }\textbf {\bibinfo {volume} {31}},\ \bibinfo {pages} {1573--1375} (\bibinfo {year} {2021})}\BibitemShut {NoStop}%
\bibitem [{\citenamefont {Abrahamsen}(1997)}]{Petter1997}%
  \BibitemOpen
  \bibfield  {author} {\bibinfo {author} {\bibfnamefont {P.}~\bibnamefont {Abrahamsen}},\ }\bibfield  {title} {\enquote {\bibinfo {title} {A review of gaussian random fields and correlation functions},}\ }\href {\doibase 10.13140/RG.2.2.23937.20325} {\  (\bibinfo {year} {1997}),\ 10.13140/RG.2.2.23937.20325}\BibitemShut {NoStop}%
\bibitem [{\citenamefont {Bart\'{o}k}\ \emph {et~al.}(2010)\citenamefont {Bart\'{o}k}, \citenamefont {Payne}, \citenamefont {Kondor},\ and\ \citenamefont {Cs\'{a}nyi}}]{Bartok2010}%
  \BibitemOpen
  \bibfield  {author} {\bibinfo {author} {\bibfnamefont {A.~P.}\ \bibnamefont {Bart\'{o}k}}, \bibinfo {author} {\bibfnamefont {M.~C.}\ \bibnamefont {Payne}}, \bibinfo {author} {\bibfnamefont {R.}~\bibnamefont {Kondor}}, \ and\ \bibinfo {author} {\bibfnamefont {G.}~\bibnamefont {Cs\'{a}nyi}},\ }\bibfield  {title} {\enquote {\bibinfo {title} {Gaussian approximation potentials: The accuracy of quantum mechanics, without the electrons},}\ }\href {\doibase 10.1103/physrevlett.104.136403} {\bibfield  {journal} {\bibinfo  {journal} {Physical Review Letters}\ }\textbf {\bibinfo {volume} {104}} (\bibinfo {year} {2010}),\ 10.1103/physrevlett.104.136403}\BibitemShut {NoStop}%
\bibitem [{\citenamefont {Bart\'{o}k}, \citenamefont {Kondor},\ and\ \citenamefont {Cs\'{a}nyi}(2013)}]{Bartok2013}%
  \BibitemOpen
  \bibfield  {author} {\bibinfo {author} {\bibfnamefont {A.~P.}\ \bibnamefont {Bart\'{o}k}}, \bibinfo {author} {\bibfnamefont {R.}~\bibnamefont {Kondor}}, \ and\ \bibinfo {author} {\bibfnamefont {G.}~\bibnamefont {Cs\'{a}nyi}},\ }\bibfield  {title} {\enquote {\bibinfo {title} {On representing chemical environments},}\ }\href {\doibase 10.1103/physrevb.87.184115} {\bibfield  {journal} {\bibinfo  {journal} {Physical Review B}\ }\textbf {\bibinfo {volume} {87}},\ \bibinfo {pages} {184115} (\bibinfo {year} {2013})}\BibitemShut {NoStop}%
\bibitem [{\citenamefont {Bart\'{o}k}\ and\ \citenamefont {Cs\'{a}nyi}(2015)}]{Bartok2015}%
  \BibitemOpen
  \bibfield  {author} {\bibinfo {author} {\bibfnamefont {A.~P.}\ \bibnamefont {Bart\'{o}k}}\ and\ \bibinfo {author} {\bibfnamefont {G.}~\bibnamefont {Cs\'{a}nyi}},\ }\bibfield  {title} {\enquote {\bibinfo {title} {Gaussian approximation potentials: A brief tutorial introduction},}\ }\href {\doibase 10.1002/qua.24927} {\bibfield  {journal} {\bibinfo  {journal} {International Journal of Quantum Chemistry}\ }\textbf {\bibinfo {volume} {115}},\ \bibinfo {pages} {1051--1057} (\bibinfo {year} {2015})}\BibitemShut {NoStop}%
\bibitem [{\citenamefont {Khatamsaz}, \citenamefont {Vela},\ and\ \citenamefont {Arróyave}(2023)}]{Khatamsaz2023}%
  \BibitemOpen
  \bibfield  {author} {\bibinfo {author} {\bibfnamefont {D.}~\bibnamefont {Khatamsaz}}, \bibinfo {author} {\bibfnamefont {B.}~\bibnamefont {Vela}}, \ and\ \bibinfo {author} {\bibfnamefont {R.}~\bibnamefont {Arróyave}},\ }\bibfield  {title} {\enquote {\bibinfo {title} {Multi-objective bayesian alloy design using multi-task gaussian processes},}\ }\href {\doibase https://doi.org/10.1016/j.matlet.2023.135067} {\bibfield  {journal} {\bibinfo  {journal} {Materials Letters}\ }\textbf {\bibinfo {volume} {351}},\ \bibinfo {pages} {135067} (\bibinfo {year} {2023})}\BibitemShut {NoStop}%
\bibitem [{\citenamefont {Chen}\ \emph {et~al.}(2017)\citenamefont {Chen}, \citenamefont {Ko}, \citenamefont {Remsing}, \citenamefont {Andrade}, \citenamefont {Santra}, \citenamefont {Sun}, \citenamefont {Selloni}, \citenamefont {Car}, \citenamefont {Klein}, \citenamefont {Perdew},\ and\ \citenamefont {Xifan}}]{Chen2017}%
  \BibitemOpen
  \bibfield  {author} {\bibinfo {author} {\bibfnamefont {M.}~\bibnamefont {Chen}}, \bibinfo {author} {\bibfnamefont {H.-Y.}\ \bibnamefont {Ko}}, \bibinfo {author} {\bibfnamefont {R.}~\bibnamefont {Remsing}}, \bibinfo {author} {\bibfnamefont {M.}~\bibnamefont {Andrade}}, \bibinfo {author} {\bibfnamefont {B.}~\bibnamefont {Santra}}, \bibinfo {author} {\bibfnamefont {Z.}~\bibnamefont {Sun}}, \bibinfo {author} {\bibfnamefont {A.}~\bibnamefont {Selloni}}, \bibinfo {author} {\bibfnamefont {R.}~\bibnamefont {Car}}, \bibinfo {author} {\bibfnamefont {M.}~\bibnamefont {Klein}}, \bibinfo {author} {\bibfnamefont {J.}~\bibnamefont {Perdew}}, \ and\ \bibinfo {author} {\bibfnamefont {W.}~\bibnamefont {Xifan}},\ }\bibfield  {title} {\enquote {\bibinfo {title} {Ab initio theory and modeling of water},}\ }\href {\doibase 10.1073/pnas.1712499114} {\bibfield  {journal} {\bibinfo  {journal} {Proceedings of the National Academy of Sciences}\ }\textbf {\bibinfo {volume} {114}},\ \bibinfo {pages} {201712499} (\bibinfo {year}
  {2017})}\BibitemShut {NoStop}%
\bibitem [{\citenamefont {Dasgupta}\ \emph {et~al.}(2021)\citenamefont {Dasgupta}, \citenamefont {Lambros}, \citenamefont {Perdew},\ and\ \citenamefont {Paesani}}]{Dasgupta2021}%
  \BibitemOpen
  \bibfield  {author} {\bibinfo {author} {\bibfnamefont {S.}~\bibnamefont {Dasgupta}}, \bibinfo {author} {\bibfnamefont {E.}~\bibnamefont {Lambros}}, \bibinfo {author} {\bibfnamefont {J.}~\bibnamefont {Perdew}}, \ and\ \bibinfo {author} {\bibfnamefont {F.}~\bibnamefont {Paesani}},\ }\bibfield  {title} {\enquote {\bibinfo {title} {Elevating density functional theory to chemical accuracy for water simulations through a density-corrected many-body formalism},}\ }\href {\doibase 10.33774/chemrxiv-2021-hstgf} {\  (\bibinfo {year} {2021}),\ 10.33774/chemrxiv-2021-hstgf}\BibitemShut {NoStop}%
\bibitem [{\citenamefont {Gillan}, \citenamefont {Alfè},\ and\ \citenamefont {Michaelides}(2016)}]{Gillan2016}%
  \BibitemOpen
  \bibfield  {author} {\bibinfo {author} {\bibfnamefont {M.~J.}\ \bibnamefont {Gillan}}, \bibinfo {author} {\bibfnamefont {D.}~\bibnamefont {Alfè}}, \ and\ \bibinfo {author} {\bibfnamefont {A.}~\bibnamefont {Michaelides}},\ }\bibfield  {title} {\enquote {\bibinfo {title} {{Perspective: How good is DFT for water?}}}\ }\href {\doibase 10.1063/1.4944633} {\bibfield  {journal} {\bibinfo  {journal} {The Journal of Chemical Physics}\ }\textbf {\bibinfo {volume} {144}},\ \bibinfo {pages} {130901} (\bibinfo {year} {2016})},\ \Eprint {http://arxiv.org/abs/https://pubs.aip.org/aip/jcp/article-pdf/doi/10.1063/1.4944633/13696839/130901\_1\_online.pdf} {https://pubs.aip.org/aip/jcp/article-pdf/doi/10.1063/1.4944633/13696839/130901\_1\_online.pdf} \BibitemShut {NoStop}%
\bibitem [{\citenamefont {Yu}\ \emph {et~al.}(2022)\citenamefont {Yu}, \citenamefont {Qu}, \citenamefont {Houston}, \citenamefont {Conte}, \citenamefont {Nandi},\ and\ \citenamefont {Bowman}}]{Yu2022}%
  \BibitemOpen
  \bibfield  {author} {\bibinfo {author} {\bibfnamefont {Q.}~\bibnamefont {Yu}}, \bibinfo {author} {\bibfnamefont {C.}~\bibnamefont {Qu}}, \bibinfo {author} {\bibfnamefont {P.~L.}\ \bibnamefont {Houston}}, \bibinfo {author} {\bibfnamefont {R.}~\bibnamefont {Conte}}, \bibinfo {author} {\bibfnamefont {A.}~\bibnamefont {Nandi}}, \ and\ \bibinfo {author} {\bibfnamefont {J.~M.}\ \bibnamefont {Bowman}},\ }\bibfield  {title} {\enquote {\bibinfo {title} {q-aqua: A many-body ccsd(t) water potential, including four-body interactions, demonstrates the quantum nature of water from clusters to the liquid phase},}\ }\href {\doibase 10.1021/acs.jpclett.2c00966} {\bibfield  {journal} {\bibinfo  {journal} {The Journal of Physical Chemistry Letters}\ }\textbf {\bibinfo {volume} {13}},\ \bibinfo {pages} {5068--5074} (\bibinfo {year} {2022})},\ \bibinfo {note} {pMID: 35652912}\BibitemShut {NoStop}%
\bibitem [{\citenamefont {Smith}\ \emph {et~al.}(2020)\citenamefont {Smith}, \citenamefont {Burns}, \citenamefont {Simmonett}, \citenamefont {Parrish}, \citenamefont {Schieber}, \citenamefont {Galvelis}, \citenamefont {Kraus}, \citenamefont {Kruse}, \citenamefont {Di~Remigio}, \citenamefont {Alenaizan}, \citenamefont {James}, \citenamefont {Lehtola}, \citenamefont {Misiewicz}, \citenamefont {Scheurer}, \citenamefont {Shaw}, \citenamefont {Schriber}, \citenamefont {Xie}, \citenamefont {Glick}, \citenamefont {Sirianni}, \citenamefont {O’Brien}, \citenamefont {Waldrop}, \citenamefont {Kumar}, \citenamefont {Hohenstein}, \citenamefont {Pritchard}, \citenamefont {Brooks}, \citenamefont {Schaefer}, \citenamefont {Sokolov}, \citenamefont {Patkowski}, \citenamefont {DePrince}, \citenamefont {Bozkaya}, \citenamefont {King}, \citenamefont {Evangelista}, \citenamefont {Turney}, \citenamefont {Crawford},\ and\ \citenamefont {Sherrill}}]{Smith2020}%
  \BibitemOpen
  \bibfield  {author} {\bibinfo {author} {\bibfnamefont {D.~G.~A.}\ \bibnamefont {Smith}}, \bibinfo {author} {\bibfnamefont {L.~A.}\ \bibnamefont {Burns}}, \bibinfo {author} {\bibfnamefont {A.~C.}\ \bibnamefont {Simmonett}}, \bibinfo {author} {\bibfnamefont {R.~M.}\ \bibnamefont {Parrish}}, \bibinfo {author} {\bibfnamefont {M.~C.}\ \bibnamefont {Schieber}}, \bibinfo {author} {\bibfnamefont {R.}~\bibnamefont {Galvelis}}, \bibinfo {author} {\bibfnamefont {P.}~\bibnamefont {Kraus}}, \bibinfo {author} {\bibfnamefont {H.}~\bibnamefont {Kruse}}, \bibinfo {author} {\bibfnamefont {R.}~\bibnamefont {Di~Remigio}}, \bibinfo {author} {\bibfnamefont {A.}~\bibnamefont {Alenaizan}}, \bibinfo {author} {\bibfnamefont {A.~M.}\ \bibnamefont {James}}, \bibinfo {author} {\bibfnamefont {S.}~\bibnamefont {Lehtola}}, \bibinfo {author} {\bibfnamefont {J.~P.}\ \bibnamefont {Misiewicz}}, \bibinfo {author} {\bibfnamefont {M.}~\bibnamefont {Scheurer}}, \bibinfo {author} {\bibfnamefont {R.~A.}\ \bibnamefont {Shaw}}, \bibinfo {author}
  {\bibfnamefont {J.~B.}\ \bibnamefont {Schriber}}, \bibinfo {author} {\bibfnamefont {Y.}~\bibnamefont {Xie}}, \bibinfo {author} {\bibfnamefont {Z.~L.}\ \bibnamefont {Glick}}, \bibinfo {author} {\bibfnamefont {D.~A.}\ \bibnamefont {Sirianni}}, \bibinfo {author} {\bibfnamefont {J.~S.}\ \bibnamefont {O’Brien}}, \bibinfo {author} {\bibfnamefont {J.~M.}\ \bibnamefont {Waldrop}}, \bibinfo {author} {\bibfnamefont {A.}~\bibnamefont {Kumar}}, \bibinfo {author} {\bibfnamefont {E.~G.}\ \bibnamefont {Hohenstein}}, \bibinfo {author} {\bibfnamefont {B.~P.}\ \bibnamefont {Pritchard}}, \bibinfo {author} {\bibfnamefont {B.~R.}\ \bibnamefont {Brooks}}, \bibinfo {author} {\bibfnamefont {I.}~\bibnamefont {Schaefer}, \bibfnamefont {Henry~F.}}, \bibinfo {author} {\bibfnamefont {A.~Y.}\ \bibnamefont {Sokolov}}, \bibinfo {author} {\bibfnamefont {K.}~\bibnamefont {Patkowski}}, \bibinfo {author} {\bibfnamefont {I.}~\bibnamefont {DePrince}, \bibfnamefont {A.~Eugene}}, \bibinfo {author} {\bibfnamefont {U.}~\bibnamefont {Bozkaya}},
  \bibinfo {author} {\bibfnamefont {R.~A.}\ \bibnamefont {King}}, \bibinfo {author} {\bibfnamefont {F.~A.}\ \bibnamefont {Evangelista}}, \bibinfo {author} {\bibfnamefont {J.~M.}\ \bibnamefont {Turney}}, \bibinfo {author} {\bibfnamefont {T.~D.}\ \bibnamefont {Crawford}}, \ and\ \bibinfo {author} {\bibfnamefont {C.~D.}\ \bibnamefont {Sherrill}},\ }\bibfield  {title} {\enquote {\bibinfo {title} {{PSI4 1.4: Open-source software for high-throughput quantum chemistry}},}\ }\href {\doibase 10.1063/5.0006002} {\bibfield  {journal} {\bibinfo  {journal} {The Journal of Chemical Physics}\ }\textbf {\bibinfo {volume} {152}},\ \bibinfo {pages} {184108} (\bibinfo {year} {2020})},\ \Eprint {http://arxiv.org/abs/https://pubs.aip.org/aip/jcp/article-pdf/doi/10.1063/5.0006002/16684807/184108\_1\_online.pdf} {https://pubs.aip.org/aip/jcp/article-pdf/doi/10.1063/5.0006002/16684807/184108\_1\_online.pdf} \BibitemShut {NoStop}%
\bibitem [{\citenamefont {Perdew}, \citenamefont {Burke},\ and\ \citenamefont {Ernzerhof}(1996)}]{PBE}%
  \BibitemOpen
  \bibfield  {author} {\bibinfo {author} {\bibfnamefont {J.~P.}\ \bibnamefont {Perdew}}, \bibinfo {author} {\bibfnamefont {K.}~\bibnamefont {Burke}}, \ and\ \bibinfo {author} {\bibfnamefont {M.}~\bibnamefont {Ernzerhof}},\ }\bibfield  {title} {\enquote {\bibinfo {title} {Generalized gradient approximation made simple},}\ }\href {\doibase 10.1103/PhysRevLett.77.3865} {\bibfield  {journal} {\bibinfo  {journal} {Phys. Rev. Lett.}\ }\textbf {\bibinfo {volume} {77}},\ \bibinfo {pages} {3865--3868} (\bibinfo {year} {1996})}\BibitemShut {NoStop}%
\bibitem [{\citenamefont {Sun}, \citenamefont {Ruzsinszky},\ and\ \citenamefont {Perdew}(2015)}]{SCAN}%
  \BibitemOpen
  \bibfield  {author} {\bibinfo {author} {\bibfnamefont {J.}~\bibnamefont {Sun}}, \bibinfo {author} {\bibfnamefont {A.}~\bibnamefont {Ruzsinszky}}, \ and\ \bibinfo {author} {\bibfnamefont {J.~P.}\ \bibnamefont {Perdew}},\ }\bibfield  {title} {\enquote {\bibinfo {title} {Strongly constrained and appropriately normed semilocal density functional},}\ }\href {\doibase 10.1103/PhysRevLett.115.036402} {\bibfield  {journal} {\bibinfo  {journal} {Phys. Rev. Lett.}\ }\textbf {\bibinfo {volume} {115}},\ \bibinfo {pages} {036402} (\bibinfo {year} {2015})}\BibitemShut {NoStop}%
\bibitem [{\citenamefont {Price}, \citenamefont {Bryenton},\ and\ \citenamefont {Johnson}(2021)}]{Price2021}%
  \BibitemOpen
  \bibfield  {author} {\bibinfo {author} {\bibfnamefont {A.~J.~A.}\ \bibnamefont {Price}}, \bibinfo {author} {\bibfnamefont {K.~R.}\ \bibnamefont {Bryenton}}, \ and\ \bibinfo {author} {\bibfnamefont {E.~R.}\ \bibnamefont {Johnson}},\ }\bibfield  {title} {\enquote {\bibinfo {title} {{Requirements for an accurate dispersion-corrected density functional}},}\ }\href@noop {} {\bibfield  {journal} {\bibinfo  {journal} {The Journal of Chemical Physics}\ }\textbf {\bibinfo {volume} {154}},\ \bibinfo {pages} {230902} (\bibinfo {year} {2021})}\BibitemShut {NoStop}%
\bibitem [{\citenamefont {Huang}, \citenamefont {Braams},\ and\ \citenamefont {Bowman}(2006)}]{Huang2006}%
  \BibitemOpen
  \bibfield  {author} {\bibinfo {author} {\bibfnamefont {X.}~\bibnamefont {Huang}}, \bibinfo {author} {\bibfnamefont {B.~J.}\ \bibnamefont {Braams}}, \ and\ \bibinfo {author} {\bibfnamefont {J.~M.}\ \bibnamefont {Bowman}},\ }\bibfield  {title} {\enquote {\bibinfo {title} {Ab initio potential energy and dipole moment surfaces of (h2o)2},}\ }\href {\doibase 10.1021/jp053583d} {\bibfield  {journal} {\bibinfo  {journal} {The Journal of Physical Chemistry A}\ }\textbf {\bibinfo {volume} {110}},\ \bibinfo {pages} {445--451} (\bibinfo {year} {2006})},\ \bibinfo {note} {pMID: 16405316},\ \Eprint {http://arxiv.org/abs/https://doi.org/10.1021/jp053583d} {https://doi.org/10.1021/jp053583d} \BibitemShut {NoStop}%
\bibitem [{\citenamefont {Duan}, \citenamefont {Fisher},\ and\ \citenamefont {Kulik}(2023)}]{Duan2023}%
  \BibitemOpen
  \bibfield  {author} {\bibinfo {author} {\bibfnamefont {C.}~\bibnamefont {Duan}}, \bibinfo {author} {\bibfnamefont {K.}~\bibnamefont {Fisher}}, \ and\ \bibinfo {author} {\bibfnamefont {H.}~\bibnamefont {Kulik}},\ }\href {\doibase 10.5281/zenodo.10215421} {\enquote {\bibinfo {title} {{kefisher98/IP\_EA\_deltaSCF: Ionization Potential, Electron Affinity, and Delta SCF for Small Organic Molecules}},}\ } (\bibinfo {year} {2023})\BibitemShut {NoStop}%
\bibitem [{\citenamefont {Duan}\ \emph {et~al.}(2020)\citenamefont {Duan}, \citenamefont {Liu}, \citenamefont {Nandy},\ and\ \citenamefont {Kulik}}]{Duan2020}%
  \BibitemOpen
  \bibfield  {author} {\bibinfo {author} {\bibfnamefont {C.}~\bibnamefont {Duan}}, \bibinfo {author} {\bibfnamefont {F.}~\bibnamefont {Liu}}, \bibinfo {author} {\bibfnamefont {A.}~\bibnamefont {Nandy}}, \ and\ \bibinfo {author} {\bibfnamefont {H.~J.}\ \bibnamefont {Kulik}},\ }\bibfield  {title} {\enquote {\bibinfo {title} {Data-driven approaches can overcome the cost–accuracy trade-off in multireference diagnostics},}\ }\href {\doibase 10.1021/acs.jctc.0c00358} {\bibfield  {journal} {\bibinfo  {journal} {Journal of Chemical Theory and Computation}\ }\textbf {\bibinfo {volume} {16}},\ \bibinfo {pages} {4373--4387} (\bibinfo {year} {2020})}\BibitemShut {NoStop}%
\bibitem [{\citenamefont {Duan}\ \emph {et~al.}(2021)\citenamefont {Duan}, \citenamefont {Chen}, \citenamefont {Taylor}, \citenamefont {Liu},\ and\ \citenamefont {Kulik}}]{Duan2021}%
  \BibitemOpen
  \bibfield  {author} {\bibinfo {author} {\bibfnamefont {C.}~\bibnamefont {Duan}}, \bibinfo {author} {\bibfnamefont {S.}~\bibnamefont {Chen}}, \bibinfo {author} {\bibfnamefont {M.~G.}\ \bibnamefont {Taylor}}, \bibinfo {author} {\bibfnamefont {F.}~\bibnamefont {Liu}}, \ and\ \bibinfo {author} {\bibfnamefont {H.~J.}\ \bibnamefont {Kulik}},\ }\bibfield  {title} {\enquote {\bibinfo {title} {Machine learning to tame divergent density functional approximations: a new path to consensus materials design principles},}\ }\href {\doibase 10.1039/D1SC03701C} {\bibfield  {journal} {\bibinfo  {journal} {Chem. Sci.}\ }\textbf {\bibinfo {volume} {12}},\ \bibinfo {pages} {13021--13036} (\bibinfo {year} {2021})}\BibitemShut {NoStop}%
\bibitem [{\citenamefont {Brémond}\ and\ \citenamefont {Adamo}(2011)}]{PBE0_DH}%
  \BibitemOpen
  \bibfield  {author} {\bibinfo {author} {\bibfnamefont {E.}~\bibnamefont {Brémond}}\ and\ \bibinfo {author} {\bibfnamefont {C.}~\bibnamefont {Adamo}},\ }\bibfield  {title} {\enquote {\bibinfo {title} {{Seeking for parameter-free double-hybrid functionals: The PBE0-DH model}},}\ }\href {\doibase 10.1063/1.3604569} {\bibfield  {journal} {\bibinfo  {journal} {The Journal of Chemical Physics}\ }\textbf {\bibinfo {volume} {135}},\ \bibinfo {pages} {024106} (\bibinfo {year} {2011})}\BibitemShut {NoStop}%
\bibitem [{\citenamefont {{Adamo}}\ and\ \citenamefont {{Barone}}(1999)}]{PBE0}%
  \BibitemOpen
  \bibfield  {author} {\bibinfo {author} {\bibfnamefont {C.}~\bibnamefont {{Adamo}}}\ and\ \bibinfo {author} {\bibfnamefont {V.}~\bibnamefont {{Barone}}},\ }\bibfield  {title} {\enquote {\bibinfo {title} {{Toward reliable density functional methods without adjustable parameters: The PBE0 model}},}\ }\href {\doibase 10.1063/1.478522} {\bibfield  {journal} {\bibinfo  {journal} {\jcp}\ }\textbf {\bibinfo {volume} {110}},\ \bibinfo {pages} {6158--6170} (\bibinfo {year} {1999})}\BibitemShut {NoStop}%
\bibitem [{\citenamefont {Lee}, \citenamefont {Yang},\ and\ \citenamefont {Parr}(1988)}]{BLYP}%
  \BibitemOpen
  \bibfield  {author} {\bibinfo {author} {\bibfnamefont {C.}~\bibnamefont {Lee}}, \bibinfo {author} {\bibfnamefont {W.}~\bibnamefont {Yang}}, \ and\ \bibinfo {author} {\bibfnamefont {R.~G.}\ \bibnamefont {Parr}},\ }\bibfield  {title} {\enquote {\bibinfo {title} {Development of the colle-salvetti correlation-energy formula into a functional of the electron density},}\ }\href {\doibase 10.1103/PhysRevB.37.785} {\bibfield  {journal} {\bibinfo  {journal} {Phys. Rev. B}\ }\textbf {\bibinfo {volume} {37}},\ \bibinfo {pages} {785--789} (\bibinfo {year} {1988})}\BibitemShut {NoStop}%
\bibitem [{\citenamefont {De}\ \emph {et~al.}(2016)\citenamefont {De}, \citenamefont {Bart{\'{o}}k}, \citenamefont {Cs{\'{a}}nyi},\ and\ \citenamefont {Ceriotti}}]{De_2016}%
  \BibitemOpen
  \bibfield  {author} {\bibinfo {author} {\bibfnamefont {S.}~\bibnamefont {De}}, \bibinfo {author} {\bibfnamefont {A.~P.}\ \bibnamefont {Bart{\'{o}}k}}, \bibinfo {author} {\bibfnamefont {G.}~\bibnamefont {Cs{\'{a}}nyi}}, \ and\ \bibinfo {author} {\bibfnamefont {M.}~\bibnamefont {Ceriotti}},\ }\bibfield  {title} {\enquote {\bibinfo {title} {Comparing molecules and solids across structural and alchemical space},}\ }\href {\doibase 10.1039/c6cp00415f} {\bibfield  {journal} {\bibinfo  {journal} {Physical Chemistry Chemical Physics}\ }\textbf {\bibinfo {volume} {18}},\ \bibinfo {pages} {13754--13769} (\bibinfo {year} {2016})}\BibitemShut {NoStop}%
\bibitem [{\citenamefont {Deringer}\ \emph {et~al.}(2021)\citenamefont {Deringer}, \citenamefont {Bart\'{o}k}, \citenamefont {Bernstein}, \citenamefont {Wilkins}, \citenamefont {Ceriotti},\ and\ \citenamefont {Cs\'{a}nyi}}]{Deringer2021}%
  \BibitemOpen
  \bibfield  {author} {\bibinfo {author} {\bibfnamefont {V.}~\bibnamefont {Deringer}}, \bibinfo {author} {\bibfnamefont {A.}~\bibnamefont {Bart\'{o}k}}, \bibinfo {author} {\bibfnamefont {N.}~\bibnamefont {Bernstein}}, \bibinfo {author} {\bibfnamefont {D.}~\bibnamefont {Wilkins}}, \bibinfo {author} {\bibfnamefont {M.}~\bibnamefont {Ceriotti}}, \ and\ \bibinfo {author} {\bibfnamefont {G.}~\bibnamefont {Cs\'{a}nyi}},\ }\bibfield  {title} {\enquote {\bibinfo {title} {Gaussian process regression for materials and modelling},}\ }\href {\doibase 10.1021/acs.chemrev.1c00022} {\bibfield  {journal} {\bibinfo  {journal} {Chemical Reviews}\ }\textbf {\bibinfo {volume} {121}},\ \bibinfo {pages} {10073--10041} (\bibinfo {year} {2021})}\BibitemShut {NoStop}%
\bibitem [{\citenamefont {Musil}\ \emph {et~al.}(2021)\citenamefont {Musil}, \citenamefont {Grisafi}, \citenamefont {Bartók}, \citenamefont {Ortner}, \citenamefont {Csányi},\ and\ \citenamefont {Ceriotti}}]{Musil2021}%
  \BibitemOpen
  \bibfield  {author} {\bibinfo {author} {\bibfnamefont {F.}~\bibnamefont {Musil}}, \bibinfo {author} {\bibfnamefont {A.}~\bibnamefont {Grisafi}}, \bibinfo {author} {\bibfnamefont {A.~P.}\ \bibnamefont {Bartók}}, \bibinfo {author} {\bibfnamefont {C.}~\bibnamefont {Ortner}}, \bibinfo {author} {\bibfnamefont {G.}~\bibnamefont {Csányi}}, \ and\ \bibinfo {author} {\bibfnamefont {M.}~\bibnamefont {Ceriotti}},\ }\bibfield  {title} {\enquote {\bibinfo {title} {Physics-inspired structural representations for molecules and materials},}\ }\href {\doibase 10.1021/acs.chemrev.1c00021} {\bibfield  {journal} {\bibinfo  {journal} {Chemical Reviews}\ }\textbf {\bibinfo {volume} {121}},\ \bibinfo {pages} {9759--9815} (\bibinfo {year} {2021})}\BibitemShut {NoStop}%
\bibitem [{\citenamefont {Forrester}\ and\ \citenamefont {A.~S\'obester}(2007)}]{Forrester2007}%
  \BibitemOpen
  \bibfield  {author} {\bibinfo {author} {\bibfnamefont {A.~I.~J.}\ \bibnamefont {Forrester}}\ and\ \bibinfo {author} {\bibfnamefont {A.~J.~K.}\ \bibnamefont {A.~S\'obester}},\ }\bibfield  {title} {\enquote {\bibinfo {title} {Multi-fidelity optimization via surrogate modelling},}\ }\href {\doibase 10.1098/rspa.2007.1900} {\bibfield  {journal} {\bibinfo  {journal} {Proceedings of Royal Society A}\ }\textbf {\bibinfo {volume} {463}},\ \bibinfo {pages} {3251--3269} (\bibinfo {year} {2007})}\BibitemShut {NoStop}%
\bibitem [{\citenamefont {Reuther}\ \emph {et~al.}(2018)\citenamefont {Reuther}, \citenamefont {Kepner}, \citenamefont {Byun}, \citenamefont {Samsi}, \citenamefont {Arcand}, \citenamefont {Bestor}, \citenamefont {Bergeron}, \citenamefont {Gadepally}, \citenamefont {Houle}, \citenamefont {Hubbell}, \citenamefont {Jones}, \citenamefont {Klein}, \citenamefont {Milechin}, \citenamefont {Mullen}, \citenamefont {Prout}, \citenamefont {Rosa}, \citenamefont {Yee},\ and\ \citenamefont {Michaleas}}]{reuther2018interactive}%
  \BibitemOpen
  \bibfield  {author} {\bibinfo {author} {\bibfnamefont {A.}~\bibnamefont {Reuther}}, \bibinfo {author} {\bibfnamefont {J.}~\bibnamefont {Kepner}}, \bibinfo {author} {\bibfnamefont {C.}~\bibnamefont {Byun}}, \bibinfo {author} {\bibfnamefont {S.}~\bibnamefont {Samsi}}, \bibinfo {author} {\bibfnamefont {W.}~\bibnamefont {Arcand}}, \bibinfo {author} {\bibfnamefont {D.}~\bibnamefont {Bestor}}, \bibinfo {author} {\bibfnamefont {B.}~\bibnamefont {Bergeron}}, \bibinfo {author} {\bibfnamefont {V.}~\bibnamefont {Gadepally}}, \bibinfo {author} {\bibfnamefont {M.}~\bibnamefont {Houle}}, \bibinfo {author} {\bibfnamefont {M.}~\bibnamefont {Hubbell}}, \bibinfo {author} {\bibfnamefont {M.}~\bibnamefont {Jones}}, \bibinfo {author} {\bibfnamefont {A.}~\bibnamefont {Klein}}, \bibinfo {author} {\bibfnamefont {L.}~\bibnamefont {Milechin}}, \bibinfo {author} {\bibfnamefont {J.}~\bibnamefont {Mullen}}, \bibinfo {author} {\bibfnamefont {A.}~\bibnamefont {Prout}}, \bibinfo {author} {\bibfnamefont {A.}~\bibnamefont {Rosa}}, \bibinfo
  {author} {\bibfnamefont {C.}~\bibnamefont {Yee}}, \ and\ \bibinfo {author} {\bibfnamefont {P.}~\bibnamefont {Michaleas}},\ }\bibfield  {title} {\enquote {\bibinfo {title} {Interactive supercomputing on 40,000 cores for machine learning and data analysis},}\ }in\ \href@noop {} {\emph {\bibinfo {booktitle} {2018 IEEE High Performance extreme Computing Conference (HPEC)}}}\ (\bibinfo {organization} {IEEE},\ \bibinfo {year} {2018})\ pp.\ \bibinfo {pages} {1--6}\BibitemShut {NoStop}%
\bibitem [{\citenamefont {Civalleri}\ \emph {et~al.}(2012)\citenamefont {Civalleri}, \citenamefont {Presti}, \citenamefont {Dovesi},\ and\ \citenamefont {Savin}}]{dft_bestdfa}%
  \BibitemOpen
  \bibfield  {author} {\bibinfo {author} {\bibfnamefont {B.}~\bibnamefont {Civalleri}}, \bibinfo {author} {\bibfnamefont {D.}~\bibnamefont {Presti}}, \bibinfo {author} {\bibfnamefont {R.}~\bibnamefont {Dovesi}}, \ and\ \bibinfo {author} {\bibfnamefont {A.}~\bibnamefont {Savin}},\ }\bibfield  {title} {\enquote {\bibinfo {title} {On choosing the best density functional approximation},}\ }in\ \href {\doibase https://doi.org/10.1039/9781849734790-00168} {\emph {\bibinfo {booktitle} {Uncertainty Quantification in Multiscale Materials Modeling}}},\ \bibinfo {editor} {edited by\ \bibinfo {editor} {\bibfnamefont {M.}~\bibnamefont {Springborg}}}\ (\bibinfo  {publisher} {RSC Publishing},\ \bibinfo {year} {2012})\ Chap.~\bibinfo {chapter} {6}, pp.\ \bibinfo {pages} {168--185}\BibitemShut {NoStop}%
\bibitem [{\citenamefont {Himanen}\ \emph {et~al.}(2020)\citenamefont {Himanen}, \citenamefont {J{\"a}ger}, \citenamefont {Morooka}, \citenamefont {Federici~Canova}, \citenamefont {Ranawat}, \citenamefont {Gao}, \citenamefont {Rinke},\ and\ \citenamefont {Foster}}]{dscribe}%
  \BibitemOpen
  \bibfield  {author} {\bibinfo {author} {\bibfnamefont {L.}~\bibnamefont {Himanen}}, \bibinfo {author} {\bibfnamefont {M.~O.~J.}\ \bibnamefont {J{\"a}ger}}, \bibinfo {author} {\bibfnamefont {E.~V.}\ \bibnamefont {Morooka}}, \bibinfo {author} {\bibfnamefont {F.}~\bibnamefont {Federici~Canova}}, \bibinfo {author} {\bibfnamefont {Y.~S.}\ \bibnamefont {Ranawat}}, \bibinfo {author} {\bibfnamefont {D.~Z.}\ \bibnamefont {Gao}}, \bibinfo {author} {\bibfnamefont {P.}~\bibnamefont {Rinke}}, \ and\ \bibinfo {author} {\bibfnamefont {A.~S.}\ \bibnamefont {Foster}},\ }\bibfield  {title} {\enquote {\bibinfo {title} {{DScribe: Library of descriptors for machine learning in materials science}},}\ }\href {\doibase 10.1016/j.cpc.2019.106949} {\bibfield  {journal} {\bibinfo  {journal} {Computer Physics Communications}\ }\textbf {\bibinfo {volume} {247}},\ \bibinfo {pages} {106949} (\bibinfo {year} {2020})}\BibitemShut {NoStop}%
\end{thebibliography}%

\appendix
\section{\label{opt}\rev{Lengthscale and Variance Optimization}}

    \begin{figure*}
        \begin{tabular}{cc}
        \includegraphics[width=0.4\linewidth]{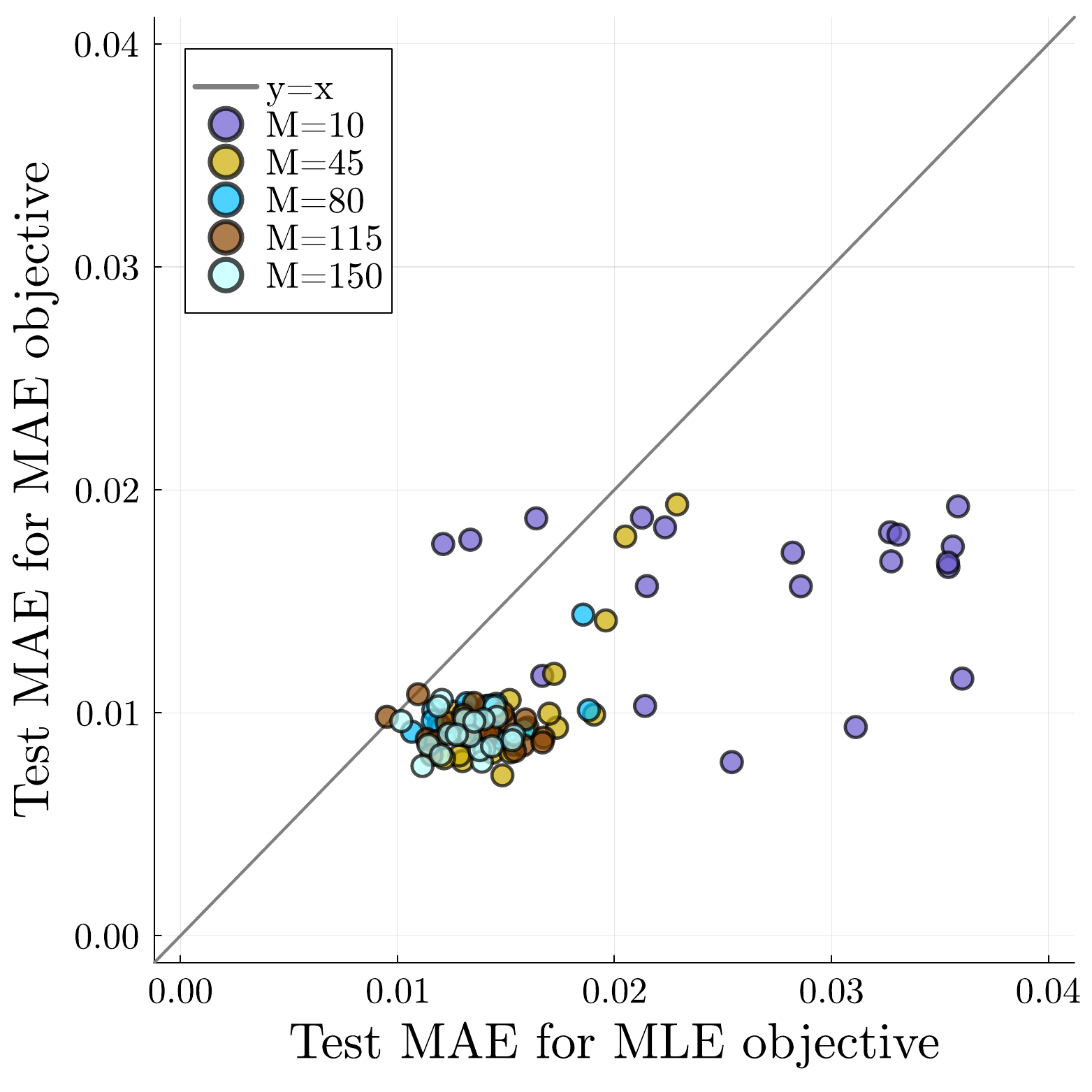}  &
        \includegraphics[width=0.4\linewidth]{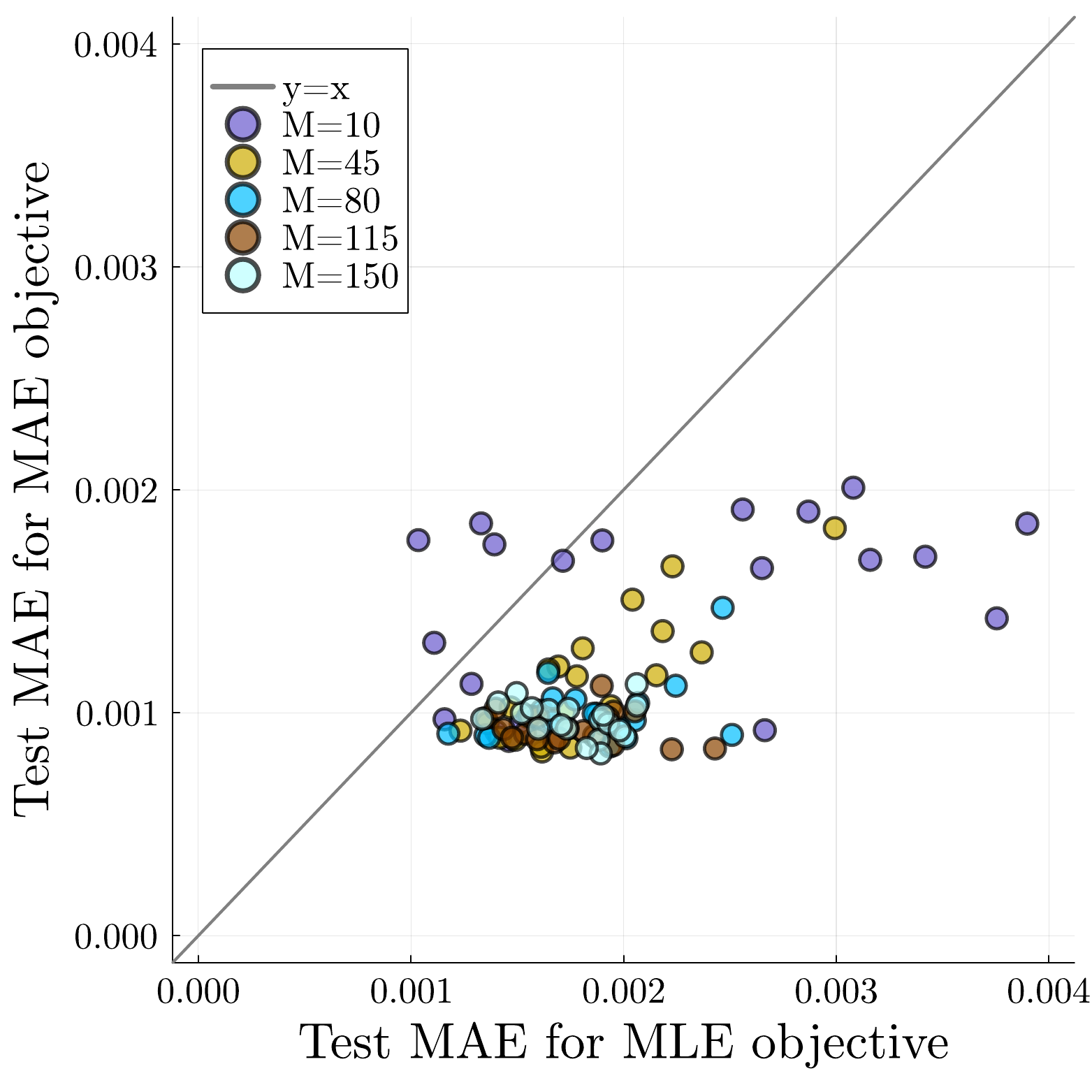} \\
        (a) Primary hyperparameters & (b) Secondary hyperparameters \\
        \end{tabular}
        \caption{\label{fig:optimization} \rev{\textbf{Water Trimer Case.} Each point compares the test MAE of a GP model which uses the MLE variance to the test MAE of a model which uses a variance estimate obtained by minimizing MAE on the optimization set. Different data points are obtained by shuffling the optimization data, and color indicates the size of the optimization data set. In plot (a) we consider a GP model for the primary regression task while in plot (b) we consider a model $\delta_{s_n}$, as defined in \eqref{eq:mt_relate}. }}
    \end{figure*}

\rev{
This section provides further detail on our procedure for hyperparameter optimization with a focus on computational tractability. We note that further improvements are possible and worth investigating in future work.
Specifically, we discuss optimization
for the test case explored in Subsections \ref{sec:num1} and \ref{sec:num2} to
estimate the hyperparameters of the prior distributions for $f_p$ and
$\delta_{s_1}$, given by \eqref{eq:multitask_prior1} and
\eqref{eq:multitask_prior2}. Both priors are constructed with the squared exponential kernel:}
\begin{eqnarray*}
    \rev{k(\bm{X},\bm{X}')}  \ \rev{=} \ \rev{v \ \exp\left( \frac{-(\bm{X}-\bm{X}')^T(\bm{X}-\bm{X}')}{2\ell} \right) }
\end{eqnarray*}
\rev{
As discussed in \ref{ss:h_opt}, we optimize hyperparameters using a data
set of $M=150$ molecular configurations. For all statistical models used in
Subsections \ref{sec:num1} and \ref{sec:num2}, the lengthscale, $\ell$, is
obtained through a maximum likelihood procedure while the variance, $v$, is
taken to be the value which minimizes MAE on the optimization set. The
latter choice was made because, for the three body energy data, variance
estimates produced by minimizing MAE achieve lower test error than the
maximum likelihood estimate of variance. 
Fig.~\ref{fig:optimization} establishes that these choices of $M$ and variance
estimate are not necessary for the multitask model to perform well; in fact,
the model appears relatively robust to reasonable optimization procedures.}

\rev{Each color in Fig.~\ref{fig:optimization} represents a different
optimization set size, $M$. We obtain multiple hyperparameter estimates for
each $M$ by shuffling the optimization data set. For each shuffled realization
of the data, we solve for estimates $\hat{\ell}_{MLE}$, $\hat{v}_{MLE}$, and
$\hat{v}_{MAE}$. The plotted points compare MAE on a test data set of $320$
molecular configurations obtained using $\hat{\ell}_{MLE}$ and $\hat{v}_{MLE}$
to MAE for the same test data set obtained using $\hat{\ell}_{MLE}$ and
$\hat{v}_{MAE}$. We can see that both approaches to estimating variance lead to
good predictive performance, but use of $\hat{v}_{MAE}$ generally produces
lower MAE than use of $\hat{v}_{MLE}$ for both $f_p$ and $\delta_{s_1}$.
Fig.~\ref{fig:optimization} also demonstrates that the optimization data set
size can be reduced without significant reductions in prediction accuracy
though there is greater variance in prediction quality for $M\leq 45$. }

\rev{ Note that given $\hat{\ell}_{MLE}$, the expression for $\hat{v}_{MLE}$ is closed form and $\hat{v}_{MAE}$ is found using $\hat{\ell}_{MLE}$. Thus, we only
optimize for one hyperparameter at a time. If the optimization procedure
reaches a threshold of 30 seconds, we halt the procedure, reshuffle the data,
and restart. In the case that $M=150$, we require on the order of $10$ seconds
on a laptop to successfully obtain MLE estimates. MAE estimates are found in
$\sim 100$ seconds for the primary task and $\sim 40$ seconds for the secondary
task. The reported amounts include any time spent on false starts which time
out at $30 s$. By comparison, a single DFT calculation for 3-b energy using a
SCAN functional approximation requires $\sim 650$ seconds on a cluster. Thus,
compared with data generation, hyperparameter optimization accounts for a small
portion of the computational budget of our numerical experiments. Our results
in Fig.~\ref{fig:optimization} suggest that we can construct successful
statistical models even when the budget for generating the optimization data
set is much smaller than in our work. For any $M$, the cost of generating the
optimization set will be dominated by CCSD(T) computations which are necessary
for all statistical models which we implement. Future work may fully evaluate
the minimum cost required to produce hyperparameters which will achieve a
specified threshold of accuracy and investigate strategies for constructing
opportunistic optimization sets from existing data.}

\section{\label{cost}Training Data Cost Model}

In our numerical results, we estimate the training cost as
\begin{align*}
    \text{model training cost} \ = \ \sum_\ell^L n_\ell \hat{c}_\ell
\end{align*}
where $n_\ell$ is the number of training systems from the $\ell^{th}$ level of theory and $\hat{c}_\ell$ is an estimate of prediction cost for that level of theory. We produce one set of estimates for $\hat{c}_1,\dots,\hat{c}_L$ for the water trimer example and a different set of estimates for the small organic molecules example. 

For the water trimers example, we randomly selected ten molecular configurations from our data set as the basis for our cost model. Using Psi4, we computed the 3-b interaction energy at the level of SCAN and CCSD(T) for each configuration. We applied a counterpoise correction to ameliorate basis set superposition error.  All computations for the water cost model where performed on physically the same Intel Xeon Platinum 8260 node of MIT SuperCloud (\href{https://supercloud.mit.edu/systems-and-software}{https://supercloud.mit.edu/systems-and-software}). All computations for a given molecular configuration where performed on the same \rev{core}. For this case study, we estimate that a CCSD(T) calculation incurs $37$ times the cost of a DFT calculation based on the average prediction time of these methods on our example molecular configurations. 
 
Several \rev{systems} in the small organic molecules data set \rev{contain more
atoms} than the water trimer configurations. The median number of electrons
\rev{per system in the organic molecules} set is $40$, the minimum is $14$, and
the maximum is $50$.
\rev{We estimate the cost of level of theory $\ell$ as the computational time
required to obtain the total energy at this level of theory averaged over} ten randomly
sampled configurations with $36$ electrons. Note that \rev{this estimate is
conservative:} CCSD(T) scales as $N_e^7$ where $N_e$ is the number of electrons
in a system while DFT scales as $N_e^3$. Therefore, an estimate of the ratio
between the cost of CCSD(T) and the cost of DFT based on $36$ electrons will be
smaller than the actual ratio for most molecular configurations in our data
set. Furthermore, our prediction of the ionization potential was performed
using ${\Delta}CC$ (delta coupled cluster)\rev{, which requires multiple
CCSD(T) calculations to reach an IP prediction.} Thus, while our reported results
(Fig.~\ref{fig:ip_levels}) estimate that CCSD(T) calculations demand $244$
times the expense of DFT, the true ratio between the cost of CCSD(T) and DFT is
even larger. \rev{Consequently the reported gains in efficiency for the multitask
method are expected to be a lower bound on the true improvement.}

\section{\label{SOAP}SOAP parameters}

To fix reasonable values for parameters required by SOAP ($r_{cut}$, $\sigma_{atom}$, $n_{max}$, $l_{max}$), we refer to established conventions as well as experimentation. The cutoff radius, $r_{cut}$ controls the size of each local neighborhood, and a small value may lead to lost geometric insight. Unfortunately, surveys on feature construction~\cite{Deringer2021,Musil2021} hold that a large cutoff radius does not necessarily provide proportionate insight: increasing radii larger than 6-8 \r{A} is rarely if ever useful. Our choice of $\sigma_{atom}$ also influences our model of each atom's neighborhood: it determines the lengthscale of the Gaussian shapes located on each of the surrounding atoms. The larger our choice of $\sigma_{atom}$, the more chance that Gaussian tails will slip passed the $r_{cut}$ border. Current literature~\cite{Deringer2021} indicates that the best practice when working with the first three rows of the periodic table is to use $\sigma_{atom}=0.3 \ \si{\angstrom}$ for systems with hydrogen and $\sigma_{atom}=0.5 \ \si{\angstrom}$ for systems without. Finally, we choose the parameters $n_{max}$ and $l_{max}$ to control the size of our expansions of the local neighborhoods. Generally~\cite{Deringer2021}, choosing $n_{max}=12$ with $l_{max}=6$ is sufficient for high accuracy. In practice, the best values for $n_{max}$ and $l_{max}$ will be sensitive to the choice of radial basis set. 

    \begin{figure}
    \centering
    \includegraphics[width=0.99\linewidth]{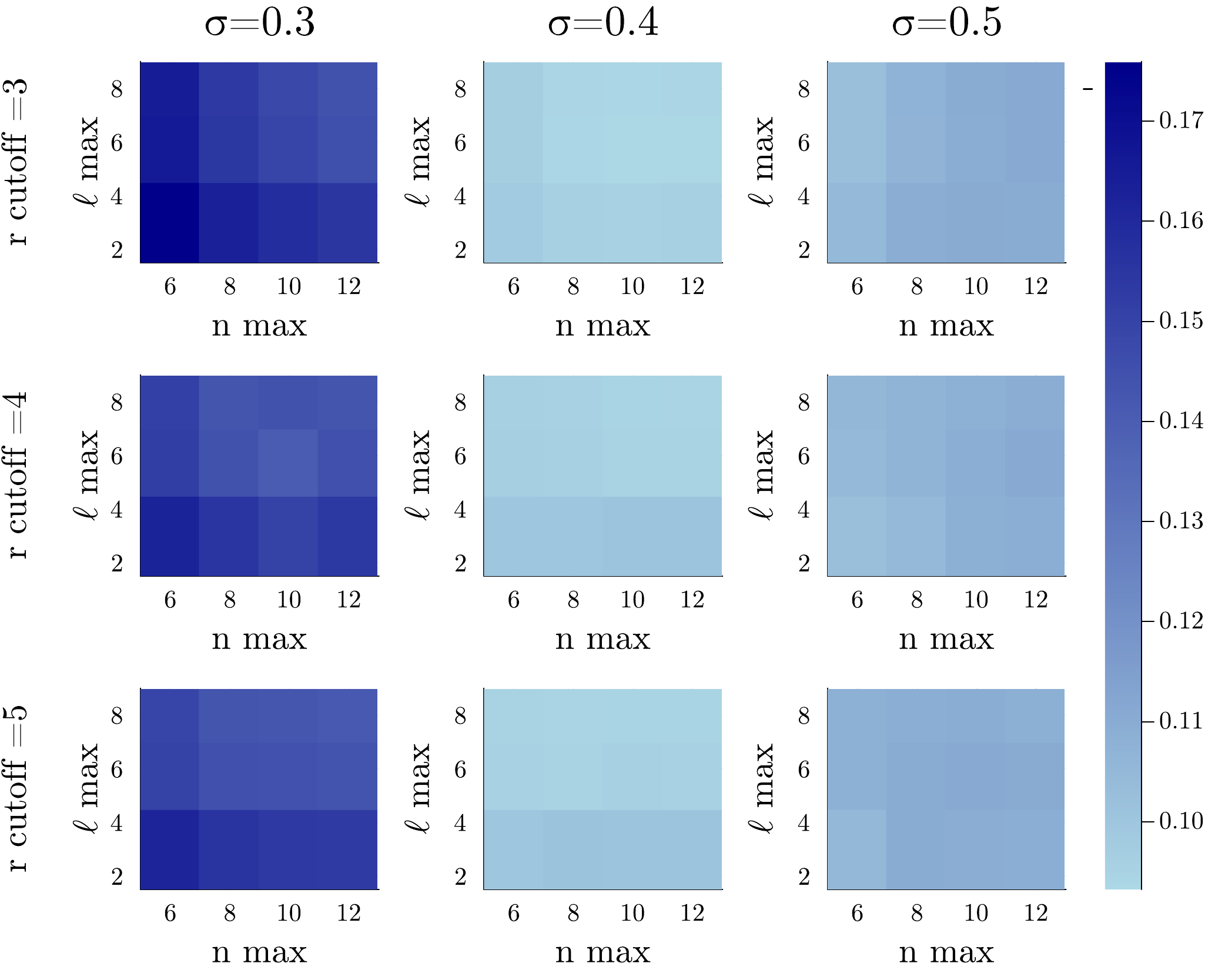}
    \caption{Example of inference results for a range of SOAP parameters. The colorbar gives the mean absolute error of predictions made for the ionization potential of small organic molecules. Both $r_{cut}$ and $\sigma_{atom}$ have units of \si{\angstrom}.}
    \label{fig:soap}
    \end{figure}

We selected reasonable candidate values for each SOAP parameter and tested the performance of the resultant features in Gaussian process regression. All features for this work were calculated using the DScribe Python package.~\cite{dscribe} Fig.~\ref{fig:soap} shows the mean absolute error (MAE) obtained with different SOAP parameters when GP regression trained with PBE level data is used to predict ionization potential (IP). Tests on CCSD(T) and PBE0 data produced similar results. Our tests include all combinations of $r_{cut}\in\{3, 4, 5\}$ , $\sigma_{atom}\in\{0.3, 0.4 ,0.5 \}$, $l_{max}\in\{2,4,6,8\}$, and $n_{max}\in\{6,8,10,12\}$. Units for $r_{cut}$ and $\sigma_{atom}$ are \si{\angstrom}. We find that our accuracy is comparable to probabilistic predictions of IP found in literature.~\cite{De_2016} Of the SOAP parameters, the choice of $\sigma_{atom}$ has the largest impact on predictive performance. 

For our numerical experiments in this work, we fix $\sigma_{atom}=0.4 \ \si{\angstrom}$ because this choice demonstrated the best overall performance across our tests. The other parameters are chosen to balance reasonable accuracy and cost. We fix both $l_{max}$ and $n_{max}$ at 8, and for the small organic molecules example, we use $r_{cut}=4 \ \si{\angstrom}$. For water trimers, we use $r_{cut}=10 \ \si{\angstrom}$ to capture all three component water molecules.

\section{\label{features}Global Features}

    \begin{figure*}
    \centering
    \includegraphics[trim={0 4.5cm 0 5cm},clip,width=0.95\linewidth]{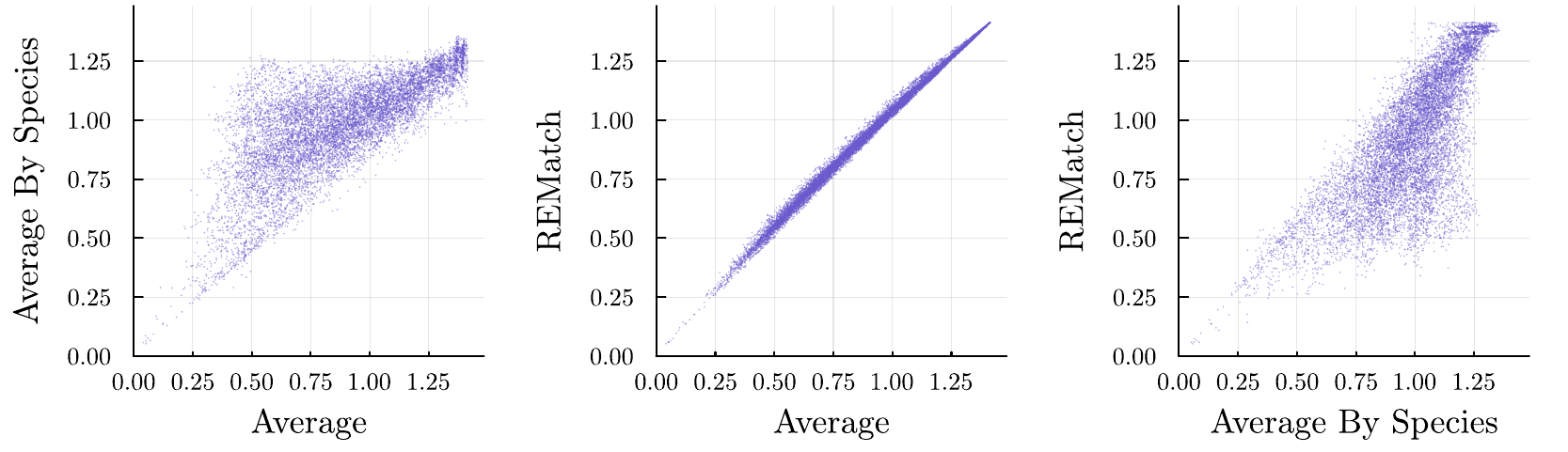}
    \caption{Correlation between different methods of constructing global features when calculating the distance between molecule pairs. \cite{De_2016}}
    \label{fig:rematch}
    \end{figure*}

When using SOAP to compare two molecular systems, we must make modeling choices to standardize our representation of entire systems. SOAP features are constructed based on the neighborhood of each atom in a system, so challenges arise when two systems contain different numbers of constituent atoms of elements. There are multiple strategies for constructing ``global'' features to capture entire systems.~\cite{De_2016} The simplest approach is to average the SOAP features for each atom in the system. It is possible the loss of information from averaging may make it challenging to distinguish between the features of similar molecular systems. A variation to this approach averages the local features corresponding to each element contained in a molecular system.~\cite{De_2016} If we wish to compare systems A and B, and system B contains more elements than A, we insert SOAP features corresponding to isolated atoms of those excess elements in the representation of A. This model suggests that an atom of such an element does not interact with the rest of the system. A Regularized entropy match (REMatch) strategy has also been proposed for constructing global features.~\cite{De_2016} This approach solves a regularized optimization problem to find the match between the sets of local features for each molecular system which maximizes information entropy.

With the same set of small organic molecules which we use to test SOAP parameters, we investigate how well the global features constructed by different approaches agree. For pairs of molecules, we construct global features using each of the three approaches outlined in the previous paragraph. We represent difference between each pair by squaring an inner product of their features. This procedure is the same as using a polynomial kernel of degree 2. By comparing the output of the kernel for different featurization strategies, we determine whether the strategies generally agree about which molecules are similar. 
    
Fig.~\ref{fig:rematch} shows the correlation in kernel outputs corresponding to each pair of molecules for different global featurization strategies. Consistently, in our tests, we find strong correlation (Pearson's coefficient $> 0.99)$ between the REMatch features and features computed by averaging all local representations. These two strategies are also both positively correlated with the ``Average by Species'' approach which produces an averaged feature for each element in the system and inserts isolated atoms to represent elements not included in the system, but these correlations are not as strong as that between the REMatch and totally averaged features. We may see this result because the REMatch and totally averaged features share the same dimension, $n$, whereas features that are constructed by species-specific averaging have dimension of $2n$ to $4n$ in the cases we considered. Because averaging features is much more computationally efficient than constructing REMatch features, we use this approach in our numerical experiments.

\end{document}